\documentclass{emulateapj}
\usepackage{times}
\usepackage[]{graphicx,epsfig,rotating}
\usepackage[pdftex]{hyperref} 

\shorttitle{ Visual Catalog of Sloan Galaxies}
\shortauthors{Nair$~$\&$~$Abraham}

\begin{document}

% Set the bibliography style here
\bibliographystyle{apj}
%\bibliography{astroshort}

\title{ A Catalog of Detailed Visual Morphological Classifications for 14034 Galaxies in the Sloan Digital Sky Survey}
%\author{\medskip Preethi B. Nair\altaffilmark{\S}\  \&\  Roberto G. Abraham}
%\affil{
%   Department of Astronomy \& Astrophysics,
%   University of Toronto, 50 St. George Street,
%   Toronto, ON, M5S~3H4.
%}
%\altaffiltext{\S}{Present address: INAF - Astronomical Observatory of Bologna, Via Ranzani 1, I - 40127 Bologna, ITALY}
\author{\medskip Preethi B Nair\footnote{Present address: INAF - Astronomical Observatory of Bologna, Via Ranzani 1, 40127 Bologna, ITALY}   \&\ Roberto G Abraham}
\affil{
   Department of Astronomy \& Astrophysics,
   University of Toronto, 50 St. George Street,
   Toronto, ON, M5S~3H4.
}
\email{preethi.nair@oabo.inaf.it,abraham@astro.utoronto.ca}
\slugcomment{Accepted by ApJ Supplement Series}

\begin{abstract}
We present a catalog of detailed visual classifications for 14034 galaxies in the Sloan Digital Sky Survey (SDSS) Data Release 4 (DR4). Our sample includes nearly all spectroscopically-targeted galaxies in the redshift range $0.01<z<0.1$ down to an apparent extinction-corrected limit of $g<16$ mag. In addition to T-Types we record the existence of bars, rings, lenses, tails, warps, dust lanes, arm flocculence and multiplicity. This sample defines a comprehensive local galaxy sample which we will use in future papers to study low redshift morphology. It will also prove useful for calibrating automated galaxy classification algorithms. In this paper we describe the classification methodology used, detail the systematics and biases of our sample and summarize the overall statistical properties of the sample, noting the most obvious trends that are relevant for general comparisons of our catalog with previously published work. 
\end{abstract}
 

\section{INTRODUCTION}

\label{sec:introduction}
Considerable progress has been made in developing techniques for automated classification of galaxies. However, the disappointing fact is that, at present, state-of-the-art automated galaxy classification is only capable of delivering crude classifications, albeit very quickly. The large numbers of classifications delivered by automated techniques have proved highly useful in their appropriate context, but they are neither as accurate, nor as comprehensive, as visual classifications made by a trained observer. 

The strengths and limitations of present machine-based galaxy classification techniques should not
come as a surprise if one reflects upon the fact that something like half the human brain is devoted to vision. Millions of years of evolution have refined our brain's capacity for image processing and analysis to the extent that a well-trained human can classify an individual image more accurately than a computer-based algorithm in essentially every field of science or industry in which comparisons have been made. For example, even state-of-the-art automated facial recognition systems cannot presently identify a human face with the accuracy routinely delivered by cursory visual inspection. Nevertheless, the basic simplicity of galaxies' structural forms suggests there is considerable room for improving automated classifications using more sophisticated tools for pattern recognition. An important step toward this goal is the development of a robust training set of digital galaxy images with detailed visual classifications. Such a training set would also prove useful for many programs of investigation which probe the systematics of galactic structure.
 
Inspired by the evident need for a catalog of detailed visual morphological classifications of local galaxies, we have undertaken to classify a $g$-band apparent magnitude limited sample of 14034 galaxies from Sloan Digital Sky Survey (SDSS) Data Release 4. Our catalog is six times larger than the catalog of visual classifications presented in \cite{Fukugita:2007p3031}. However our goal is to not just increase the size of our sample relative to previous work, but also to take visual classifications in the SDSS to a more detailed level reminiscent of earlier generations of catalogs (e.g. the Hubble Atlas \citep{Sandage:1961p5247}, the Revised Shapley-Ames Catalog \citep{Sandage:1981p5045}, and the de Vaucouleurs Atlas \citep{deVaucouleurs:1963p9392,deVaucouleurs:1991p4597}). In addition to T-Types, our catalog attempts to record subtle morphological features such as the existence and prominence of bars, rings, lenses, tails, shells, warps and dust lanes, and the nature of spiral structure (arm flocculence and multiplicity).

Our catalog is presented with two main goals in mind:
\begin{enumerate}

\item We seek to provide a generally useful comprehensive local sample with highly detailed morphological classifications. Future papers in this series will use this catalog to explore the local abundance and evolutionary properties of rings, bars, ansae, spiral structure, dust
and tidal features in galaxies.

\item The catalog is a starting point for future attempts to improve automated galaxy classification by incorporating detection algorithms for subtle morphological features. With the exception of some simple attempts to automate the detection of galactic bars, to date no automated classification method attempts to characterize the `fine structure' of galaxy morphology.

\end{enumerate}

A plan for this paper follows. In Section 2 we describe our sample, and note its strengths and limitations. 
Section 3 describes our classification methodology and Section 4
presents a number of image montages which illustrate our classification scheme. Section 5 compares our
classifications against those from other studies in order to estimate the reliability of our catalog. 
Systematic effects introduced by seeing, inclination and distance are explored in Section 6.
The catalog itself is presented in Section 7, which is the heart of this paper. Summary statistics for our sample as a whole are presented in Section 8. We conclude in Section 9. 
Throughout this paper we assume a flat $\Lambda$-dominated cosmology with 
$h=0.7$, $\Omega_M=0.3$ and $\Omega_\Lambda=0.7$.\\

\begin{figure*}[htbp]
\unitlength1cm
\hspace{1cm}
\vspace{0.3cm}
\begin{minipage}[t]{4.0cm}
\rotatebox{0}{\resizebox{7cm}{5cm}{\includegraphics{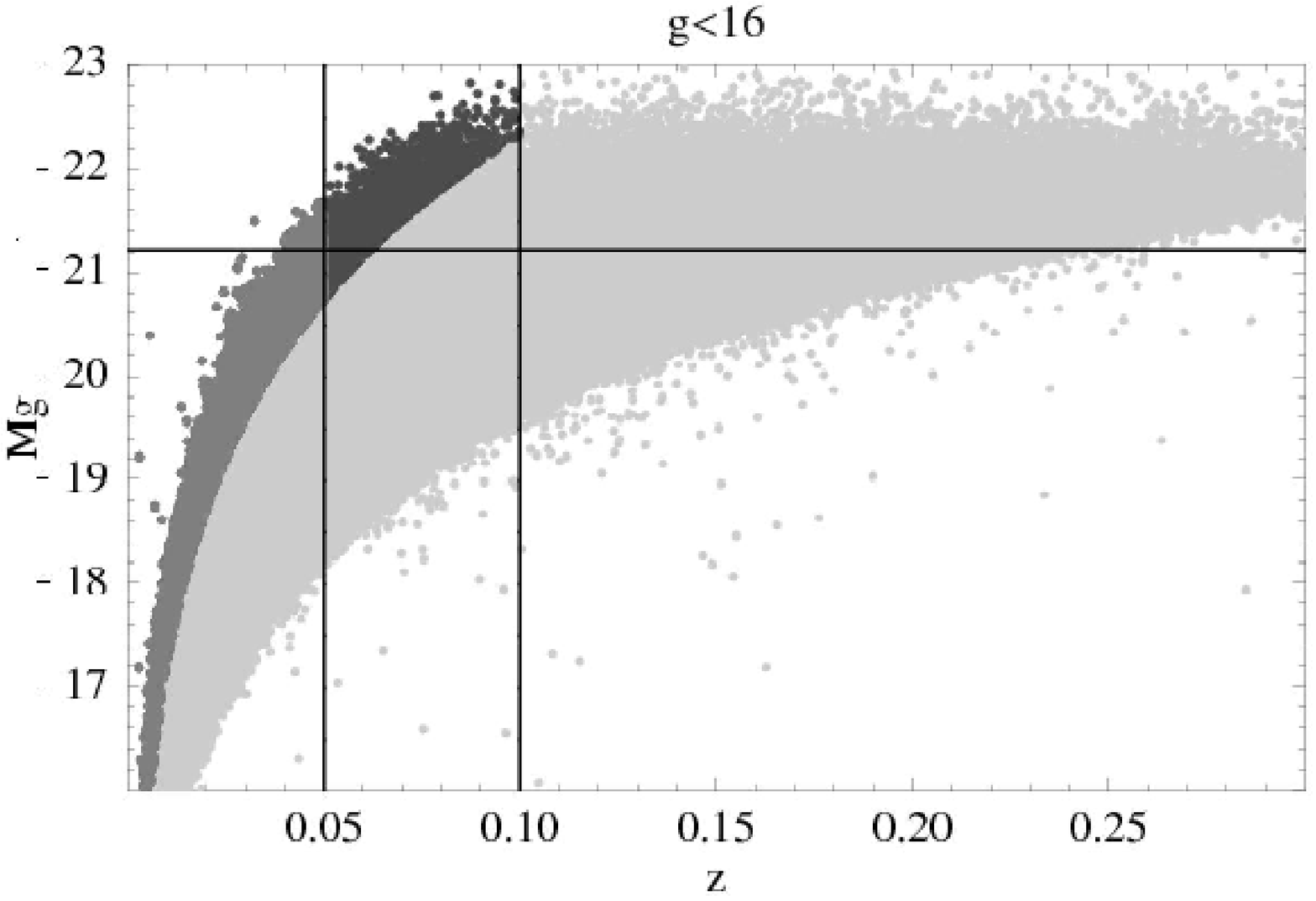}}}
\end{minipage}
\hspace{4cm}
\begin{minipage}[t]{4.0cm}
\rotatebox{0}{\resizebox{7cm}{5cm}{\includegraphics{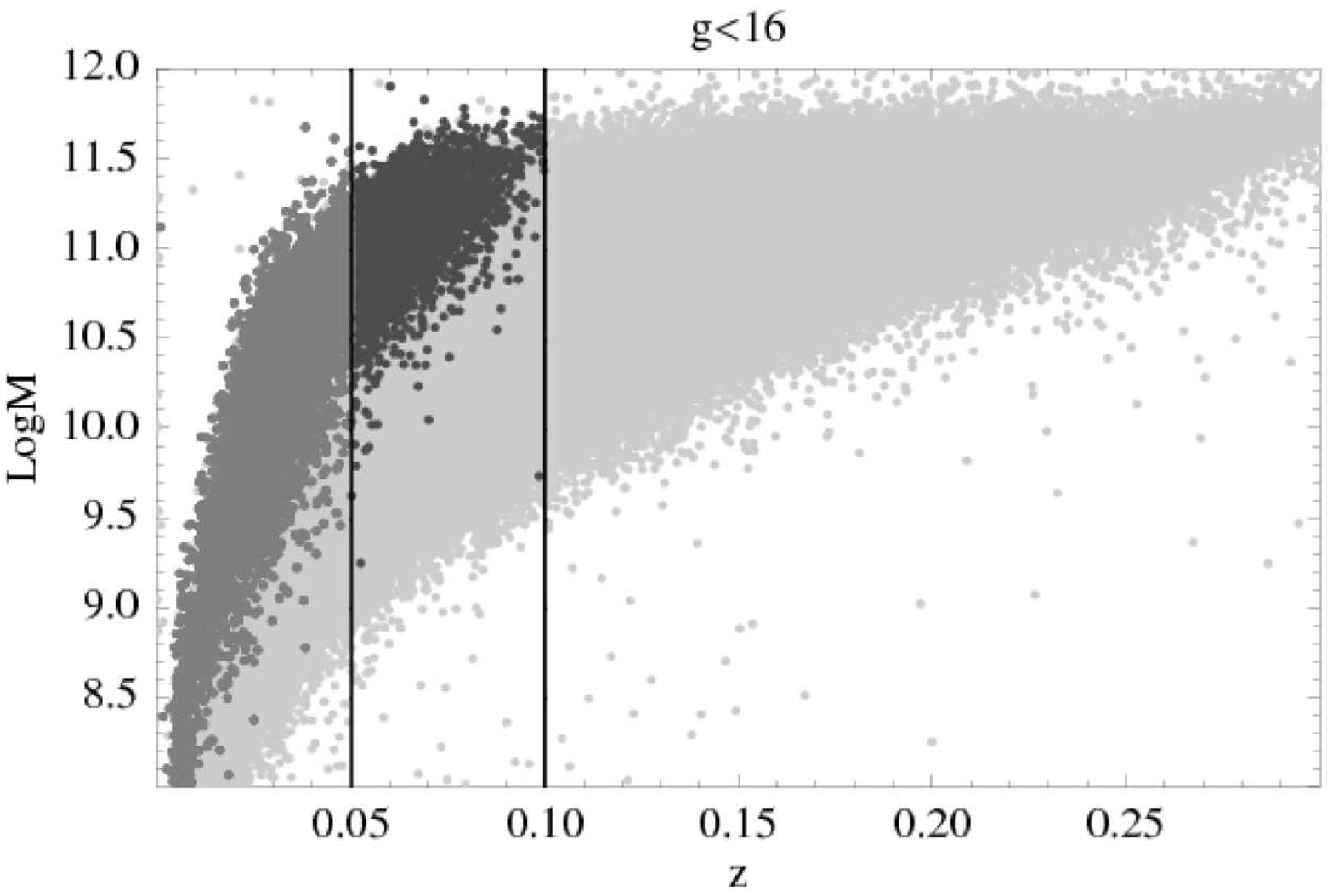}}}
\end{minipage}
\caption{\label{fig:AbsMagRedshiftDistribution} 
[Left] Absolute magnitude vs. redshift for our sample of 14034 SDSS galaxies with $g<16$ mag. The horizontal line corresponds to the
characteristic magnitude $M_\star$ of the SDSS luminosity function determined by \cite{Blanton:2003p2277}. Vertical lines are the redshift cuts at $z = 0.05$ and $z = 0.1$ described in the text. The light grey points show the distribution of the entire Data Release 4 main galaxy sample up to $z<0.3$. The intermediate grey points show the distribution of the sample with $z<0.05$ while the dark grey points show the distribution with $0.05<z<0.1$. [Right] Logarithm of the stellar mass (from \cite{Kauffmann:2003p97}) as a function of redshift.
As expected from the left-hand panel, we see the brightest and most massive galaxies dominate our sample at z$>$0.05.
}
\end{figure*}

\begin{figure*}[htbp]
\begin{center}
\includegraphics[width=3.0in]{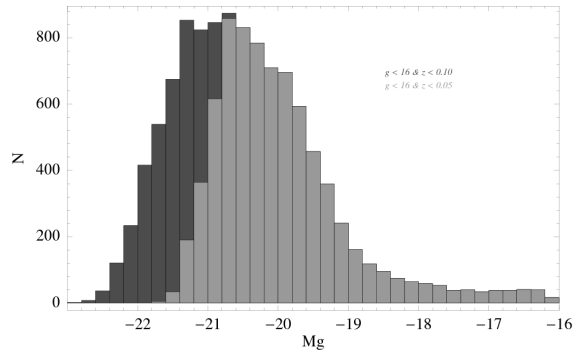} 
\caption{\label{fig:gAbsMagDistribution} 
Number of galaxies vs. absolute g-band magnitude for our sample of 14034 SDSS galaxies with 0.01$<$z$<$0.1 and $g<16$ mag (extinction corrected). The light grey area shows the distribution of the sample with $z<0.05$ while the dark grey area shows the distribution with $0.05<z<0.1$. Note that the bulk of the bright galaxy population is missed with a $z<0.05$ cut. See text for details on sample selection.
}
\end{center}
\end{figure*}

\section{Sample}
\label{sec:sample}

\subsection{Sample Selection}
The local sample presented here is derived from the SDSS \citep{Stoughton:2002p1611,York:2000p3192,AdelmanMcCarthy:2006p4486} DR4 release which covers 6670 sq. deg in the \em{u',g',r',i',z'} \rm filters. (The central wavelengths of these filters are 3553 \AA, 4686 \AA, 6166 \AA, 7480 \AA, and 8932 \AA, respectively). The galaxies are selected from the spectroscopic main sample described in \cite{Strauss:2002p6172}. We use the DR4 photometry catalogs to select all objects with an extinction corrected {\em g'} band magnitude brighter than 16 mag at redshifts $z$ less than $z=0.1$, which gives us a sample of 14700 galaxies. Objects mistakenly classified as galaxies 
%or too close to a star such that diffraction spikes run through them 
have been removed. The current catalog also excludes all objects at redshifts below $z \thicksim 0.01$, because extraction of extended galaxies for quantitative analysis at such low redshifts is difficult on the basis of SDSS Atlas pipeline images which leaves us with a sample of 14034 galaxies.

The upper redshift limit of $z=0.1$ in our sample is chosen in order to exclude objects where morphological detail is difficult to detect because of distance effects. As will be discussed next, to some extent the choice of this upper redshift limit is a balance between seeking to include a fair sample of massive galaxies while at the same time recognizing the inherent limitations on the classification of faint galaxies imposed by variable seeing and relatively low signal-to-noise in the SDSS imaging data.

Figure~\ref{fig:AbsMagRedshiftDistribution} shows the absolute magnitude and stellar mass distribution of galaxies (see \S2.2 for a discussion of these mass estimates) as a function of redshift for SDSS DR4 (light gray) along with our sub-sample of DR4 galaxies (dark gray). The public-domain {\tt K-Correct} software (version 4; \citet[]{Blanton:2005p79}) was used to derive the K-corrections needed to produce this figure. At a redshift of $z\sim 0.05$ the SDSS images probe a physical resolution similar to that probed by the {\em Hubble Space Telescope} (HST) viewing galaxies at $z \thicksim1$. We hope our classifications will be useful as a complement to HST evolutionary studies out to at least $z\sim1$, so in some sense $z\sim0.05$ (instead of $z\sim0.1$ as we have chosen) would have been a sensible upper limit for our classifications, because beyond this redshift the rest-frame resolution of the SDSS data is actually poorer than that of HST probing $z=1$ galaxies\footnote{
The HST Advanced Camera for Survey's image scale of  0.04\arcsec/pixel corresponds to a physical scale of
0.32 kpc/pixel at $z=1$.
The SDSS telescope's 0.4\arcsec/pixel corresponds to 0.74 kpc/pixel at z$\sim$0.1. 
}
However, Figure~\ref{fig:AbsMagRedshiftDistribution} shows that making a cut at $z=0.05$ would have resulted in our missing most of the bright, massive galaxy population (systems brighter than $M_*$, the characteristic magnitude of the luminosity function). This is further emphasized by Figure~\ref{fig:gAbsMagDistribution} which shows the distribution of absolute magnitudes in our sample for $z\le0.05$ (light gray) and $z\le0.1$ (dark gray). Below $ z = 0.05$ we do not probe enough cosmic volume to include many $M_*$ galaxies in the sample, but such bright objects are the focus of many high-redshift studies. We therefore decided to increase our redshift limit to $z=0.1$ at the expense losing some resolution in the rest-frame.

The catalog presented in this paper is statistically more representative for detailed morphological studies than any other sample published to date. For example, there are essentially no galaxies fainter than $M_g \thicksim -19$ mag in the analyses presented by \cite{Frei:1996p4556} and \cite{vandenBergh:2002p9582}. Recent morphological catalogs by \cite{Fukugita:2007p3031} and \cite{Driver:2006p17183} do not provide the visual detail presented here. However, it need hardly be emphasized that because of the redshift and magnitude cuts just described, our sample is by no means volume-limited, and something like the $V_{max}$ formalism should be incorporated when using the catalog in this paper to calculate space densities. Evolutionary
changes in morphology are perceptible at $z\sim0.5$  ({\em e.g.} a change in the fraction of massive galaxies with bars; Abraham et al. 1999, Sheth et al. 2008),
but we expect such changes to be imperceptible for objects at $z < 0.1$, and therefore treat the
entire sample as `local'.

\subsection{Supplementary data}
For the convenience of the reader, we have augmented our catalog of morphological classifications
with published supplementary data derived by 
%the NYU group and the Garching SDSS 
other groups. Sersic profile parameters and environmental density estimates  from the NYU group (Blanton et al. 2003) have been included.  Environment estimates from \cite{Baldry:2006p103} and the \cite{Yang:2007p19054} SDSS group catalog have also been included. We also include derived stellar masses, ages, and star formation rates from the Garching group \citep{Kauffmann:2003p97,Kauffmann:2003p7199,Brinchmann:2004p3060}. The reader is referred to the original papers cited above for details on how these parameters were derived. 

\section{Classification Methodology}
\label{sec:classification}

The galaxy classification scheme used here is primarily based on the Carnegie Atlas of Galaxies \citep{Sandage:1994p4888} in consultation with the Third Reference Catalog of Bright Galaxies \citep{deVaucouleurs:1963p9392}, RC3, along with images for many fiducial objects obtained using the IPAC NED database. Traditional methods for classification involve printing galaxy images at various contrast ratios and manually inputting the classification back into a table. For the $\thicksim$ 14000 objects in our sample this would be very cumbersome. We therefore developed web-based graphical display software to allow us to inspect images at multiple contrast levels and record our classifications directly into a database.

After experimenting with a large subsample of objects in all 5 bands to determine in a general way the most efficient path forward, we adopted the following classification procedure. The registered SDSS fpC reduced frames was used to re-extract the galaxies in all bands with SExtractor \citep{Bertin:1996p6150}, using the $r'$-band image as the template. The segmentation of each galaxy was checked visually to ensure `parent' and `child' objects were correctly separated. Small `postage stamp' images (of width 50 $h^{-1}$ kpc and 100 $h^{-1}$ kpc) were then created for each object in each band at 5 contrast ratios determined by the flux range spanned by the galaxy. For any object where the automatically-chosen contrast ratios were not appropriate for identification purposes, we used ds9 manually to inspect images at a range of contrast ratios. The T-Types for the entire sample was classified twice by the first author, with a mean deviation of less than 0.5 T-Types.

While we found that a clear distinction between the various Hubble subclasses was possible to make
for most of our objects, not all classifications are clear-cut. Therefore some classifications are suffixed 
by flags using a notation which indicates that the classification is somewhat doubtful (?), or that the galaxy is peculiar (p), or simply that the classification is highly uncertain (:). This notation is adopted from the RC3 scheme. It should be noted that a (p) implies a peculiarity in a galaxy such as the presence of shells or tidal tails or some other feature (for which the secondary flags need to be checked) and not a `Peculiar' galaxy. Peculiar galaxies, or galaxies which could not be classified as a standard Hubble type have been assigned a T-Type of `99' (unknown). Secondary characteristic flags may be set for both normal T-Types and unknown galaxies, such as the `bulge-like', or `disk-like` flags or one of the \cite{Elmegreen:2005p117} types (`clump-clusters', `tadpoles', `doubles', and `chains') which have entered into the lexicon of high-redshift galaxy forms. It should be noted that our purpose in adding the latter secondary flags to our catalog is not only to identify probable local counterparts to these high redshift types, but also to identify objects which might be confused with such objects if seen at higher redshifts. For example, we flagged a number of objects that would seem to us to be rapidly star-forming clumpy spirals that would resemble clump clusters if seen further to the UV and with slightly poorer resolution. Since we have been fairly liberal in assigning \cite{Elmegreen:2005p117} types, due care should be taken when analyzing these subsamples, and we feel this categorization in our catalog is mainly useful for identifying samples worthy of follow-up observations.

As has already been emphasized, a major goal of our catalog is to capture information on a range of morphological properties beyond a galaxy's basic type. Henceforth we will refer to properties such as bars, rings, lenses, ansae, tidal tails and the details of spiral structure as a galaxy's {\em fine structure}. We noted the existence of a broad range of fine structure. Stellar bars were classified as the following types: strong, intermediate, weak, ansae and peanut. Rings were classified as nuclear, inner, outer, pseudo-outer (R1/R2), and collisional. Lenses were classified as inner or outer. Arm types from grand design to flocculent were recorded, as well as arm length and multiplicity. We recorded the presence of dust lanes, galaxy orientation, interaction features such as tails, warps, shells and bridges, the morphology of the nearest interacting galaxy and the merger orientation. Classifications were done using the $g'$-band images, though to confirm the presence of some features, like bars, we opted to check the $r'$ and $i'$ band data as well. T-Types are based solely on the $g'$-band images. We ultimately decided to exclude information determined from the $u'$-band and $z'$-band images completely due to poor signal-to-noise ratios.

\section{Illustrations of Representative Galaxies}

In this section we present color montages constructed by assigning RGB colors to $g'$, $r'$, $i'$ data channels. These color images were taken from the SDSS Imaging Server. In order to allow the reader to compare our classifications to those from other sources, we have tried to choose objects in these montages that are part of the small sample of objects in our catalog that overlap with other published samples (e.g. the RC3, or the catalog of Fukugita et al. 2007). Aside from this constraint, the galaxies in these montages have been selected at random from the catalog.

\subsection{Galaxies with Classical Forms (Standard T-Types)}

\begin{figure*}[b!]
\unitlength1cm
\hspace{1cm}
%\vspace{1.5cm}
\begin{minipage}[t]{4.0cm}
\rotatebox{0}{\resizebox{16cm}{7.8cm}{\includegraphics{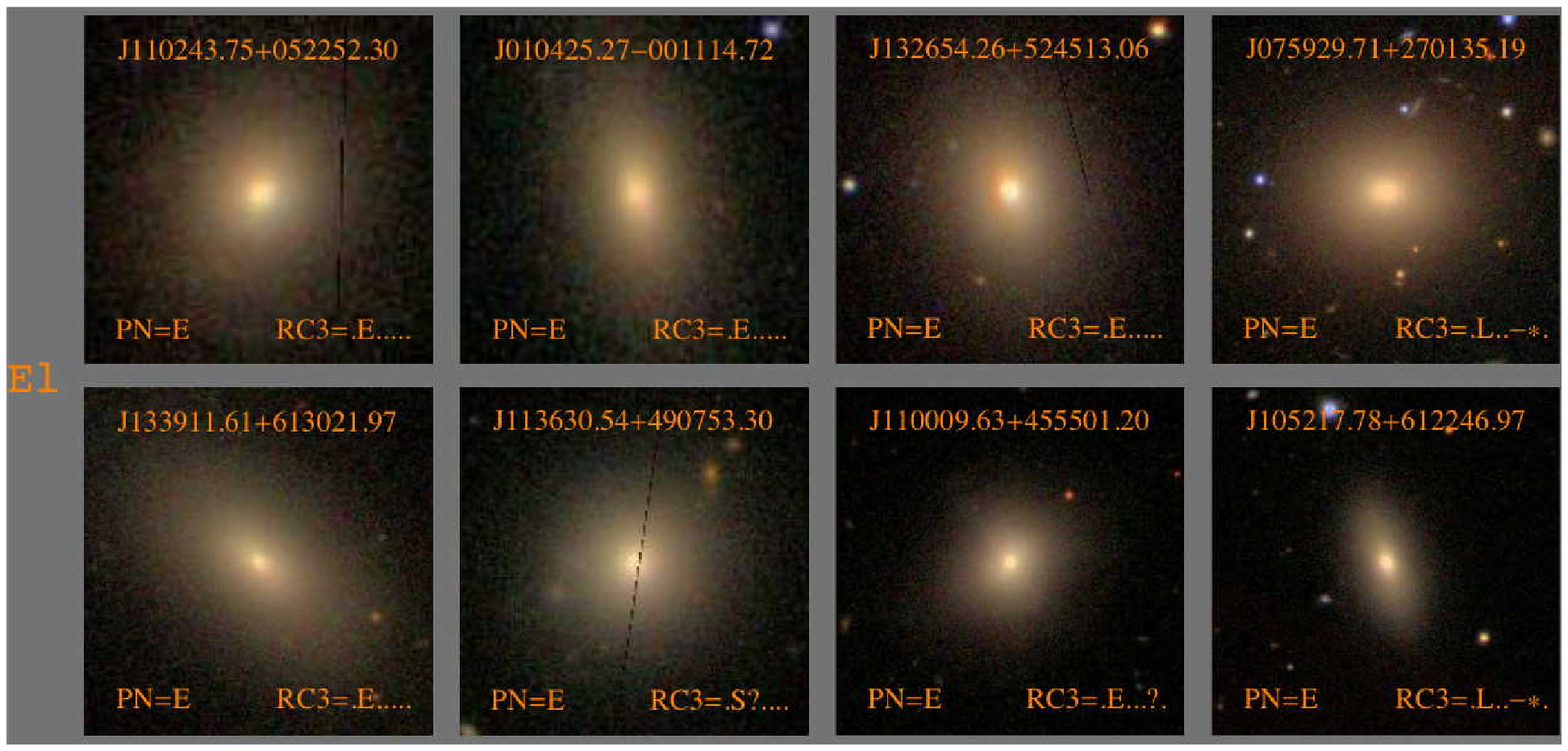}}}
\rotatebox{0}{\resizebox{16cm}{7.8cm}{\includegraphics{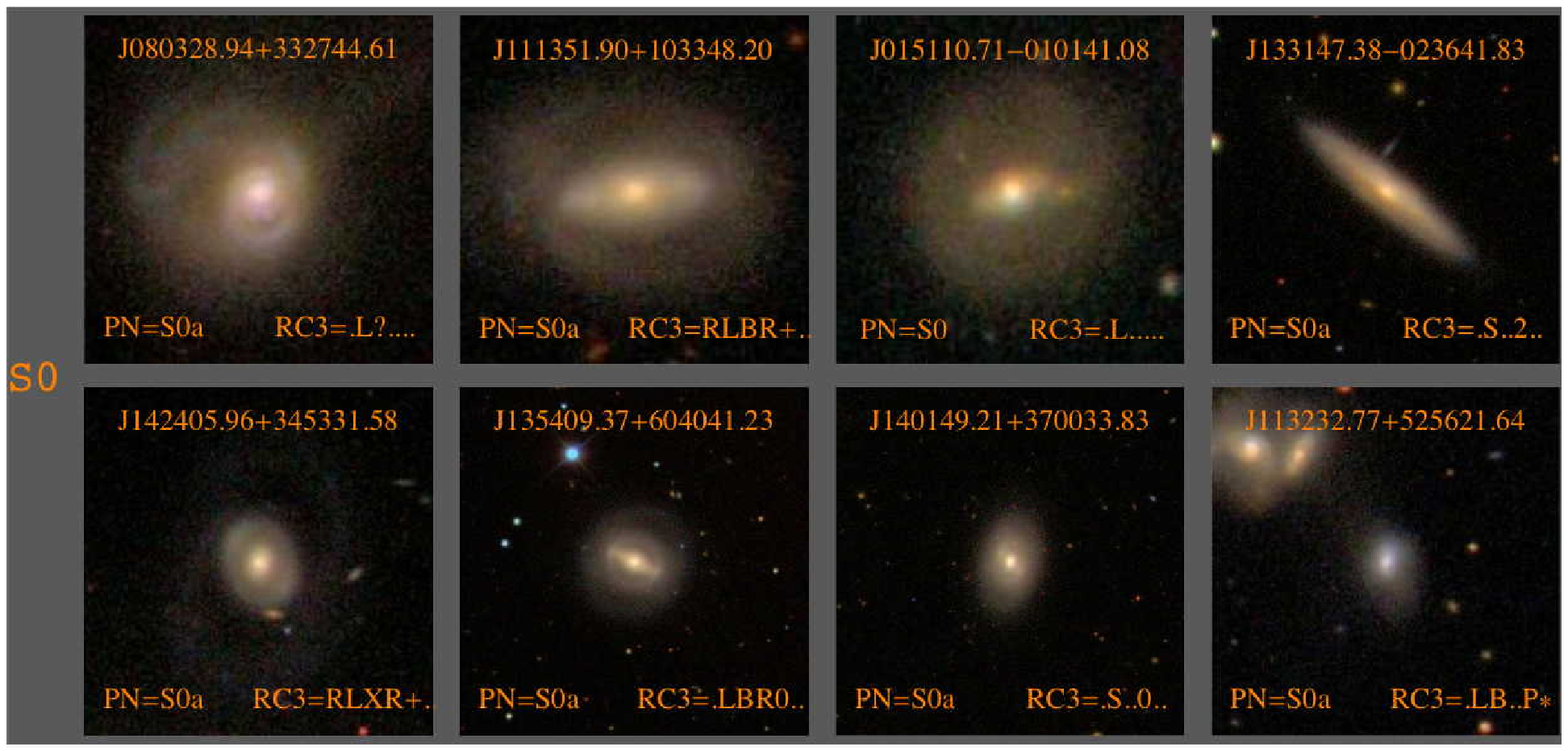}}}
\rotatebox{0}{\resizebox{16cm}{7.8cm}{\includegraphics{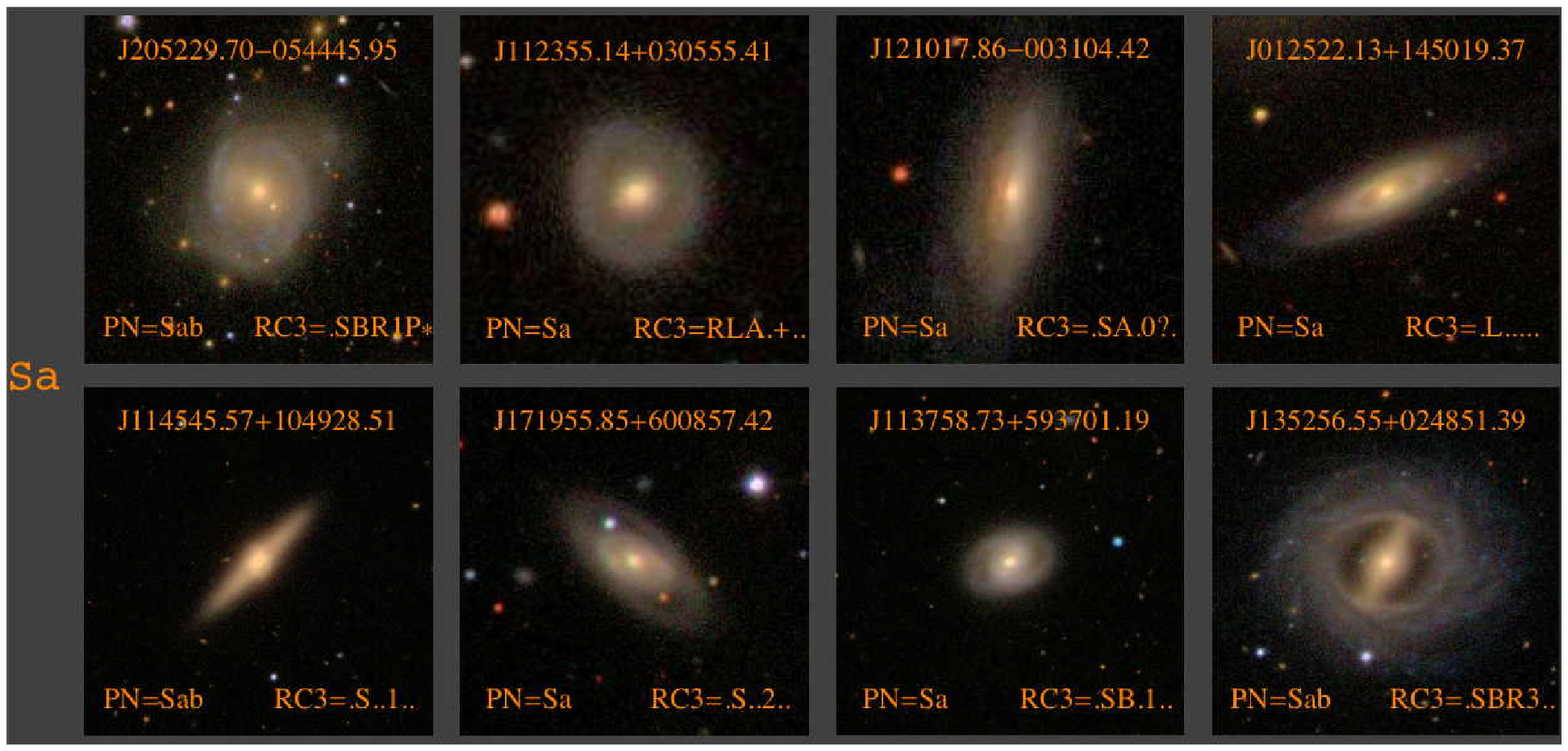}}}
\end{minipage}
\caption[Montage of E, S0 and Sa galaxies]{\label{fig:EMontage} A montage of images representing {\bf Top}: E + E/S0 galaxies, {\bf Middle}: S0+ S0a galaxies, and {\bf Bottom}: Sa + Sab galaxies (2 rows for each) as classified by the first author (PN). In each of these categories, galaxies are arranged in order of decreasing stellar mass. The J2000 object identifier is listed at the top with the redshift and the RC3 classification at the bottom. The seven possible letters starting from the left in the RC3 designation identify (1) peculiarities such as outer rings 'R', (2) Type: E for elliptical, L for lenticular or S for spiral (3) bar class: B for strong bar, X for weak bar, A for no bar, (4) inner rings, (5) T-Type eg 3 for Sb. Fields (6) and (7) are used for additional flags, like unsure (?). Each stamp is 50 $h^{-1}$ kpc on a side.}
\end{figure*}

\begin{figure*}[htbp!]
%\begin{center}
%\unitlength1cm
\hspace{1cm}
%\vspace{1.5cm}
\begin{minipage}[t]{4.0cm}
\rotatebox{0}{\resizebox{16cm}{7.8cm}{\includegraphics{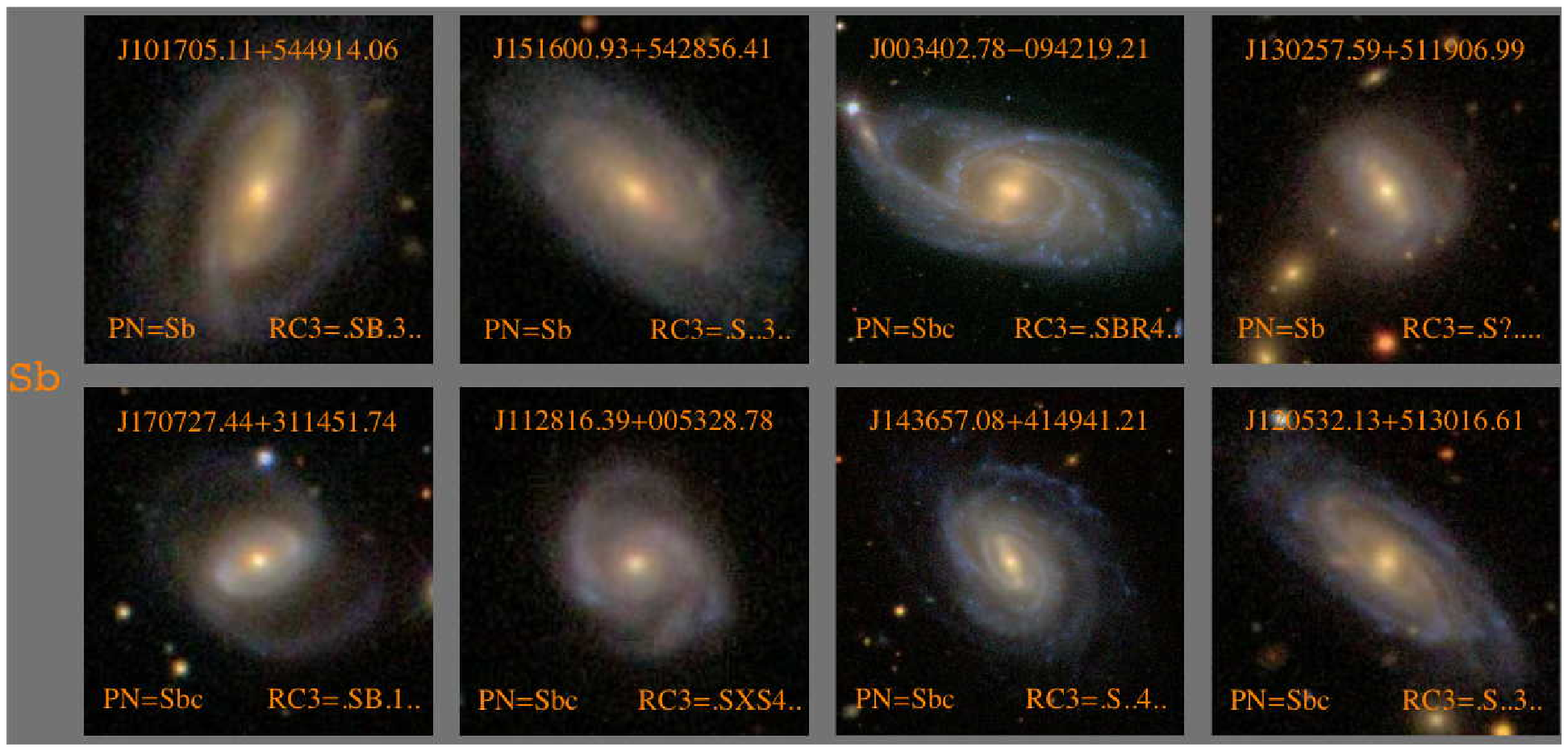}}}
\rotatebox{0}{\resizebox{16cm}{7.8cm}{\includegraphics{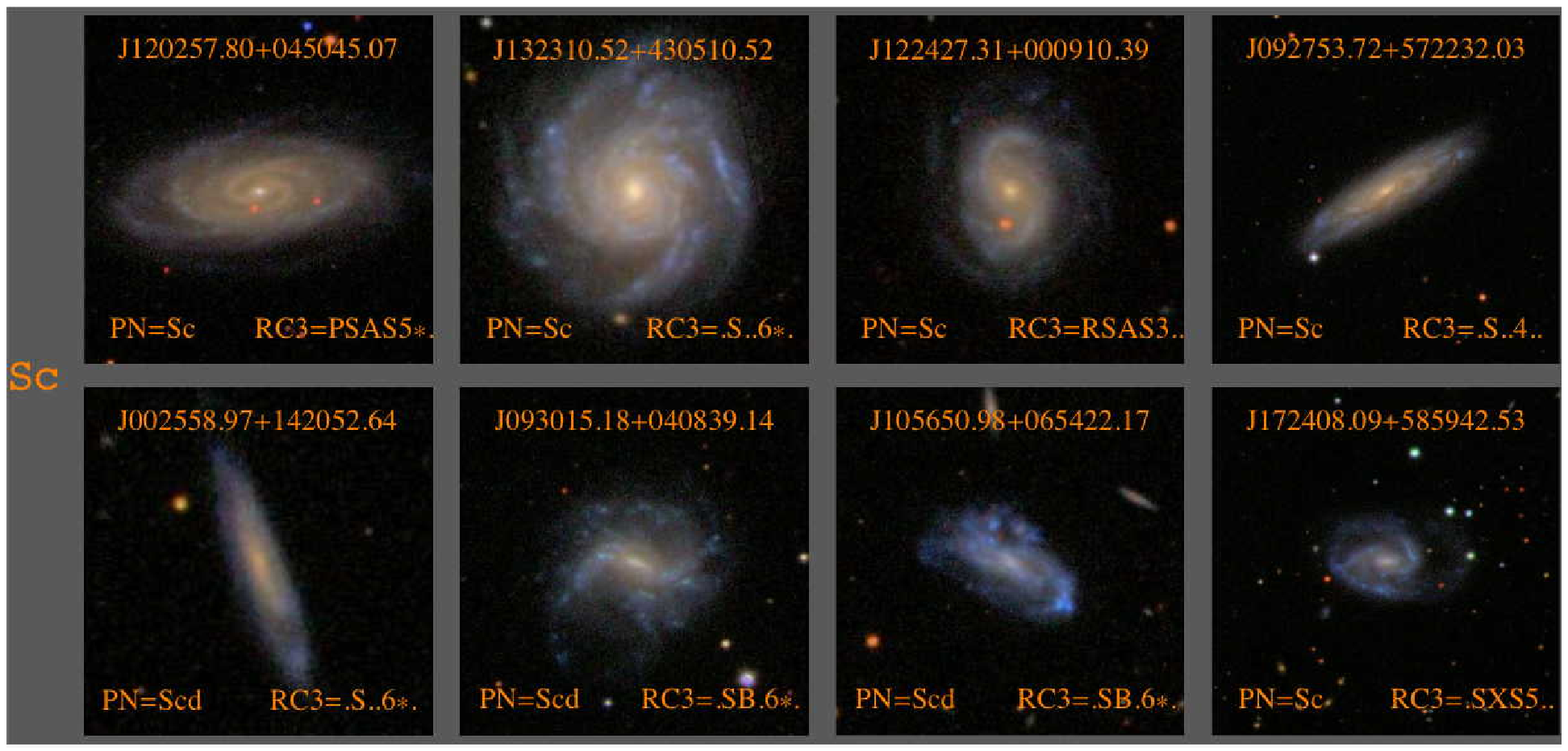}}}
\rotatebox{0}{\resizebox{16cm}{7.8cm}{\includegraphics{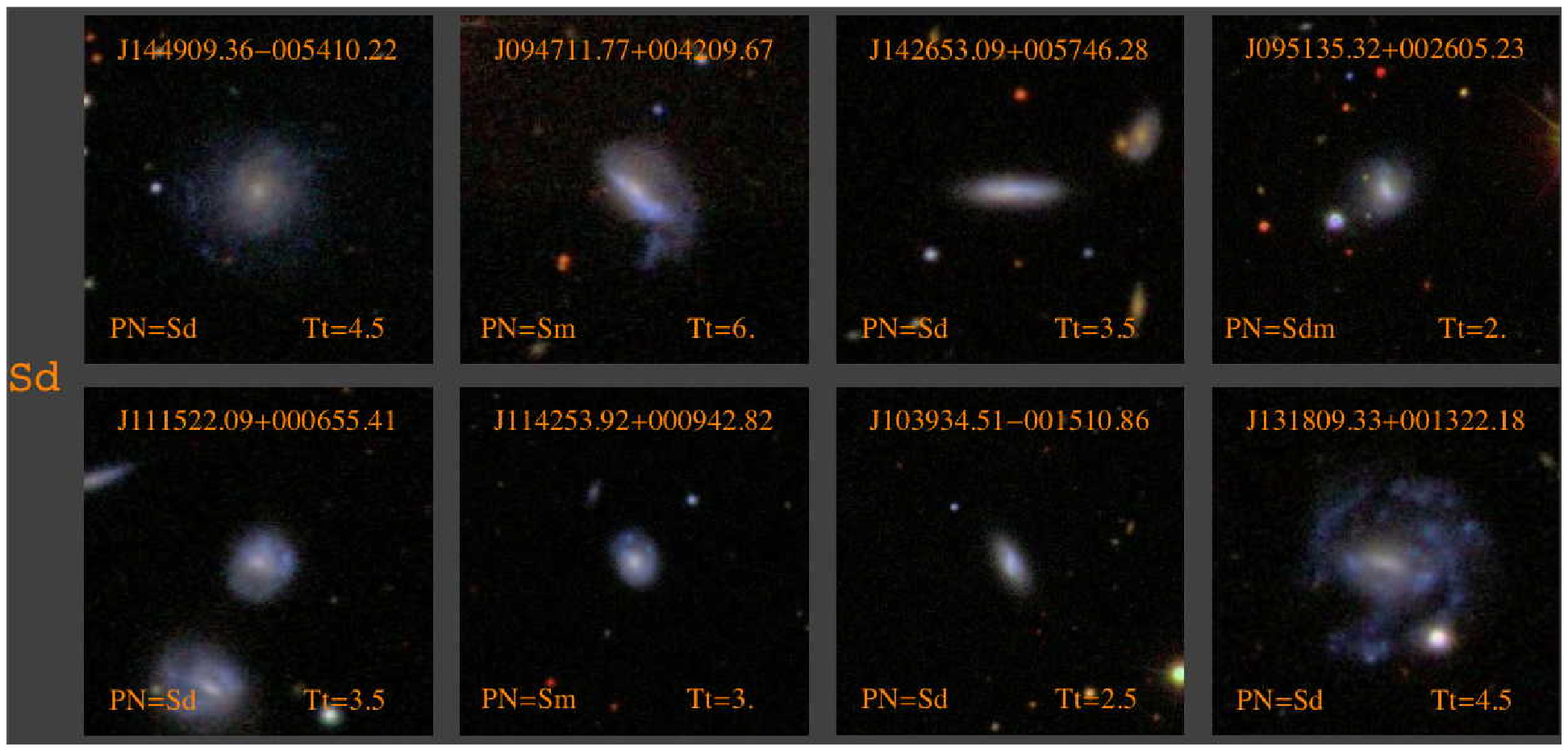}}}
\end{minipage}
\caption[Montage of Sb, Sc and Sd galaxies]{\label{fig:SMontage} 
A montage of images representing Sb, Sc and Sd galaxies (2 rows for each) as classified by the first author (PN).
In each of these categories, galaxies are arranged in order of decreasing stellar mass. The J2000 object identifier is listed at the top with the redshift and the RC3 classification at the bottom (when available). The seven possible letters starting from the left in the RC3 designation identify (1) peculiarities such as outer rings 'R', (2) Type: E for elliptical, L for lenticular or S for spiral (3) bar class: B for strong bar, X for weak bar, A for no bar, (4) inner rings, (5) T-Type eg 3 for Sb (6 and 7) flags like unsure(?). Each stamp is 50 $h^{-1}$ kpc on a side. Although color images are shown, the classification was carried out on the g-band images. See text for details.}
%\end{center}
\end{figure*}

\begin{figure*}[htbp!]
\unitlength1cm
\hspace{1cm}
%\vspace{1.5cm}
\begin{minipage}[t]{4.0cm}
\rotatebox{0}{\resizebox{16cm}{7.8cm}{\includegraphics{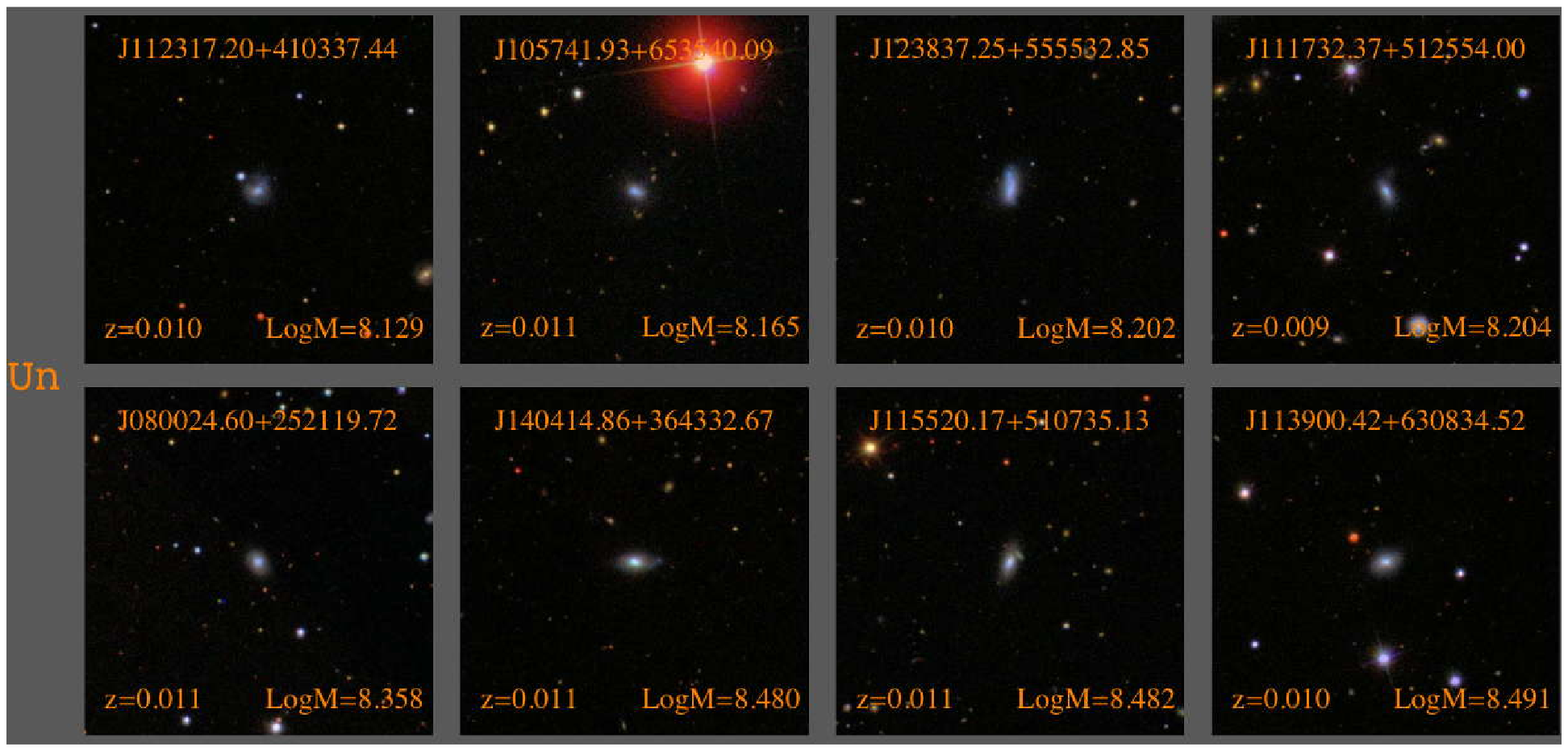}}}
\rotatebox{0}{\resizebox{16cm}{7.8cm}{\includegraphics{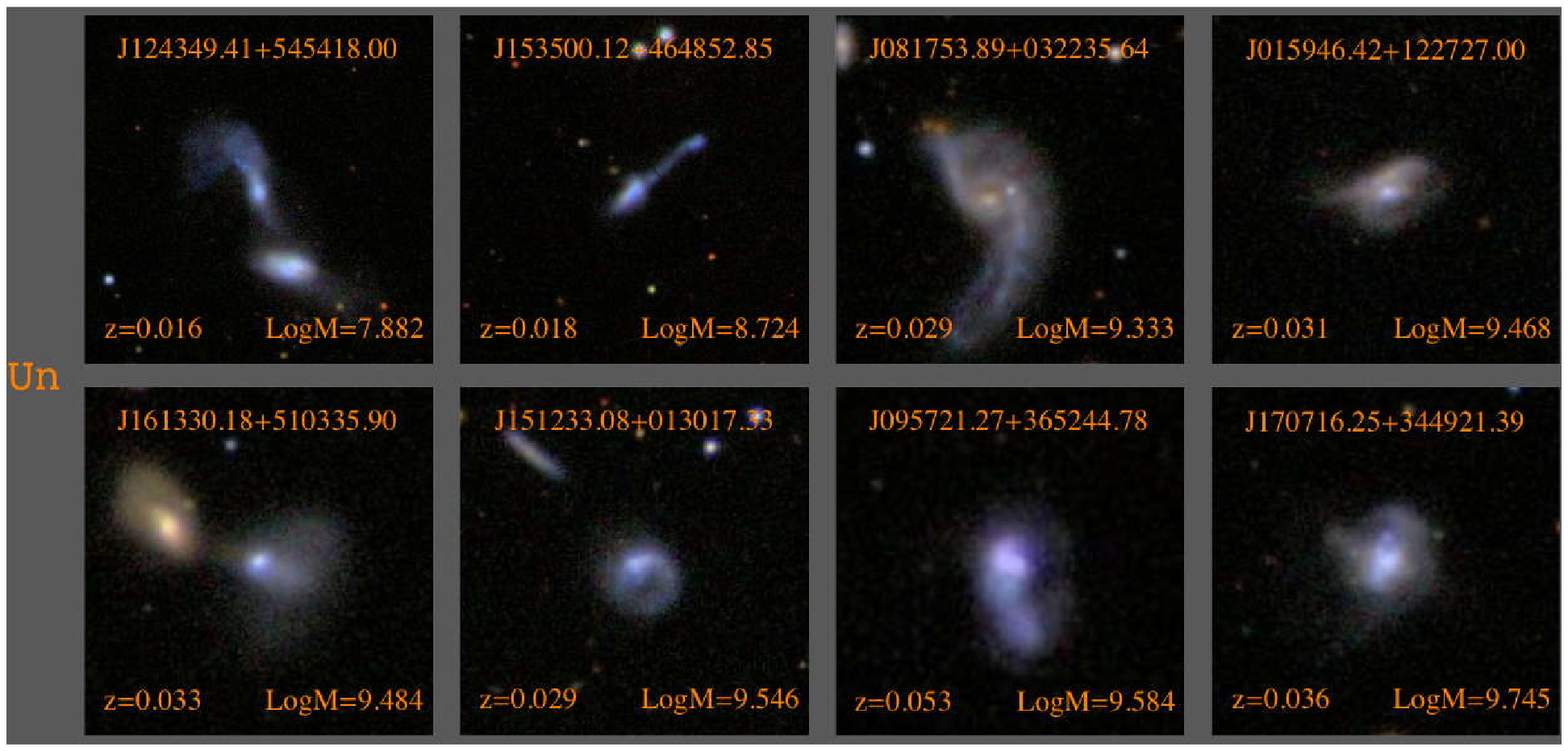}}}
\rotatebox{0}{\resizebox{16cm}{7.8cm}{\includegraphics{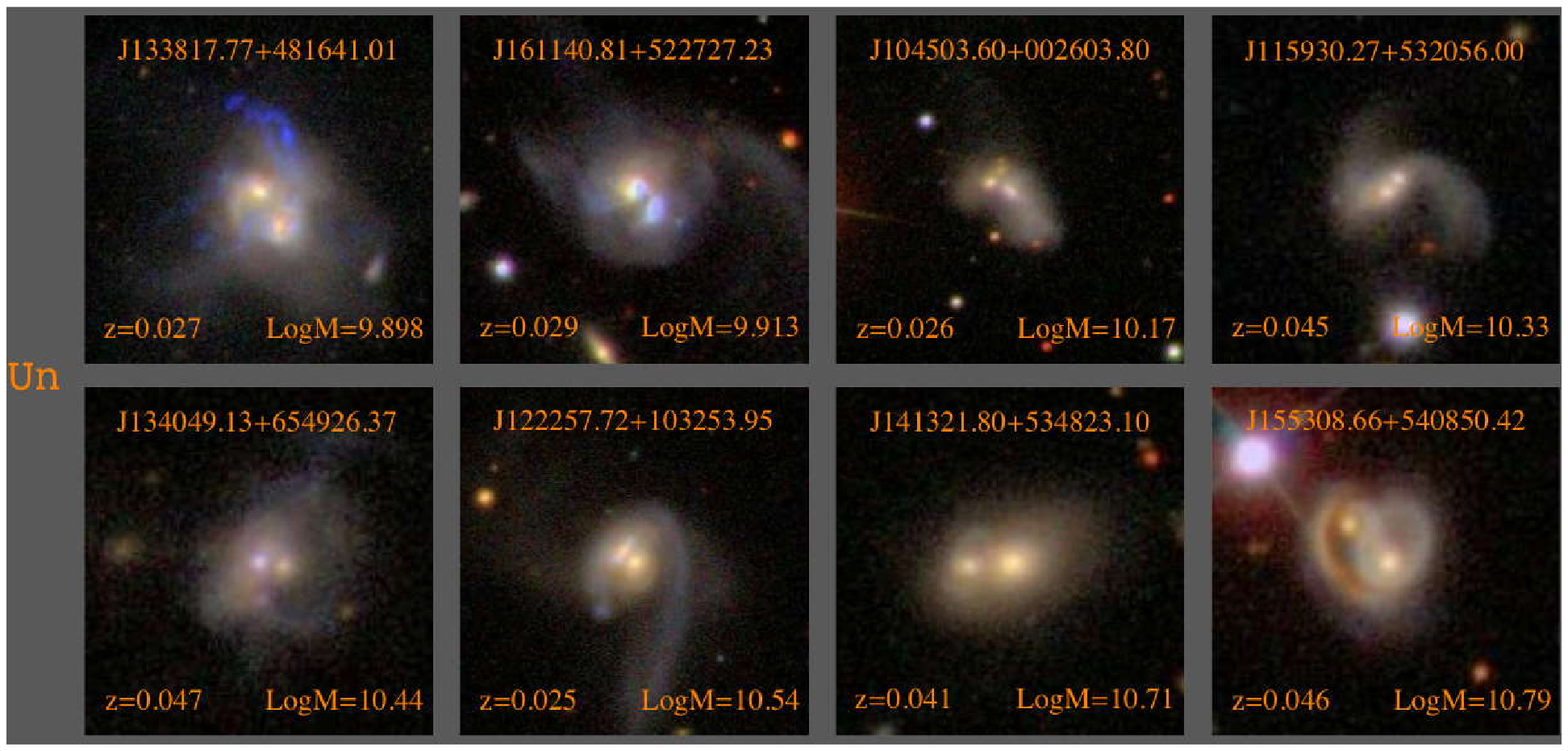}}}
\end{minipage}
\caption[Montage of Unknown galaxies ]{\label{fig:UnknownMontage} 
A montage of images representing unknown (T-Type=99) galaxies. {\bf Top}: Examples of objects deemed too small
for reliable classification. These are dwarf galaxies, perhaps 
dwarf spheroidals or dwarf spirals. {\bf Middle}: Objects grossly distorted by merging/interactions. {\bf Bottom}: Likely end stages or remnants of mergers. Objects are arranged in order of increasing mass. The J2000 object identifier is listed at the top with the redshift and the author classification at the bottom. Each stamp is 50 $h^{-1}$ kpc on a side. Although color images are shown, the classification was carried out on the g-band images. See text for details. The color images are taken from the SDSS Imaging Server. 
 }
\end{figure*}

\begin{figure*}[htbp!]
\unitlength1cm
%\vspace{0.7cm}
\hspace{1cm}
%\vspace{1.5cm}
\begin{minipage}[t]{4.0cm}
\rotatebox{0}{\resizebox{16cm}{8cm}{\includegraphics{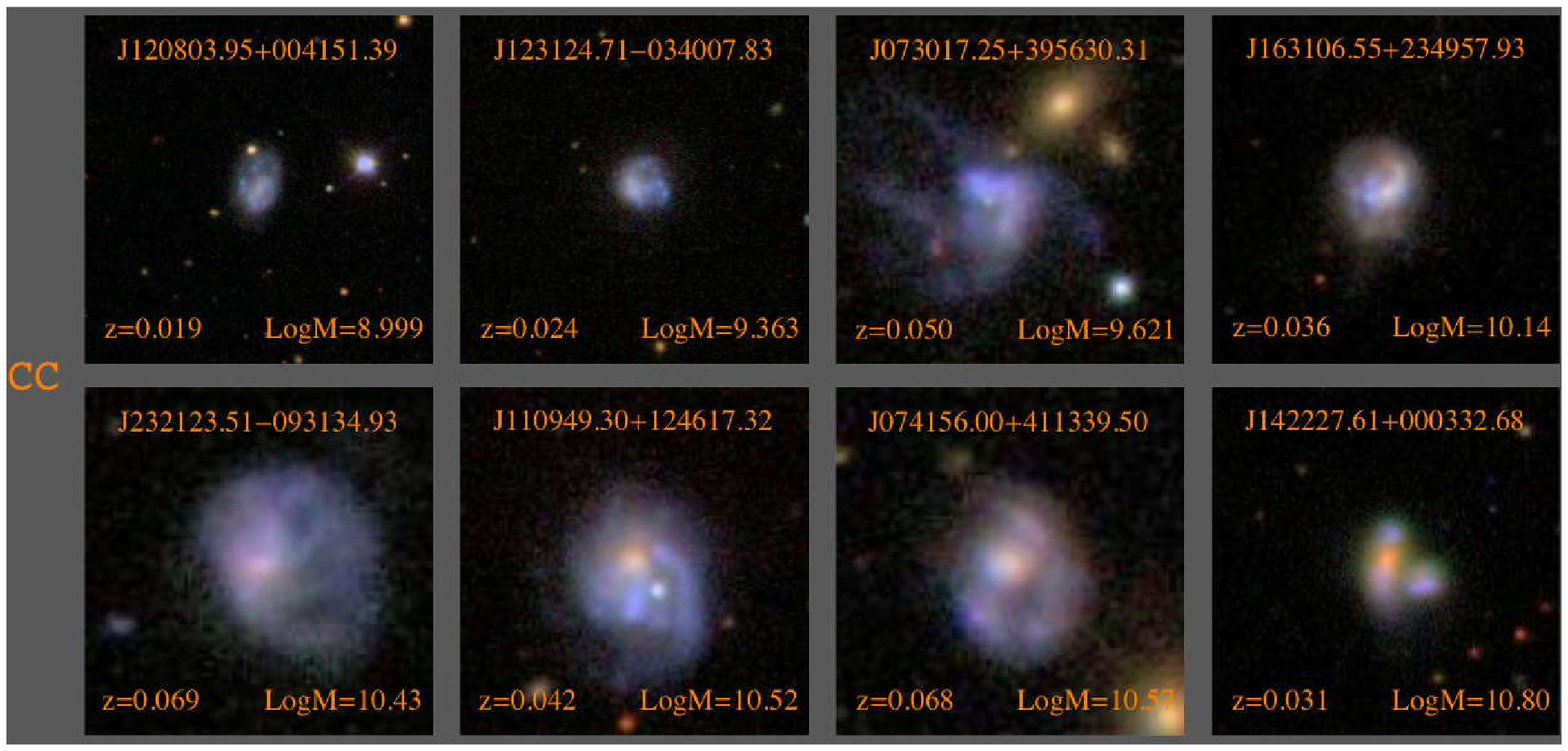}}}
\rotatebox{0}{\resizebox{16cm}{8cm}{\includegraphics{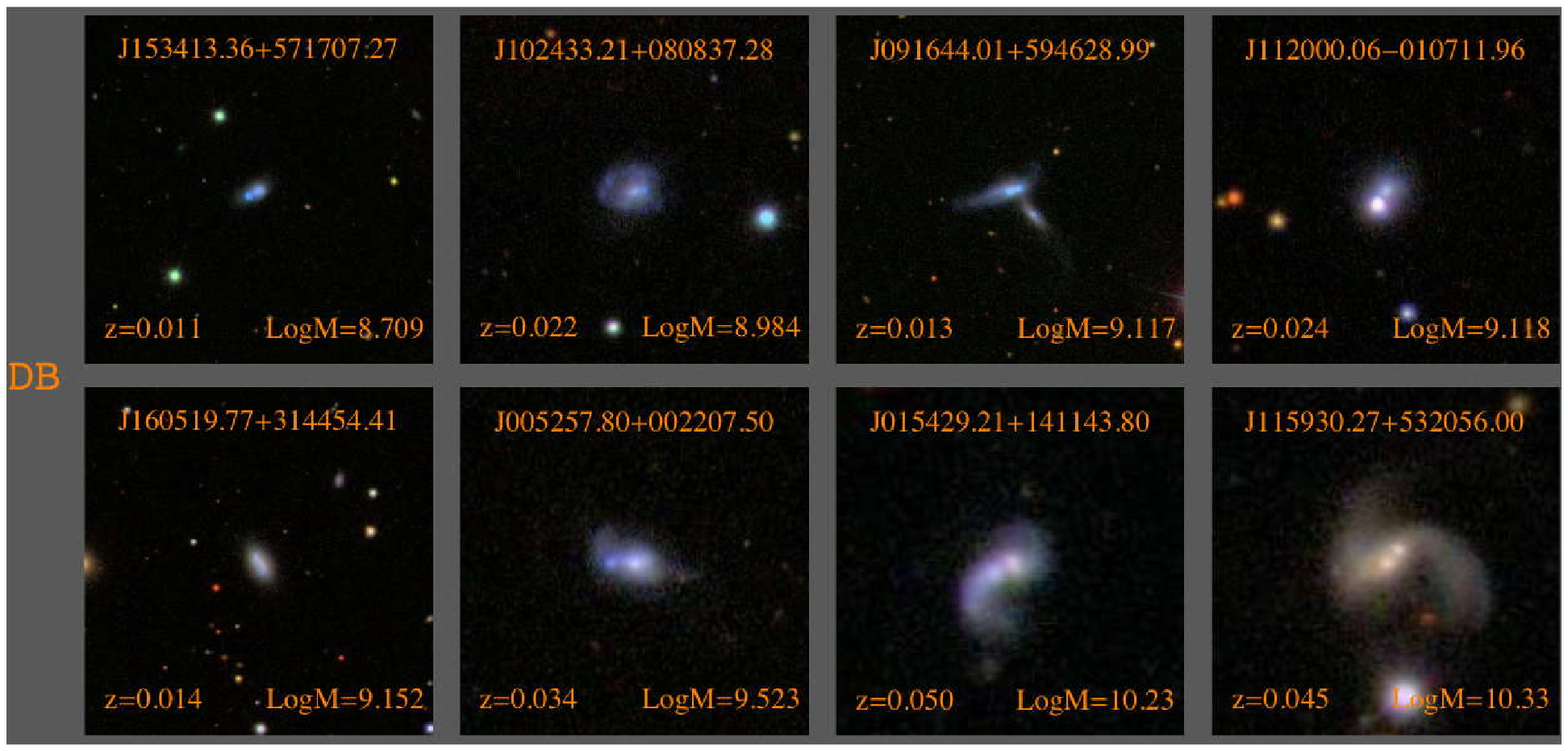}}}
\rotatebox{0}{\resizebox{16cm}{4cm}{\includegraphics{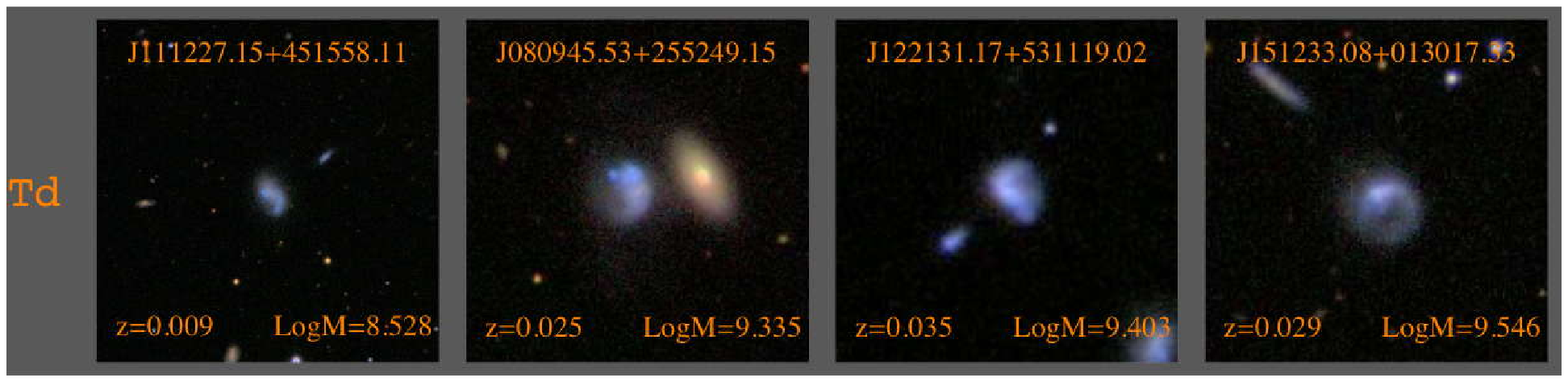}}}
\end{minipage}
\caption[Montage of Peculiar galaxies ]{\label{fig:UnknownMontage2} 
A montage of images representing galaxies which may be classified as (or easily mistaken for, at higher redshift) 
the \cite{Elmegreen:2005p117} designations of {\bf [Top]} Clump Cluster (CC), {\bf [Middle]} Double galaxies (DB) and {\bf [Bottom]} Tadpole galaxies (Td). As described in the text, in a few cases these secondary classifications have been assigned to objects mainly because we felt that, if seen at higher redshift and in the rest-frame UV, these galaxies would resemble systems such as clump clusters. The J2000 object identifier is listed at the top with the redshift and the author classification at the bottom. Each stamp is 50 $h^{-1}$ kpc on a side. Although color images are shown, the classification was carried out on the g-band images. See text for further details. The color images are taken from the SDSS Imaging Server.
}
\end{figure*}

Figures~\ref{fig:EMontage} and \ref{fig:SMontage} show color composite images of a number of 
E through Sd galaxies in our catalog which have RC3 or Fukugita et al. (2007) classifications. In \S5 we will present a detailed comparison of our classifications against those given in earlier work, but at this stage it is perhaps worthwhile to foreshadow this discussion by referring the reader to Table~\ref{tab:TTypeClassification}, which shows the relation between RC3 T-Types, our T-Types, and those of Fukugita et al (2007)\footnote{We also remind the reader that an easy way to remember the numeric equivalent for a T-Type is to note that all major classes are odd numbers (e.g. Sa galaxies have T-Type 1, and Sb galaxies have T-Type 3), while the finer separation are even numbers (e.g. Sab has T-Type 2).}. For illustrative purposes in these figures we have grouped sub-categories of galaxies into broader classes: E and E/S0 galaxies are shown together, as are S0 and S0/a galaxies, Sa and Sab galaxies, Sb and Sbc galaxies, Sc and Scd galaxies and galaxies with T-Types later than Sd. The J2000 object identifier is listed at the top of each galaxy's panel, along with our own T-Type (PN T-Type) and the RC3 type\footnote{The RC3 short form classification is made up of 7 characters which starting from the left identify (1) peculiarities such as outer rings: 'R' for full, 'P' for pseudo, (2) Crude class: E for elliptical, L for lenticular or S for spiral (3) bar class: B for strong bar, X for weak bar, A for no bar, (4) inner rings, R or lenses, L, (5) T-Type for example 3 for Sb, (6) flags like unsure(?) and (7) flags.} or Fukugita et al. (2007) T-Type. The objects are arranged in order of decreasing stellar mass. We remind the reader that the presence or absence of a dust lane is not used as a classification criterion.

\subsection{Galaxies with Unusual Forms}

There are 353 objects which do not have a regular T-Type classifications in this catalog. These objects have had T-Types of `99' assigned to them, and the great majority of these objects represent objects that do not find a natural home in the Hubble Sequence, although a few are simply galaxies that we felt were too small to reliably classify. Wherever possible galaxies with unusual forms have been assigned flags in our catalog which correspond to secondary classifications, such as `bulge-like' or `disk-like' and in some cases the high redshift sub-classifications of `clump cluster', `double', `tadpole' and `chain' introduced by \cite{Elmegreen:2005p117}. Galaxies where a sub-class could not be assigned are more likely to be mergers and the appropriate merger flags have been set.

Figure~\ref{fig:UnknownMontage} shows a montage of galaxies with unusual forms. Since this category of objects is rather broad, the reader will probably wish to subdivide it to isolate interesting objects using the secondary indicators noted earlier. For example, the reader can easily filter the catalog for objects that are unusual because they are somewhat too small to be reliably classified (top two rows), or objects that are unusual with structure suggestive of being at an early stage of merging (most objects in the middle two rows), and objects that appear to be near the end stages of a merger (most objects in the bottom two rows). The reader should note that while secondary flags in the catalog can be used to assign most objects into these three bins (`too small to classify', `early-phase merger', and `late-phase merger'), many interesting objects cannot be classified as any of these. For example, the last object in the last row, J155308.66+540850.42, appears to be a double collisional ring system, which we have been calling `Preethi's Cross-Eyed Galaxy'. The rarity of collisional rings makes it unlikely that this system is an optical superposition of two collisional rings, but on the other hand numerical simulations of `bullseye' collisions do not (at present) generate multiple rings.

Figure~\ref{fig:UnknownMontage2} shows galaxies which we have sub-classified as being (possible) low-redshift equivalents of high-z clump clusters (top panel), doubles (middle panel) or tadpole galaxies (bottom panel) as defined by \cite{Elmegreen:2005p117}. As has already been noted, in some cases (for example the clump cluster J074156.00+411339.50 in the top panel, second row, third column) these
objects appear to be bona-fide counterparts to high-redshift galaxy forms, while in other cases we show objects which could easily be mistaken for these forms at high redshifts (e.g. J232123.51-093134.93, which is a very clumpy asymmetric spiral galaxy). Note that we found no examples of chain galaxies at z$>$0.01. 

\subsection{Fine Structures}

Figures~\ref{fig:BarMontage} -- \ref{fig:Shells} show representative examples of the fine structures we have classified in our objects. Once again, we show systems with the J2000 object identifier listed at the top, the galaxy redshift at the bottom left and our T-Type classification at the bottom right. A comparison of the fine structures recorded in these figures with those captured by the RC3 classifications for these objects will be shown in \S5. 

\subsubsection{Bars}

\begin{figure*}[htbp!]
\unitlength1cm
\hspace{1cm}
%\vspace{1.5cm}
\begin{minipage}[t]{4.0cm}
\rotatebox{0}{\resizebox{16cm}{7.8cm}{\includegraphics{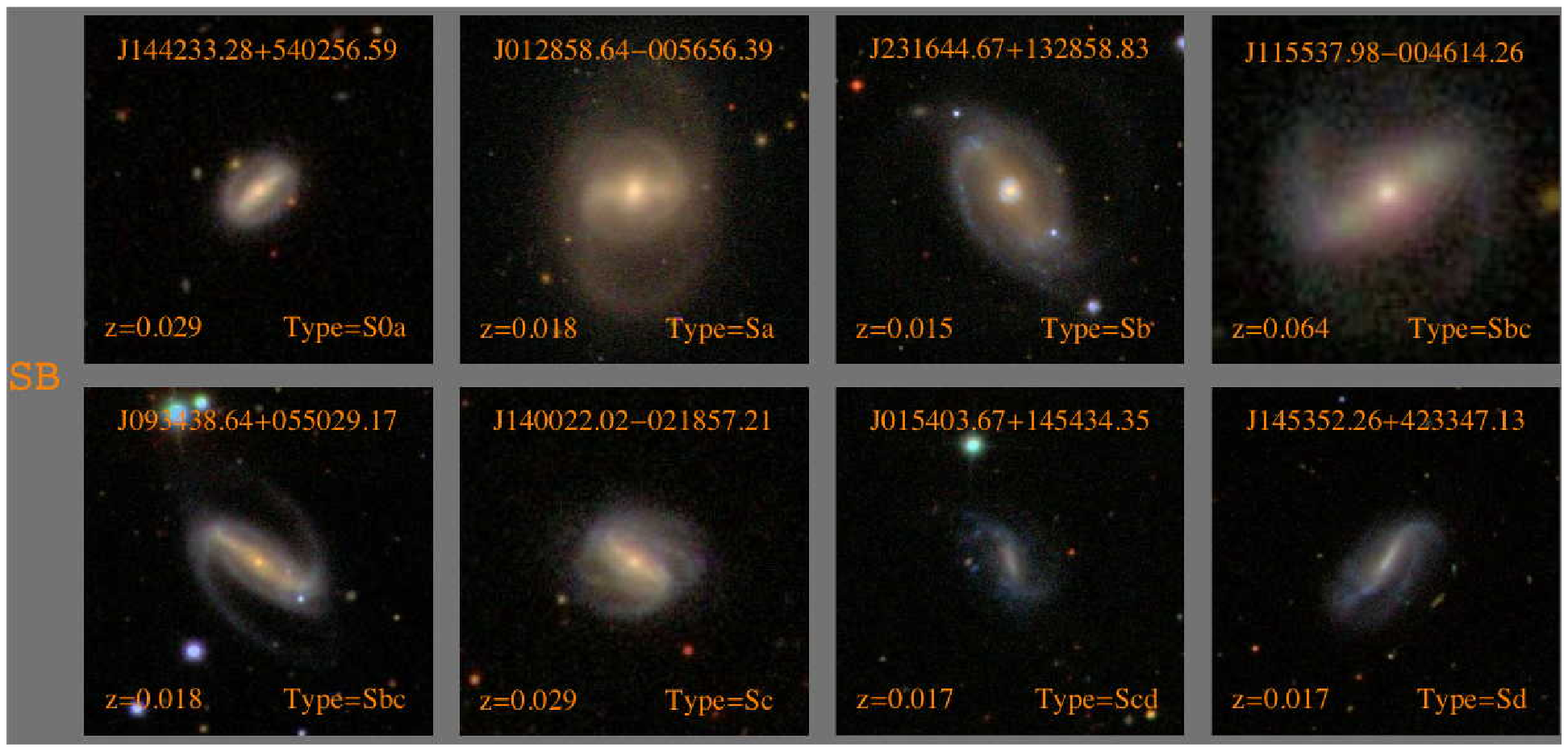}}}
\rotatebox{0}{\resizebox{16cm}{7.8cm}{\includegraphics{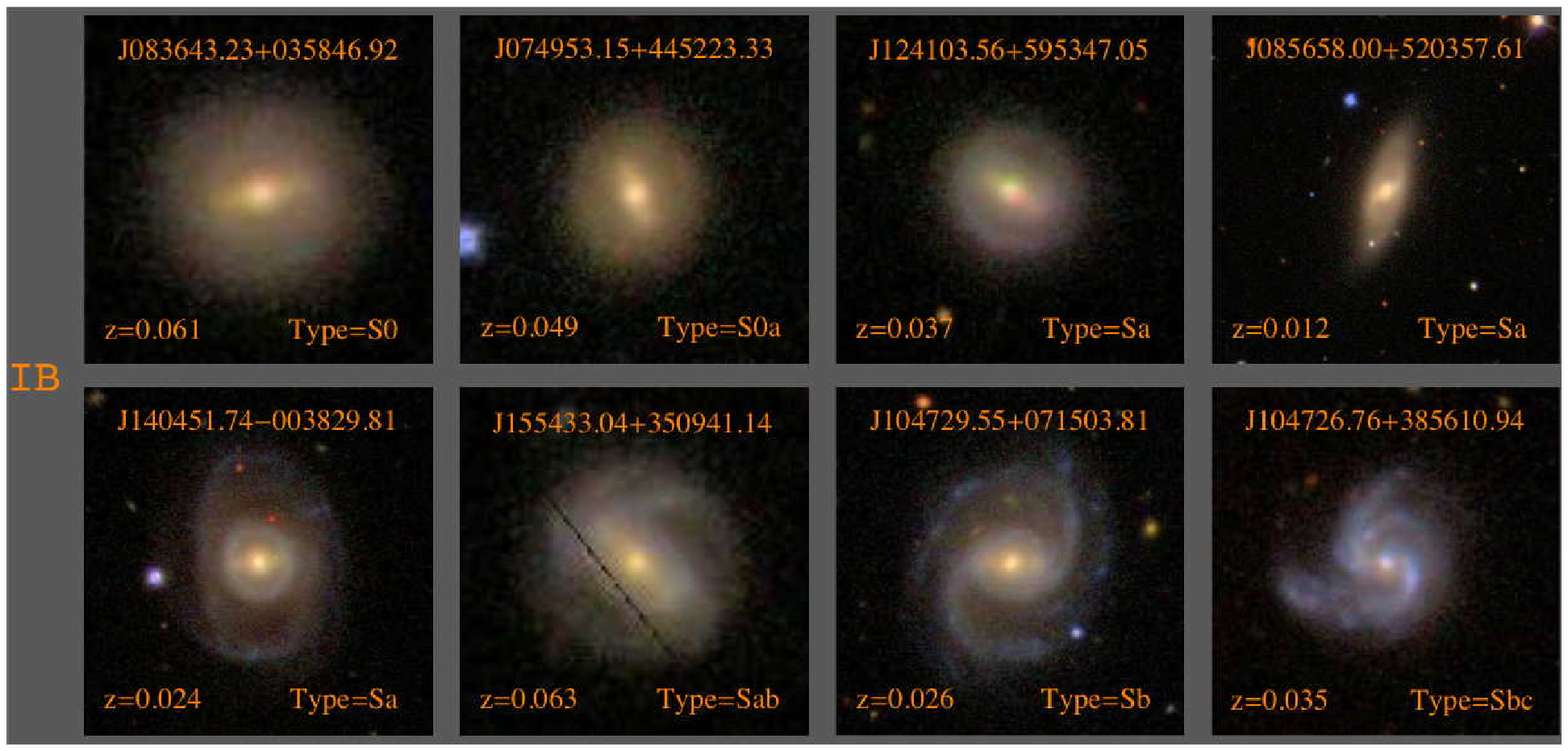}}}
\rotatebox{0}{\resizebox{16cm}{7.8cm}{\includegraphics{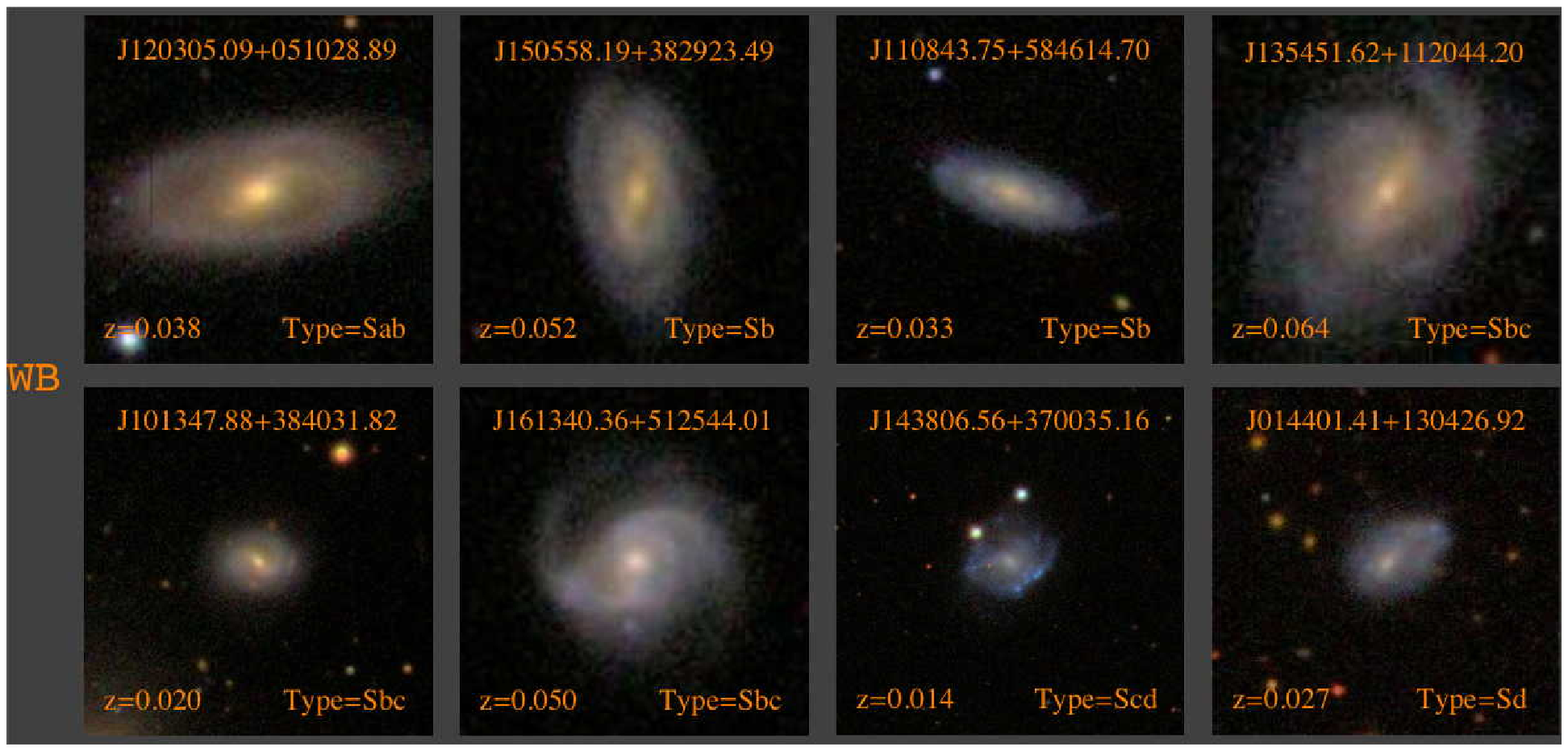}}}
\end{minipage}
\caption[Montage of Barred galaxies ]{\label{fig:BarMontage} 
A montage of images representing {\bf Top}: Strong bars, {\bf Middle}: Medium bars and {\bf Bottom}: Weak bars (2 rows for each) as classified by the author arranged in order of T-Type. Strong bars are comparable in size to the galaxy and have a significant amount of the total flux of the galaxy. Weak bars are smaller in size and contain a small fraction of the total flux of the galaxy. Intermediate bars span the range between strong and weak bars. The J2000 object identifier is listed at the top with the redshift and the author classification at the bottom. Each stamp is 50 $h^{-1}$ kpc on a side. Our classification of bar strength is based on inspection of the $g'$-band images but the determination of a bar's existence was based on studying each galaxy in all three bands ($g', r', i'$). Color images were not used to determine the `strength' of the bar. See text for details. The color images are taken from the SDSS Imaging Server. 
 }
\end{figure*}

\begin{figure*}[htbp!]
\unitlength1cm
\hspace{1cm}
%\vspace{1.5cm}
\begin{minipage}[t]{4.0cm}
\rotatebox{0}{\resizebox{16cm}{7.8cm}{\includegraphics{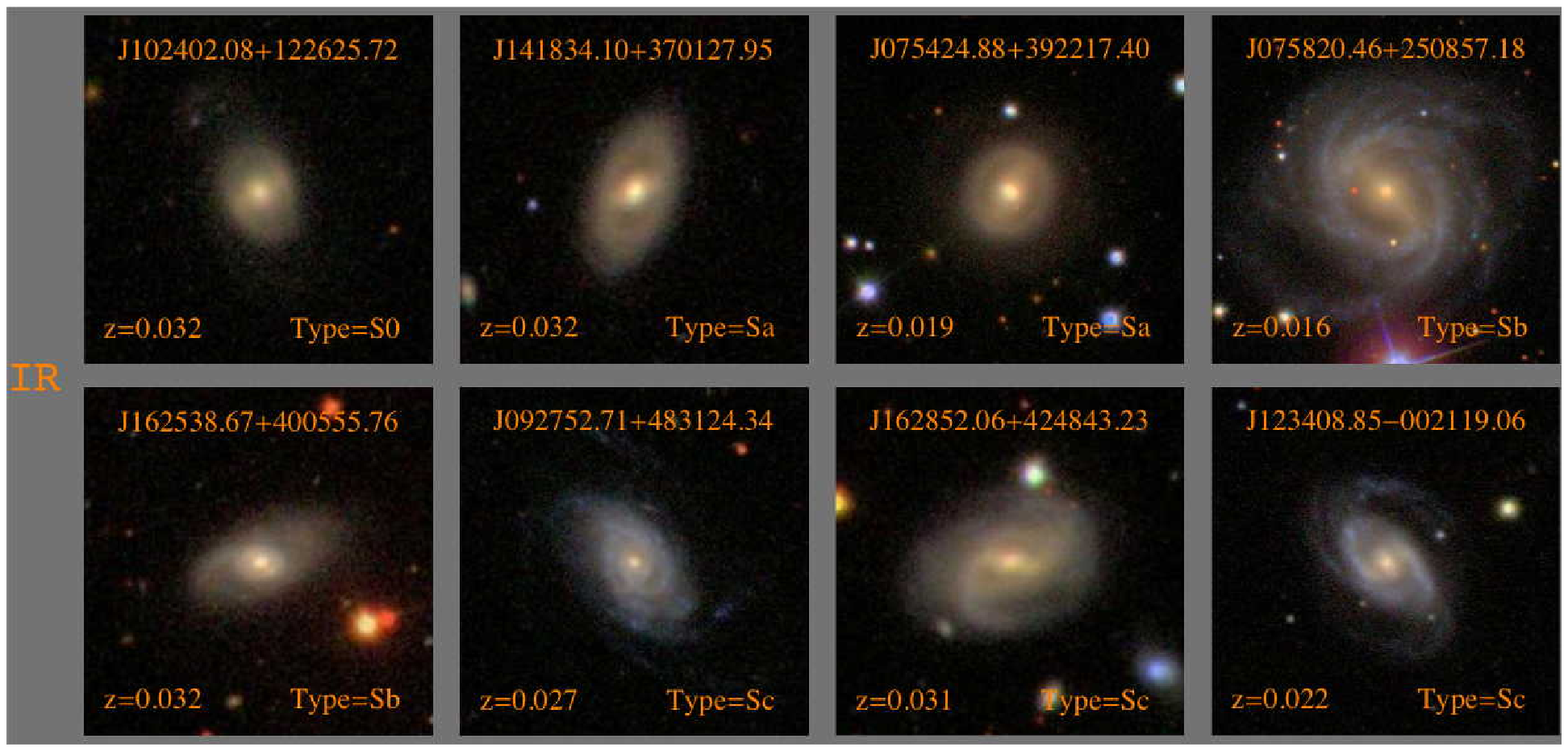}}}
\rotatebox{0}{\resizebox{16cm}{7.8cm}{\includegraphics{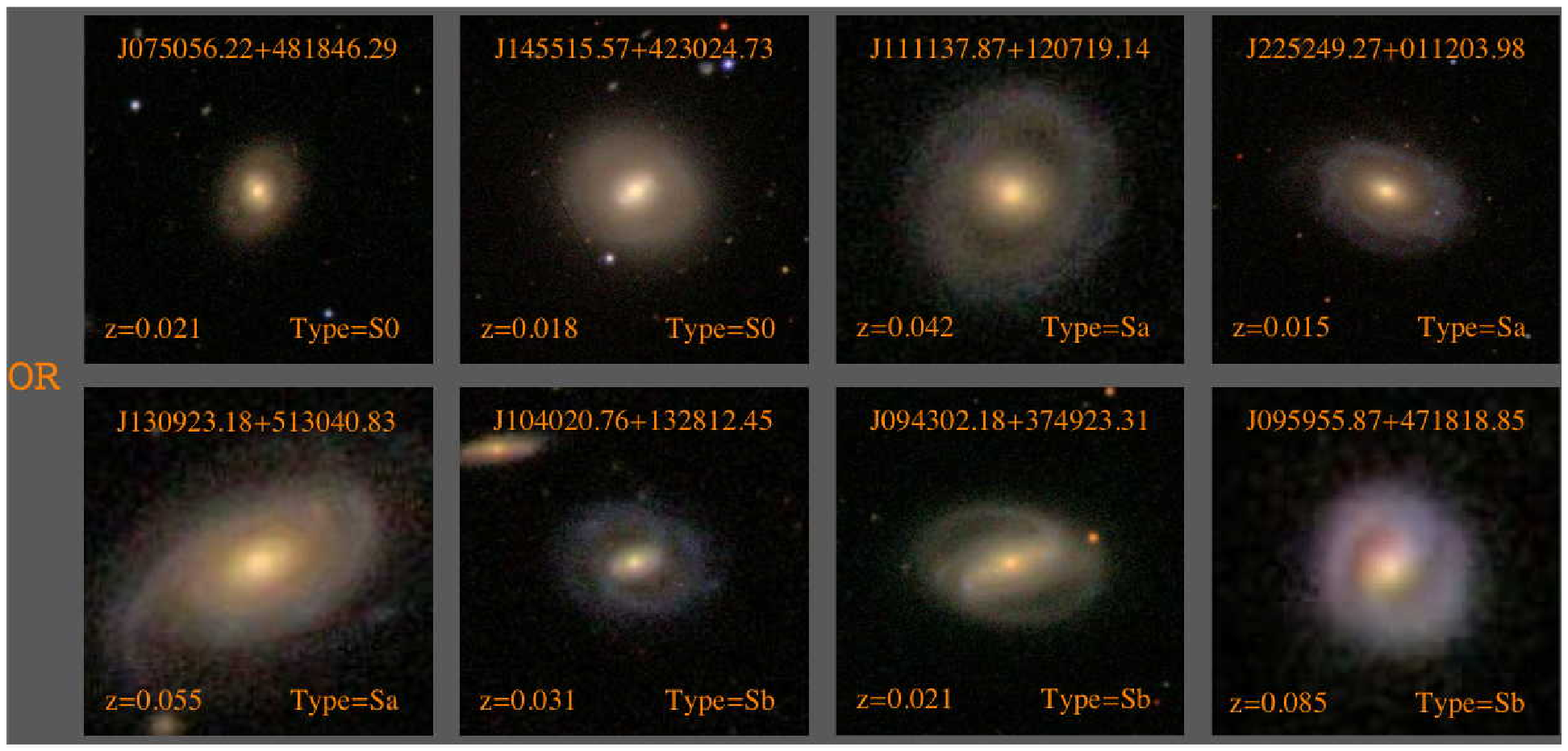}}}
\rotatebox{0}{\resizebox{16cm}{7.8cm}{\includegraphics{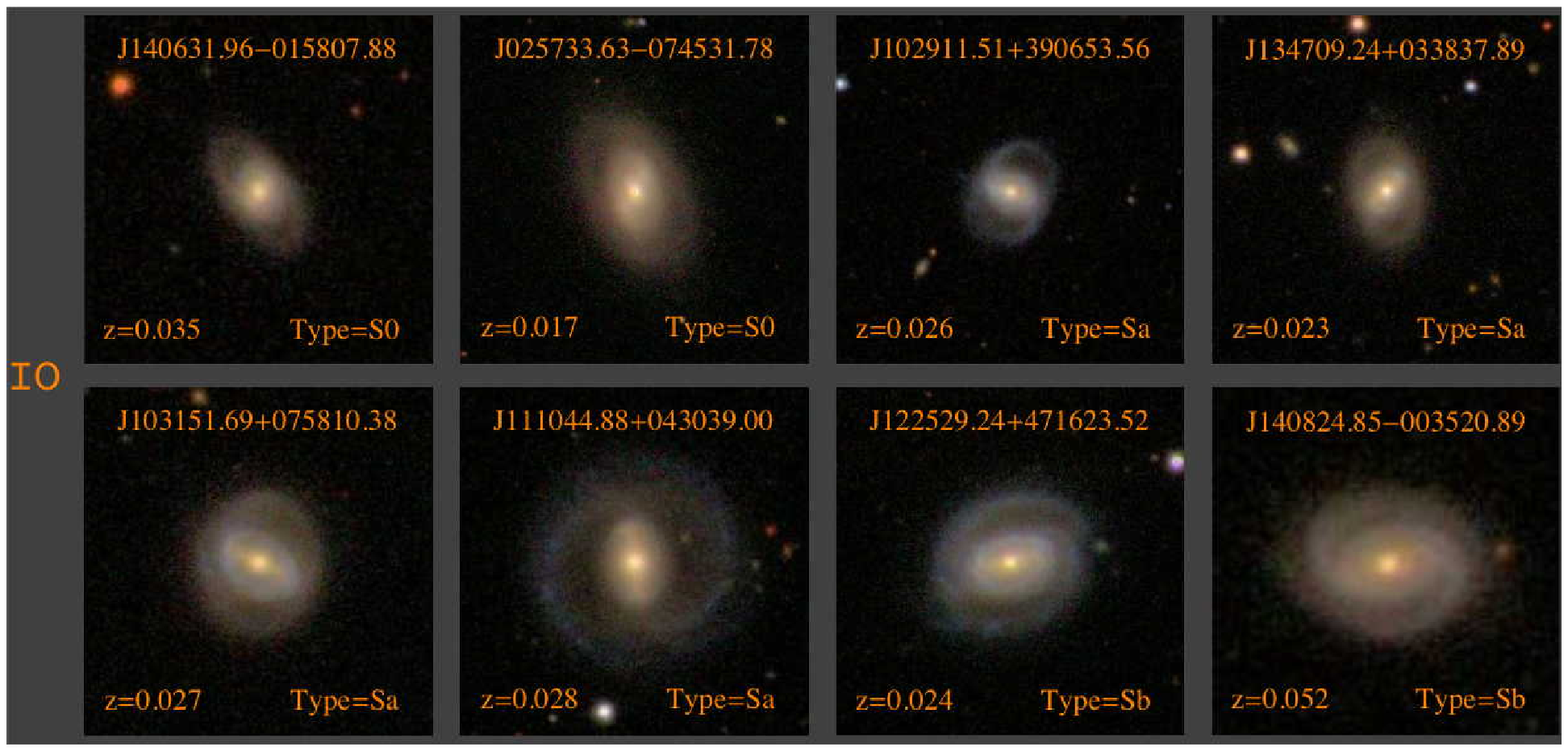}}}
\end{minipage}
\caption[Montage of Ringed galaxies]{\label{fig:RingMontage} 
A montage of images representing inner rings, outer rings and objects with both (2 rows for each) as classified by the author. Inner rings are more easily identified in barred object where they begin near where the bars end. Partial rings have also been included in this category. Outer rings are fairly easily identified in most systems. Pseudo-Rings as defined by \cite{Buta:1990p8843} are included in this category. Confusion can arise in systems with no bars and only one ring as well as with collisional ring systems. The J2000 object identifier is listed at the top with the redshift and the author classification at the bottom. Each stamp is 50 $h^{-1}$ kpc on a side. Although color images are shown, the classification was carried out on the g-band images. See text for details. The color images are taken from the SDSS Imaging Server. 
 }
\end{figure*}

\begin{figure*}[htbp!]
\unitlength1cm
\hspace{1cm}
%\vspace{1.5cm}
\begin{minipage}[t]{4.0cm}
\rotatebox{0}{\resizebox{16cm}{11.5cm}{\includegraphics{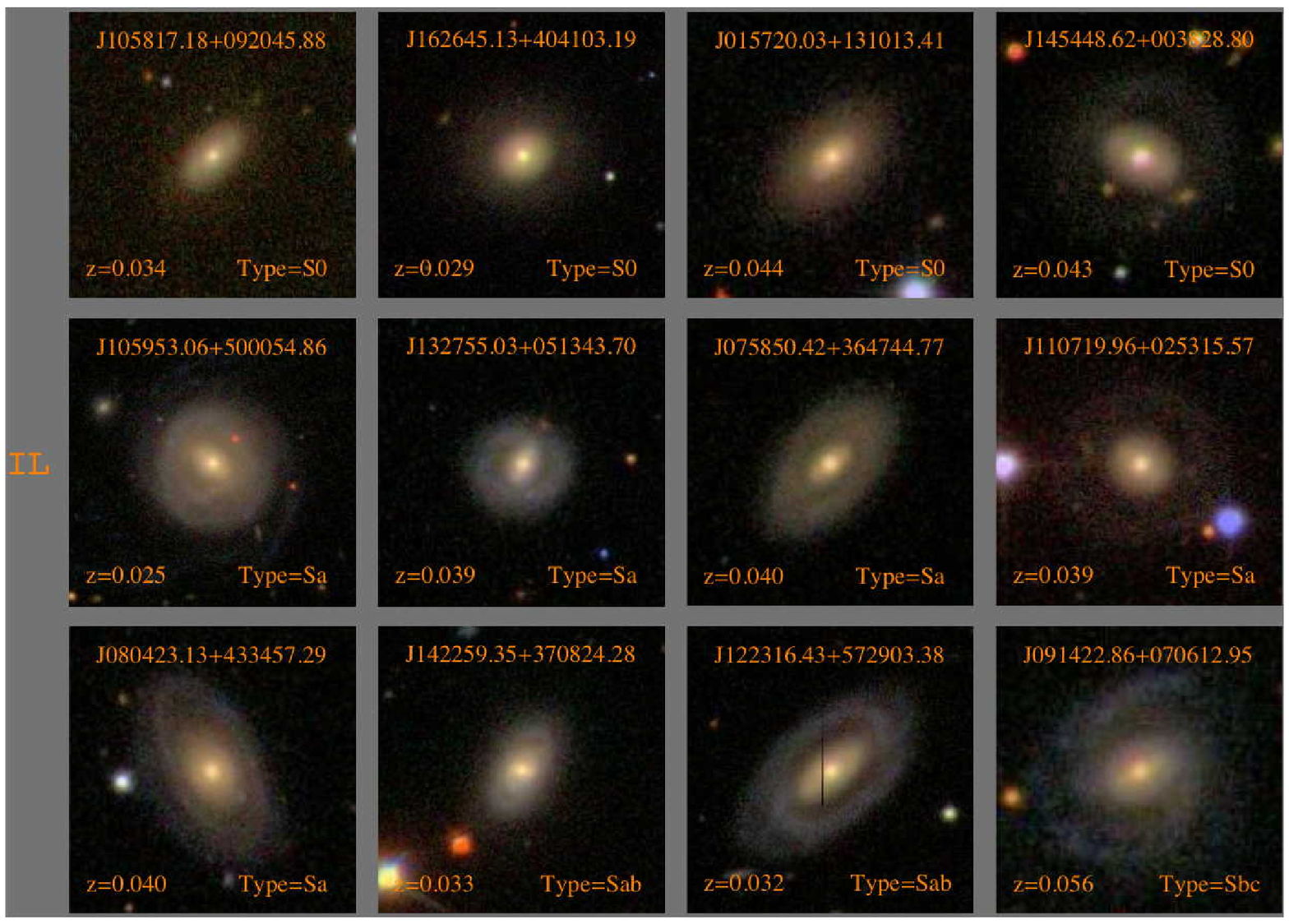}}}
%\vspace{0.5cm}
\rotatebox{0}{\resizebox{16cm}{11.5cm}{\includegraphics{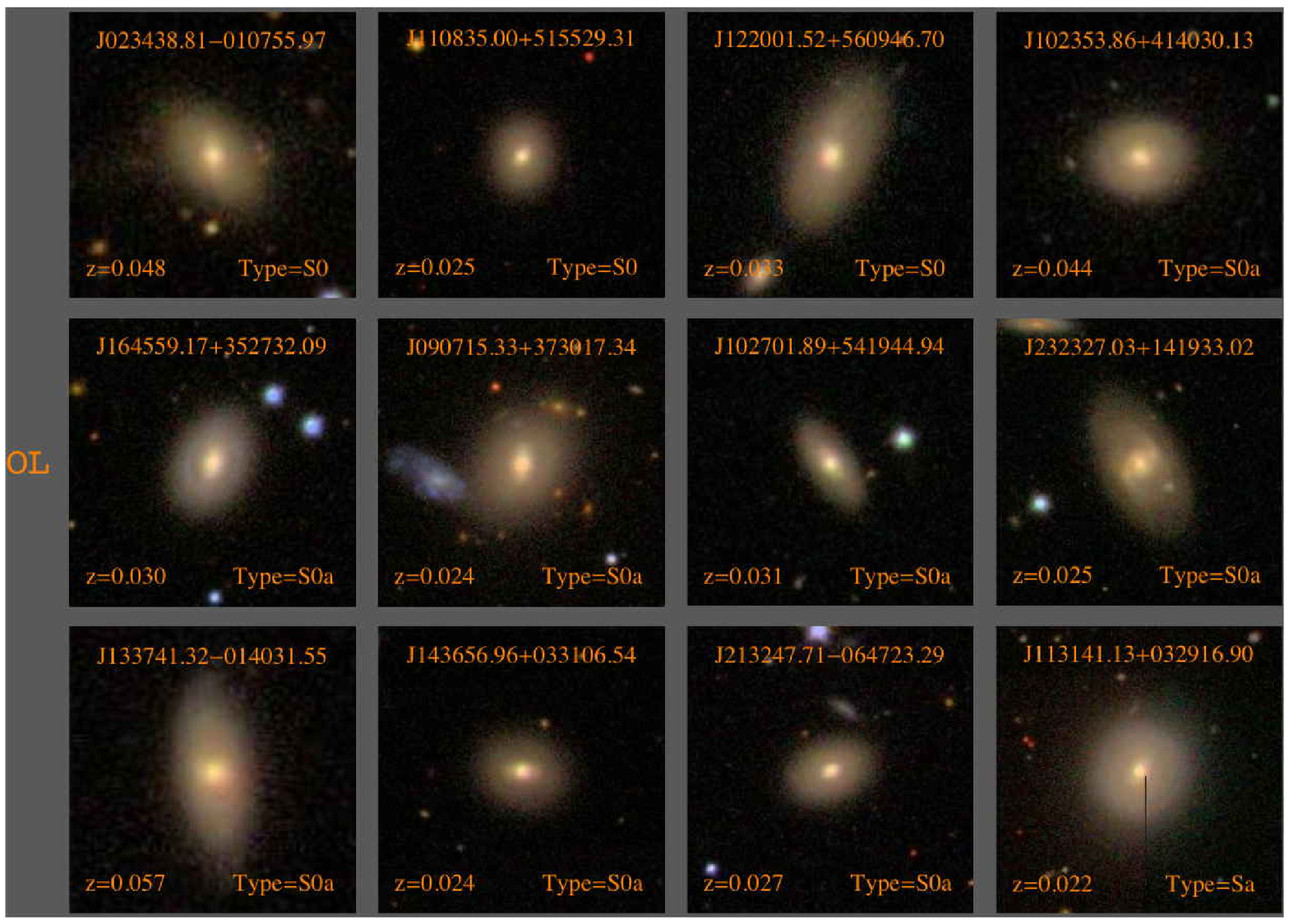}}}
\end{minipage}
\caption[Montage of lenses in galaxies]{\label{fig:LensMontage} 
A montage of images representing Inner and Outer lenses (3 rows for each) as classified by the author. The lenses are seen as regions of near constant surface brightness with very little variation with radius. Inner lenses are most easily identified when they have an outer ring. Outer lenses can also lead to outer rings. The J2000 object identifier is listed at the top with the redshift at the bottom left and the author classification at the bottom right. Each stamp is 50 $h^{-1}$ kpc on a side. Although color images are shown, the classification was carried out on the g-band images. See text for details. The color images are taken from the SDSS Imaging Server. 
 }
\end{figure*}

\begin{figure*}[htbp!]
\unitlength1cm
\hspace{1cm}
%\vspace{1.5cm}
\begin{minipage}[t]{4.0cm}
\rotatebox{0}{\resizebox{16cm}{7.8cm}{\includegraphics{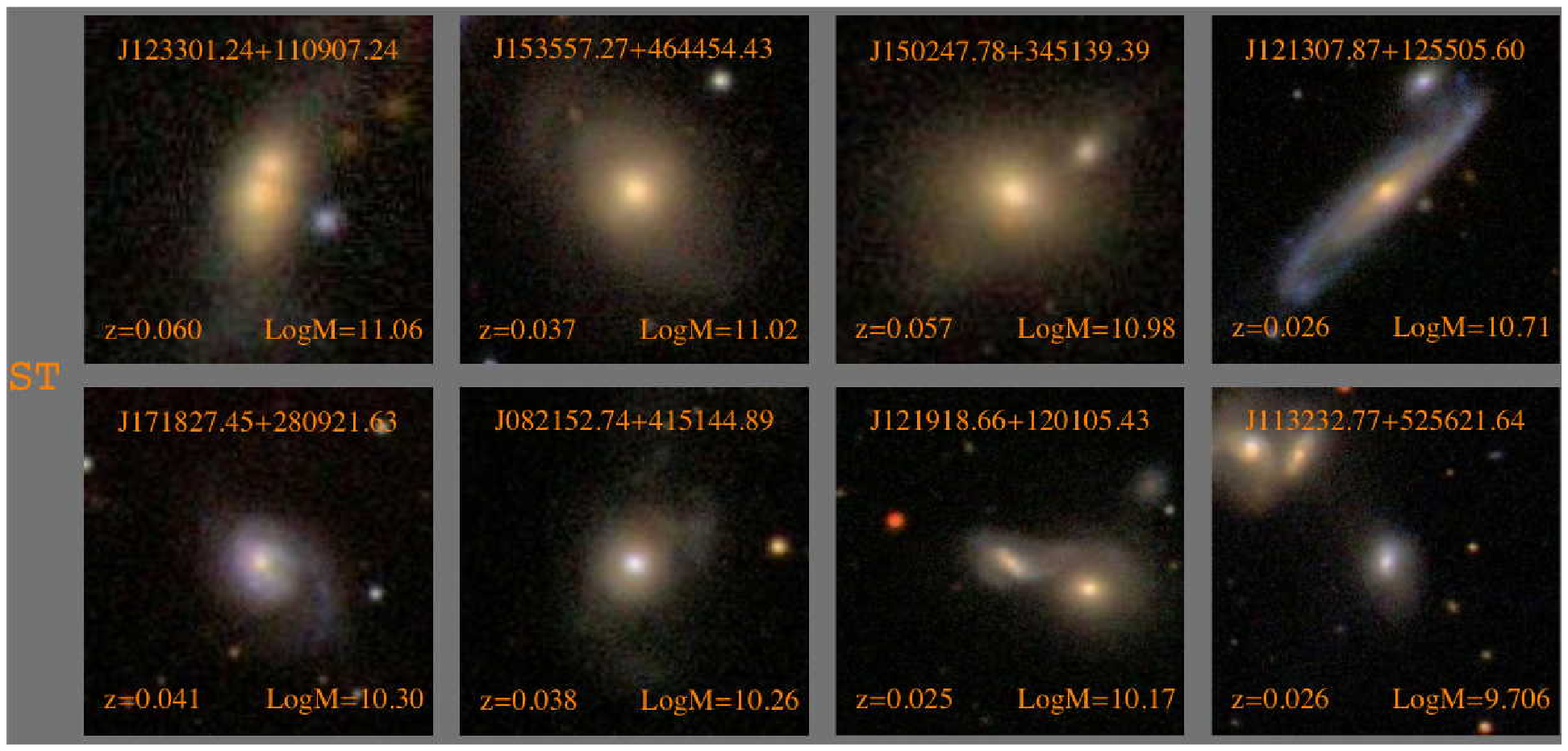}}}
\rotatebox{0}{\resizebox{16cm}{7.8cm}{\includegraphics{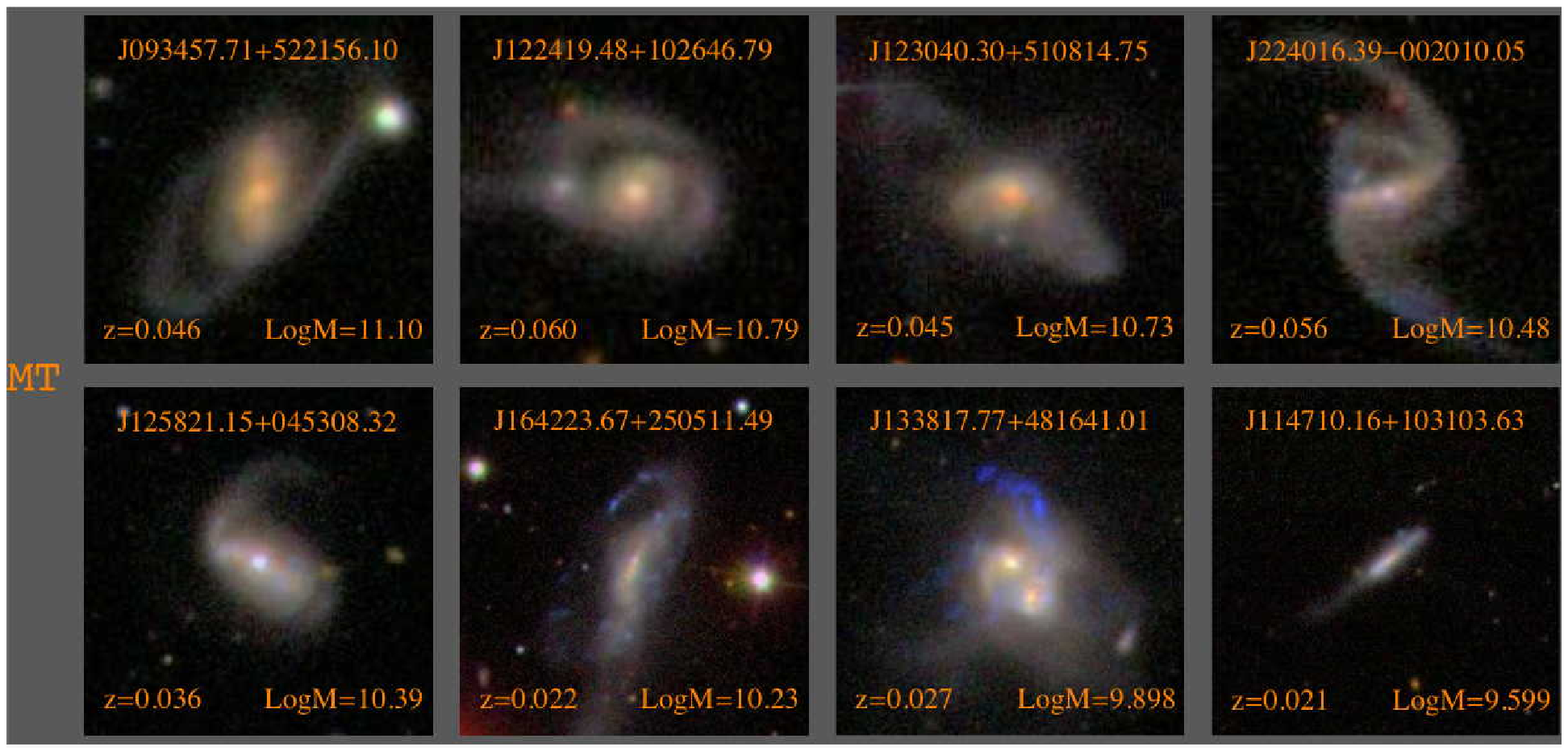}}}
\rotatebox{0}{\resizebox{16cm}{7.8cm}{\includegraphics{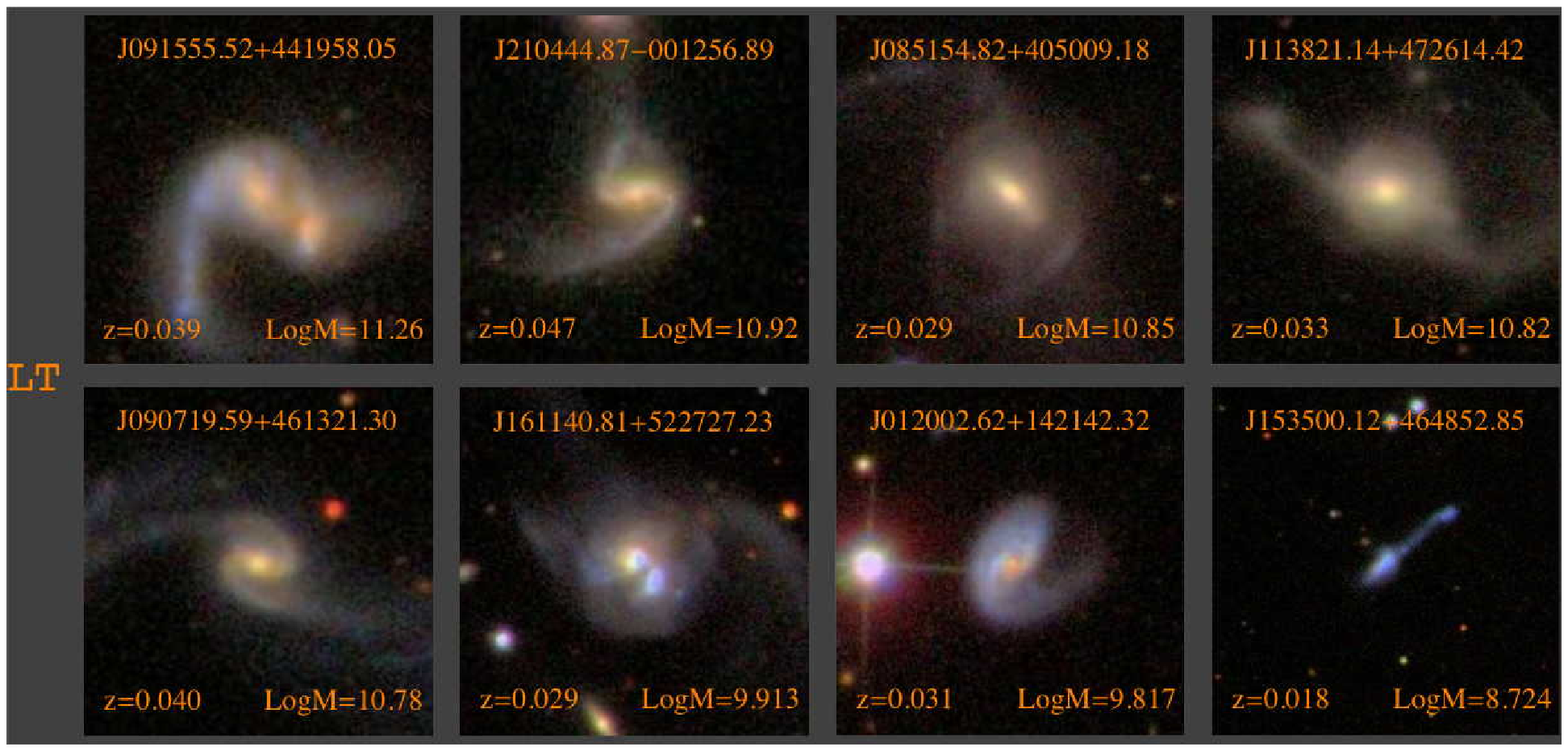}}}
\end{minipage}
\caption[Montage of Interacting Tails]{\label{fig:TailsMontage} 
A montage of images representing short, medium and long tidal tails (2 rows for each) as classified by the author. Tidal tails are classified into the three categories based on comparison with the host galaxy size. If the tails are much larger than the galaxy, they are classified as long tails. Tails comparable in size to the galaxy are classified as medium tails while those much smaller than the galaxy are classified as short tails. Many objects display multiple tails. The J2000 object identifier is listed at the top with the redshift at the bottom left and the author classification at the bottom right. Each stamp is 50 $h^{-1}$ kpc on a side. Although color images are shown, the classification was carried out on the g-band images. See text for details. The color images are taken from the SDSS Imaging Server. 
 }
\end{figure*}

\begin{figure*}[htbp!]
\unitlength1cm
\hspace{1cm}
%\vspace{1.5cm}
\begin{minipage}[t]{4.0cm}
\rotatebox{0}{\resizebox{16cm}{12cm}{\includegraphics{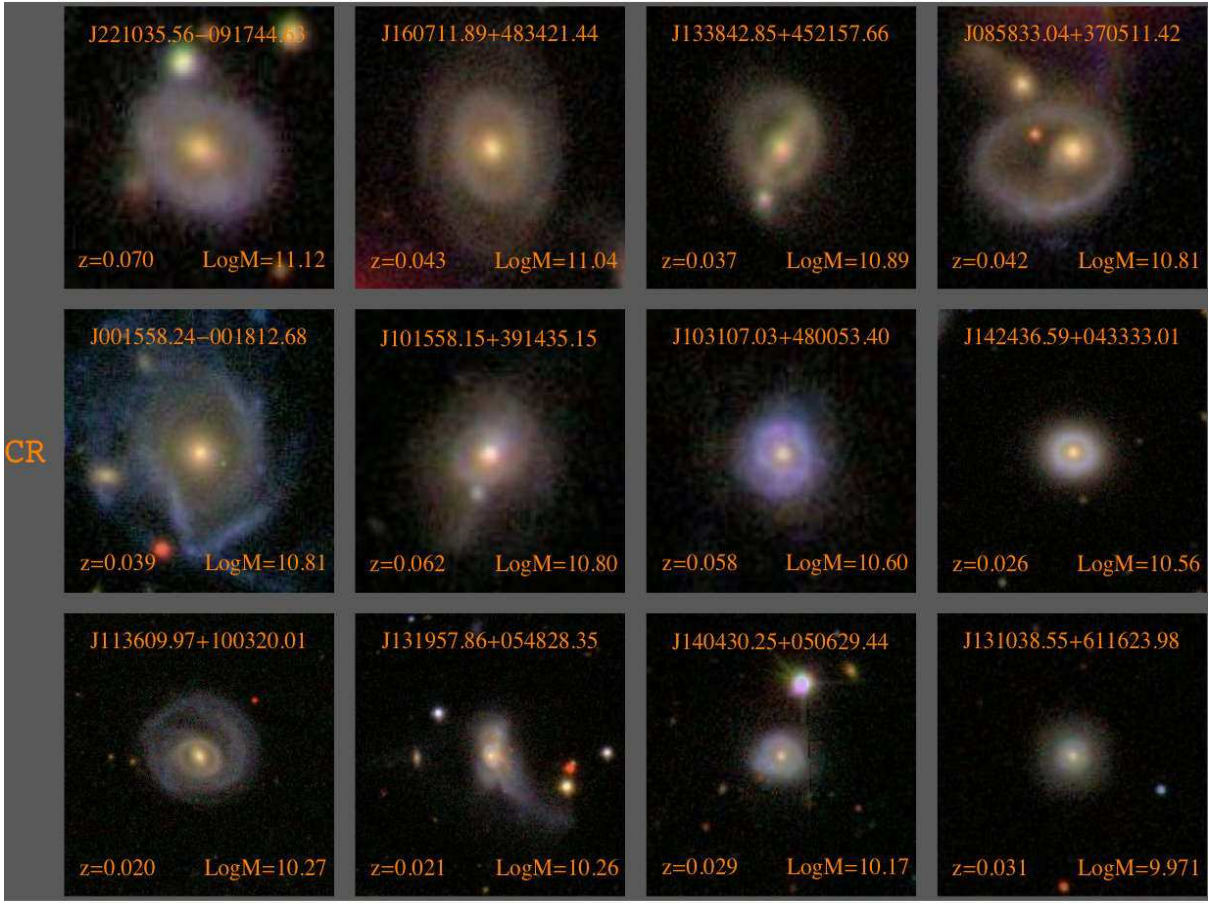}}}
\end{minipage}
\caption[Collisional Ring galaxies]{\label{fig:CollisionalRinged} 
A mosaic of images representing collisional ring systems, which are ringed galaxies formed by bulls-eye collision between two galaxies, as classified by the author. The classification into collisional rings is based on the color stamp of the galaxy (as the rings are normally blue), the shape of the ring and on the presence of spokes. The J2000 object identifier is listed at the top with the redshift and the author classification at the bottom. Each stamp is 50 $h^{-1}$ kpc on a side. The color images are taken from the SDSS Imaging Server.
 }
\end{figure*}

\begin{figure*}[htbp!]
\unitlength1cm
\hspace{1cm}
%\vspace{1.5cm}
\begin{minipage}[t]{4.0cm}
\rotatebox{0}{\resizebox{16cm}{8cm}{\includegraphics{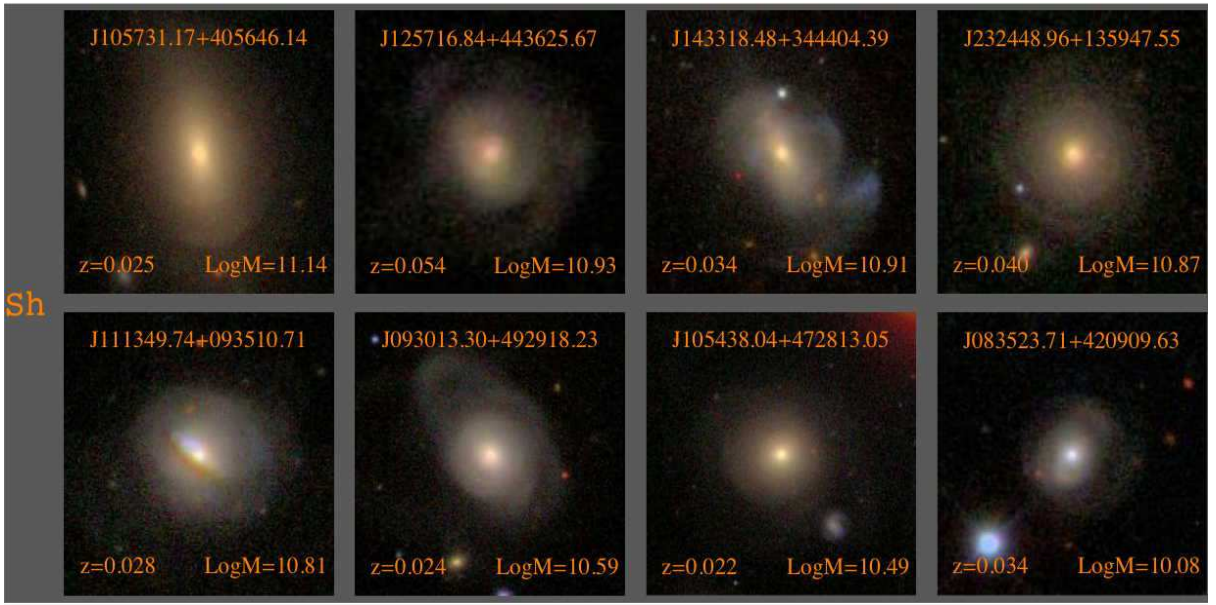}}}
\end{minipage}
\caption[Shells]{\label{fig:Shells} 
A mosaic of images representing shells as classified by the author. The J2000 object identifier is listed at the top with the redshift and the author classification at the bottom. Each stamp is 50 $h^{-1}$ kpc on a side. Although color images are shown, the classification was carried out on the g-band images. See text for details. The color images are taken from the SDSS Imaging Server. 
 }
\end{figure*}

\begin{figure*}[htbp!]
\unitlength1cm
\hspace{1cm}
%\vspace{1.5cm}
\begin{minipage}[t]{4.0cm}
\rotatebox{0}{\resizebox{16cm}{7.8cm}{\includegraphics{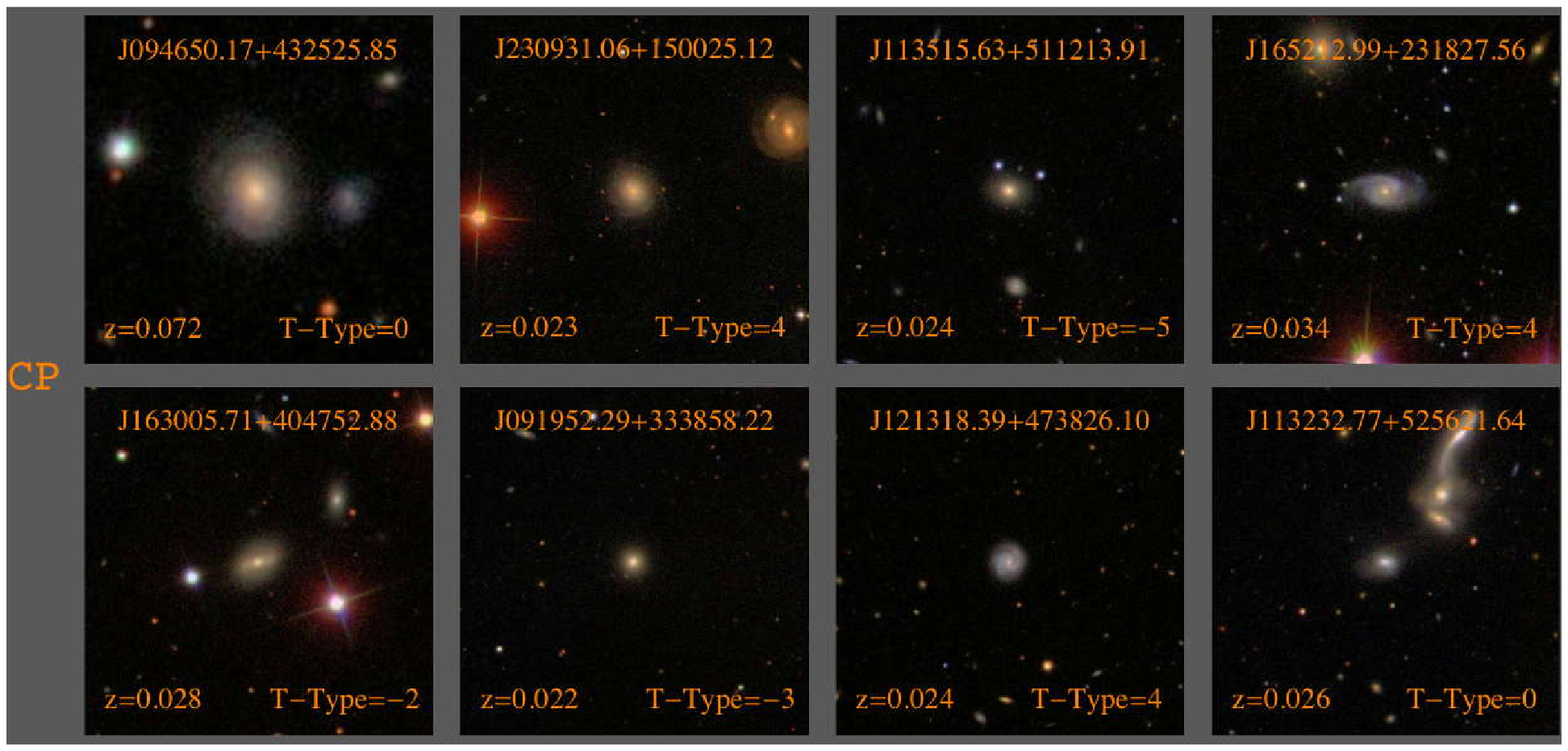}}}
\rotatebox{0}{\resizebox{16cm}{7.8cm}{\includegraphics{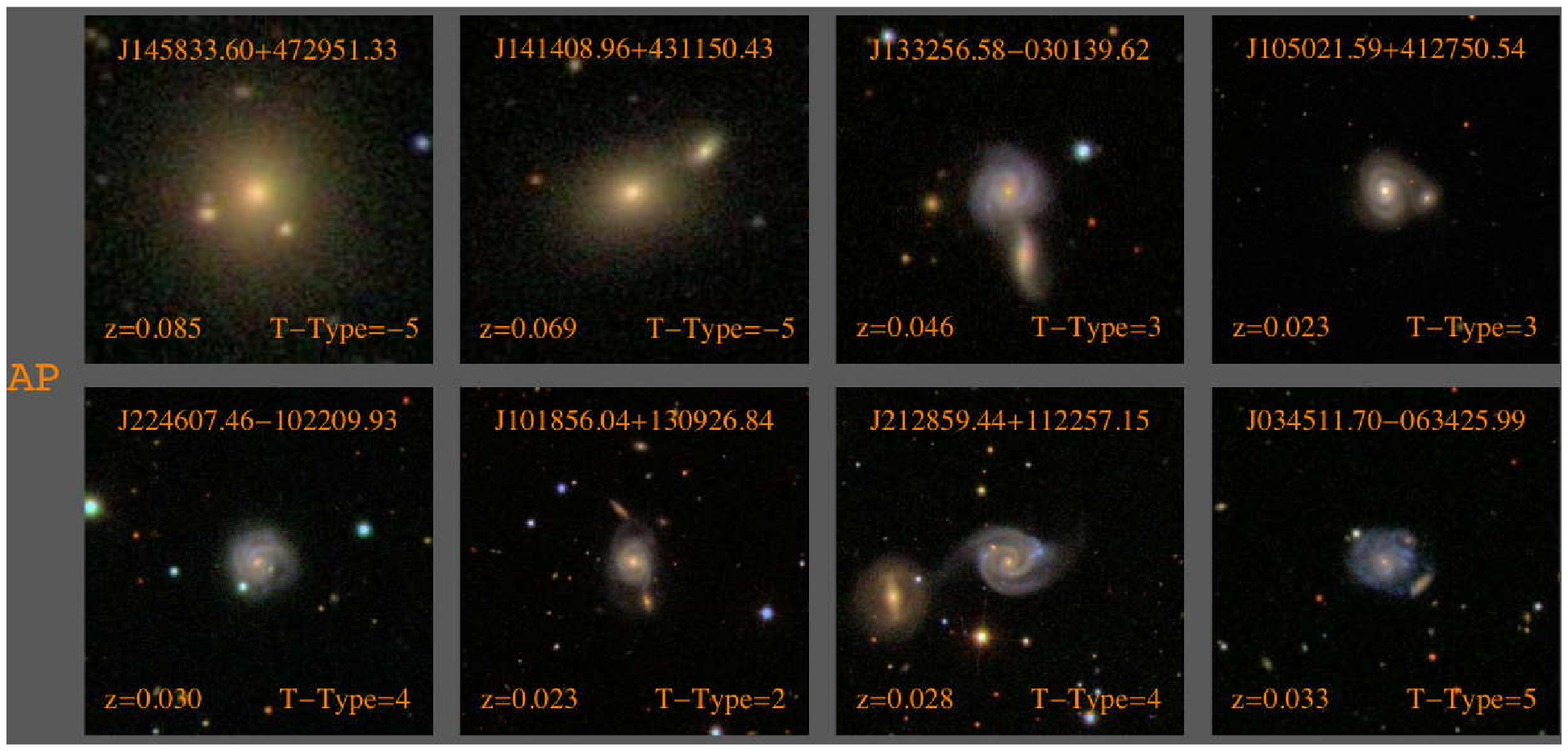}}}
\rotatebox{0}{\resizebox{16cm}{7.8cm}{\includegraphics{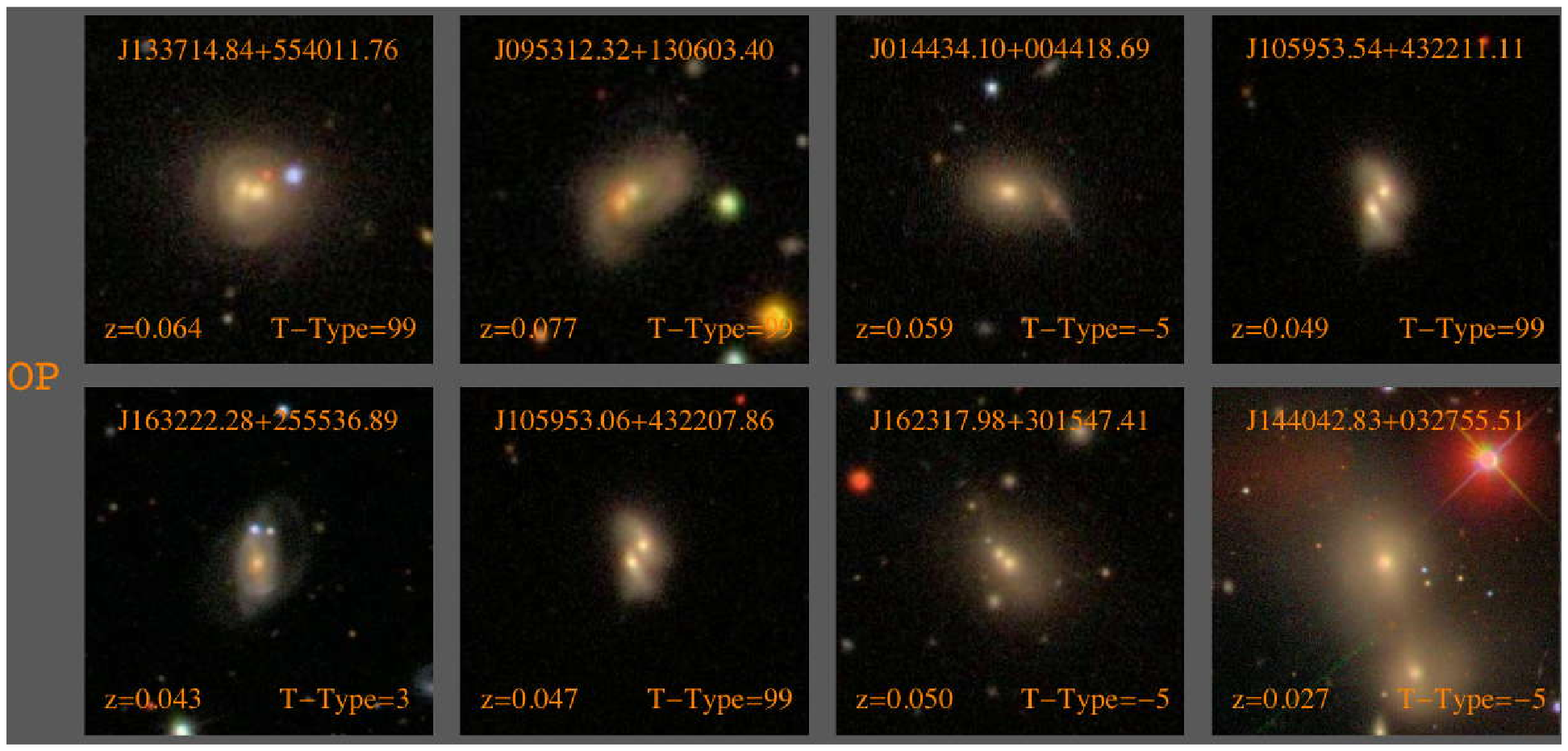}}}
\end{minipage}
\caption[Paired galaxies]{\label{fig:PairedGalaxies} 
A mosaic of images representing paired systems. The J2000 object identifier is listed at the top with the redshift and the author classification at the bottom. Each stamp is 100 $h^{-1}$ kpc on a side. The color images are taken from the SDSS Imaging Server. {\bf Top}: Close pairs, which may or may not be interacting, {\bf Middle}: Adjacent pairs and {\bf Bottom}: Overlapping pairs. The pairs may not necessarily be real and are not necessarily galaxies. In some cases they may be stars which are very close to the galaxy of interest. The pair type segmentation flag gives the type of object in the pair. Refer to the text and Section 7 for the catalog description. 
 }
\end{figure*}

Figure~\ref{fig:BarMontage} shows a random sample of galaxies with strong, intermediate and weak bars. The strength of the bar is defined in terms of the size of the bar compared to the galaxy diameter and its prominence. In our system, we refer to those bars that dominate that light distribution as strong bars. Weak bars are smaller in size and contain only a small fraction of the total flux of the galaxy. Intermediate bars span the range between strong and weak bars. {\em It is important to note that in our scheme all the bar types are viewed as definite bars --- this is unlike the RC3 scheme, where systems classed weakly barred include objects that possibly contain bars.} In this sense our system is more conservative than that of the RC3, and (as will be shown in the next section) our bar fraction is lower. It is also worth noting that our classification of bar strength is based on inspection of the $g'$-band images but the determination of a bar's existence was based on studying each galaxy in all three bands ($g', r', i'$). However, in nearly every case (98\% of the time) bars observed in the $r'$ and $i'$ bands are also observed in $g'$. As expected, many of our barred objects have a ring or lens component. We will investigate the connection between rings and bars in a later paper in this series.

\subsubsection{Rings and Lenses}

Figure~\ref{fig:RingMontage} shows a montage of inner (top panel), outer (middle panel) and combination (bottom panel) ringed galaxies, and Figure~\ref{fig:LensMontage} shows a montage of galaxies with inner lenses (top panel) or outer lenses (bottom panel). Inner rings are more easily identified when bars are present. In galaxies without bars, inner ring classifications are much harder due to confusion with outer rings. Outer rings and pseudo outer rings as defined by \cite{Buta:1996p2930} are also distinguishable. Confusion can occur with lens galaxies as in Figure~\ref{fig:LensMontage} and collisional ring systems. Lens galaxies (or galaxies with regions of constant surface brightness) can exhibit either an inner or an outer lens. Both lens types can lead to the corresponding rings. There may be systems where outer rings eventually form lenses \citep{Buta:1996p2930}.
%So there are almost certainly galaxies with outer lenses that eventually form outer rings.

\subsubsection{Interaction signatures (tails, shells, collisional rings)}
We have also classified objects based on interaction signatures. We identify objects with tidal tails, collisional rings, and shells, all of which seem to be nearly fool-proof signatures of interactions. Figure~\ref{fig:TailsMontage} shows examples of galaxies with short, intermediate and long tails. The strength of the tail is defined in terms of the size of the tail compared to the size of the galaxy. Tails larger than the diameter of the galaxy are classified as large tails. Tails comparable to the diameter of the galaxy are classified as medium (or intermediate) tails, while tails much smaller than the diameter of the galaxy are classified as short tails. Multiple tails can exist in a single system. Figure~\ref{fig:CollisionalRinged} shows a sample of collisional ring systems which are ringed galaxies caused by bulls-eye collisions between two galaxies. The rings formed are bluer than what would be seen in a normal-ringed galaxy and can also be asymmetric. Our classification into collisional rings is based on the color stamp of the galaxy, the shape of the ring, on the presence of spokes (as in the cartwheel galaxy ESO 350-G040) and/or the presence of a nearby companion galaxy with which the ringed galaxy may be interacting. Figure~\ref{fig:Shells} shows examples of galaxies which have shells and are most probably the end stage of a merger. They are predominantly disturbed E, E/S0, and S0 galaxies. %The `Interacting type' flag should be used to select objects.

%XXX:-
In addition to these fool-proof signatures of interaction, we identify if objects are in pairs, i.e. with a companion object within a 50kpc radii based on the 100kpc color stamps. Specifically objects are flagged as `close pairs', `adjacent pairs' or `overlapping pairs' with a fourth `projected pair' flag set if the paired object seems to be a projection effect or if it is completely enclosed by the light profile of the primary object. It is important to note that the paired object is not necessarily a galaxy but can also be a star. There are 38 galaxies with diffraction spikes from a nearby star running through the image. The `Pair' flag (see \S \ref{sec:catalog})should be used in conjunction with the `Pair-type' flag to select/exclude specific companion objects.  The `Interaction-type' flag should be used to select pairs with merger signatures such as tidal tails. We use the `Pair-type'  flag to select clean samples of galaxies with no contamination by companions (real or projected). Figure~\ref{fig:PairedGalaxies} shows representative examples of `close pairs', `adjacent pairs' and `overlapping pairs'.

\section{COMPARISON WITH PREVIOUS WORK}
\label{sec:reliability}
\begin{deluxetable*}{lrrrrrrrcccccccccccr}
\tabletypesize{\scriptsize}
\tablecaption{T-Type Classification Schemes \label{tab:TTypeClassification}}
\tablecolumns{19}
\tablewidth{0pt}
\tabletypesize{\scriptsize}
\tablehead{
\colhead{Class } & 
\colhead{c0 } &
\colhead{E0 } &
\colhead{E+ } &
\colhead{S0-} &
\colhead{S0 } &
\colhead{S0+ } &
\colhead{S0/a } &
\colhead{Sa} &
\colhead{Sab } &
\colhead{Sb} &
\colhead{Sbc } &
\colhead{Sc} &
\colhead{Scd } &
\colhead{Sd} &
\colhead{Sdm } &
\colhead{Sm} &
\colhead{ Im} &
\colhead{?}
}
\startdata
RC3          & -6   & -5 & -4 & -3  & -2  & -1 & 0  & 1  & 2 & 3 &  4 &   5 & 6 &  7 &   8 &   9 &   10 & : \\  
Fukugita   &  0   &   0 &  0  &  1  &  1  &  1 & 1 & 2  & 2 & 3 &  3 &   4 & 4  &  5  & 5  &   6 &    6  & -1\\  
PN (this work)             &  -5   & -5 & -5 & -3  & -2  & -2 & 0  & 1  & 2 & 3 &  4 &   5 & 6 &  7 &   8 &   9 &   10 & 99\\ 
\enddata
\end{deluxetable*}

\begin{figure*}[t!]
\unitlength1cm
\hspace{0.5cm}
\begin{minipage}[t]{5.0cm}
\rotatebox{0}{\resizebox{7.5cm}{7.5cm}{\includegraphics{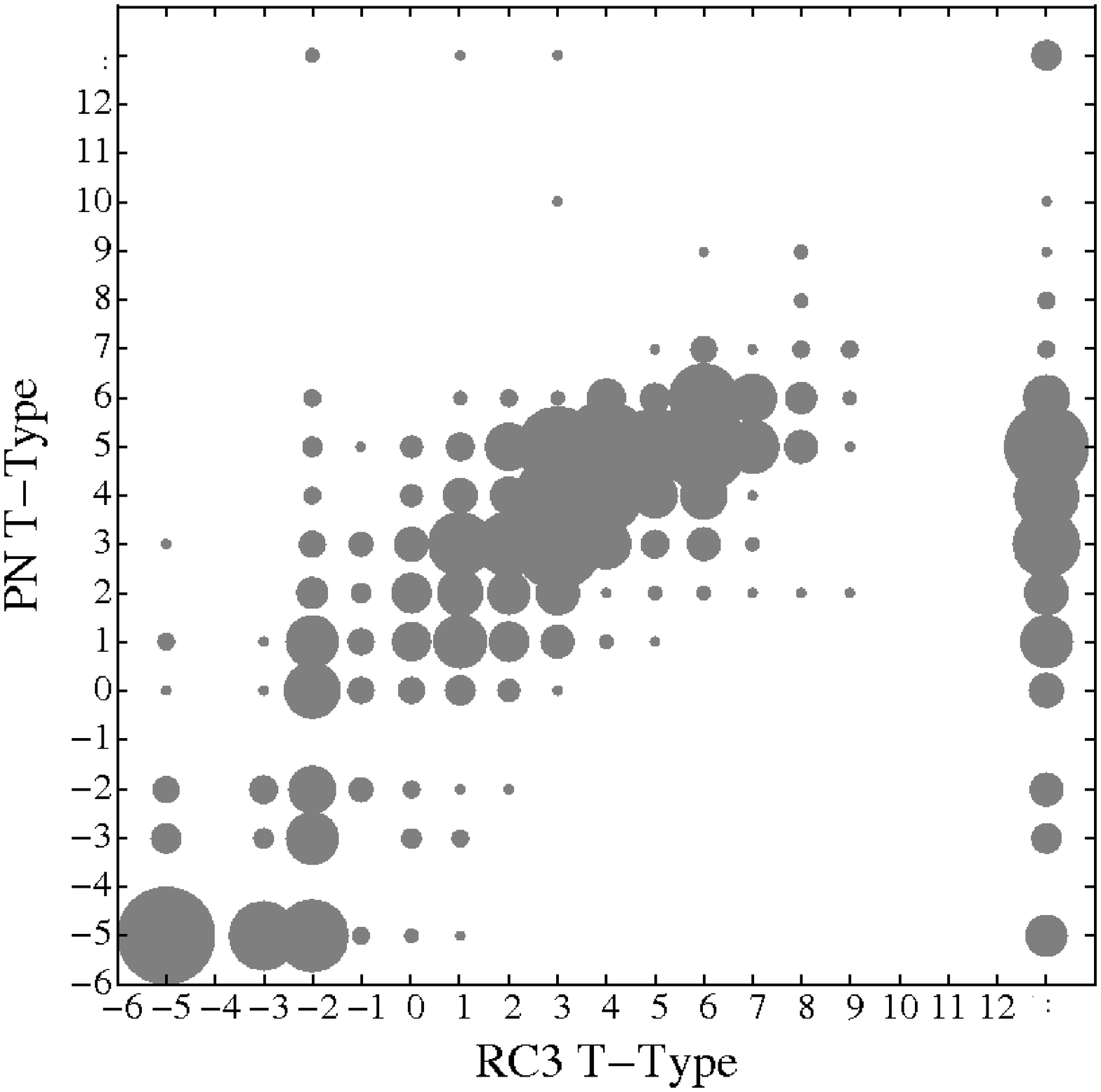}}}
\end{minipage}
\hspace{4cm}
\begin{minipage}[t]{5.0cm}
\rotatebox{0}{\resizebox{7.5cm}{7.5cm}{\includegraphics{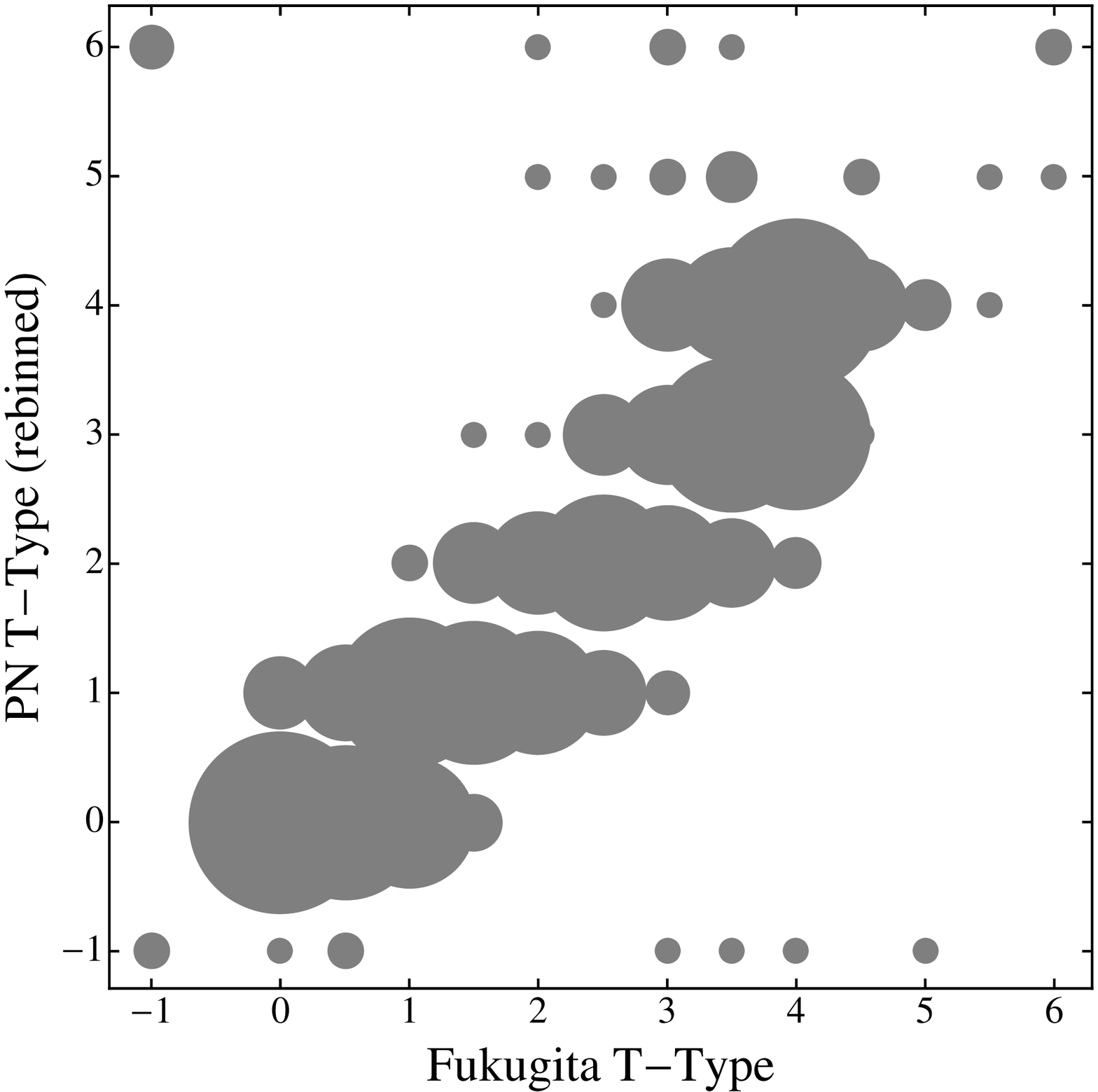}}}
\end{minipage}
\caption[Comparison of Classifications.]{\label{fig:ClassificationComparison} 
Point size is keyed to number of objects. (a) Comparison of our (PN) T-Types vs. RC3 classification for 325 objects in common. The mean deviation is $\thicksim$1.5 T-Types (b) Comparison of our (PN) T-Types vs \cite{Fukugita:2007p3031} classifications for 450 objects in common to both samples. The mean deviation is 0.8 Fukugita type bins.
}
\end{figure*}

\subsection{Comparison of T-Type classifications}

The left panel of Figure~\ref{fig:ClassificationComparison} shows a two-dimensional histogram comparing classifications for the 1793 objects in our catalog which overlap with the RC3. 
Point sizes in this figure correspond to the number of galaxies in each 2D histogram cell. The RC3 T-Types cE and cD (T-Type = -6$/$-4) and S0- (-1) are incorporated into our T-Types -5 and 0 respectively. There are a number of objects in the RC3 which are not defined explicitly but are stated to be either E (elliptical), L (lenticular) or S (disk) and highly doubtful. Objects classified as E or L alone without any subclass have been assigned to T-Types -5 and -2. Objects assigned an S only have been given a T-Type of `:' (or unknown) in this figure.  

Inspection of the left panel of Figure~\ref{fig:ClassificationComparison} shows that, overall, our classifications are quite consistent with those of the RC3. There is a weak trend for our classifications to be slightly later overall, but the mean deviation in classification is only about 1.2 T-Types, when excluding objects classified as unknown. This is about what is expected for expert classifier inter-comparisons (Naim et al. 1995). A significant difference with respect to RC3 is seen when one includes objects classed as `unknown' (T-Type=13) in the RC3. This is because our classification scheme incorporates more freedom to account for morphological peculiarities, such as mergers and signatures of interactions, and indeed most disagreements between RC3 classifications and our classifications occur in systems that exhibit morphological peculiarities.

The right panel of Figure~\ref{fig:ClassificationComparison} shows our classification compared to the 584 objects which overlap with the \cite{Fukugita:2007p3031} sample. We have re-binned our classification to match the scheme used by \cite{Fukugita:2007p3031}. Again the agreement between T-Types is fairly consistent but our classifications are slightly earlier than the Fukugita et al. (2007) classification scheme. The mean deviation in classification between our classifications and the Fukugita et al. (2007) classifications is $< 0.8$ bins, although we note that Fukugita et al. (2007) used rather coarse T-Type bins.

\subsection{Comparison of bar classifications}

For objects later than E/S0 we find bars, rings and lenses are 26\% $\pm0.5\%$, 25\% $\pm0.5\%$ and 5\% $\pm0.5\%$ of our sample population respectively. 
These fractions are lower limits which do not include objects for which we were not completely confident with the fine-classification. Inclusion of these objects increases the bar and ring fractions by 5\% each, and the lens fraction by 3\%. The bar fractions are still low compared to previous local studies which quote bar fractions higher than 60\% \citep{deVaucouleurs:1963p9392}, though consistent with the RC3 visual strong bar fractions in the local universe. 

\begin{figure*}[htbp!]
%\begin{center}
\unitlength1cm
\hspace{1cm}
%\vspace{1.5cm}
\begin{minipage}[t]{4.0cm}
\rotatebox{0}{\resizebox{16cm}{7.7cm}{\includegraphics{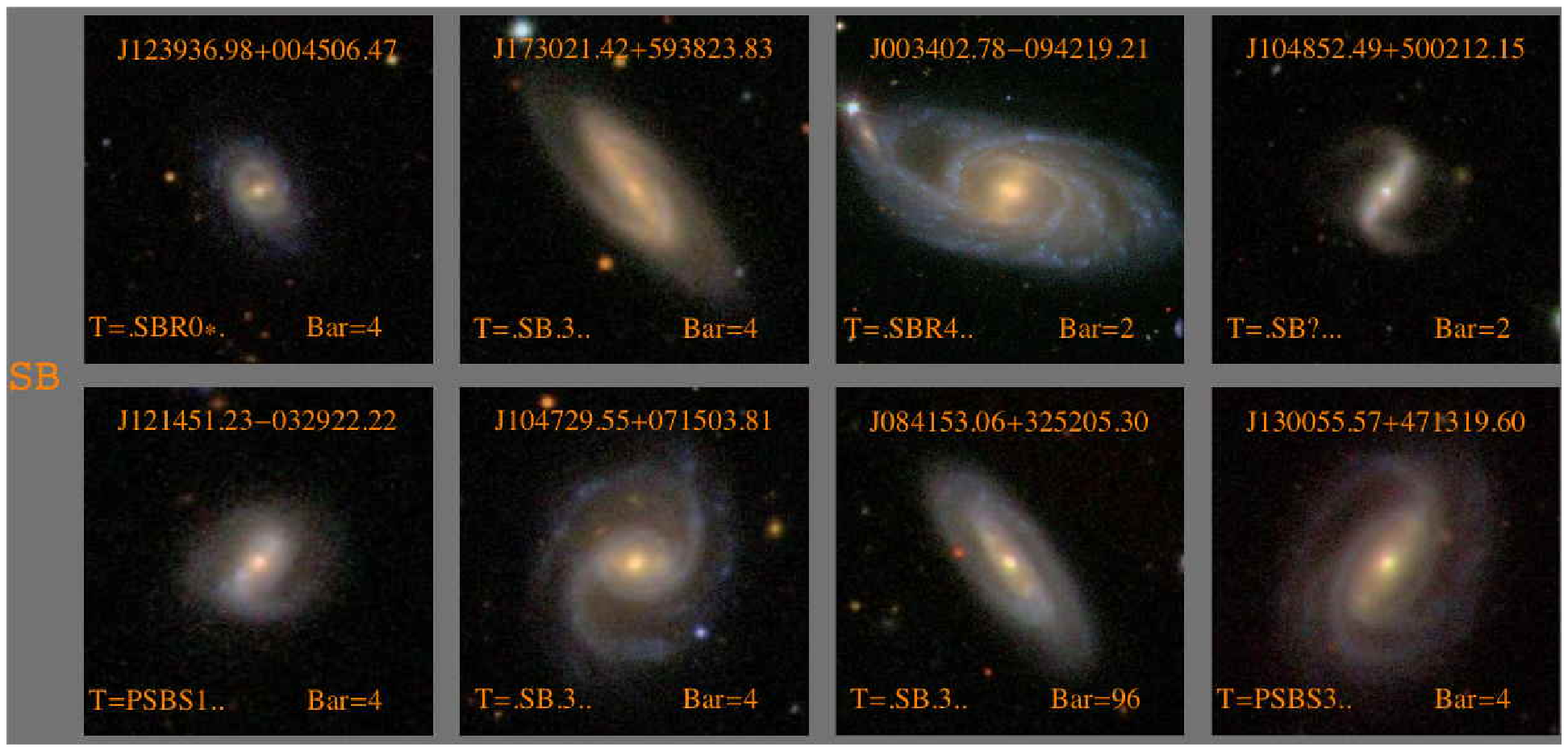}}}
\rotatebox{0}{\resizebox{16cm}{7.7cm}{\includegraphics{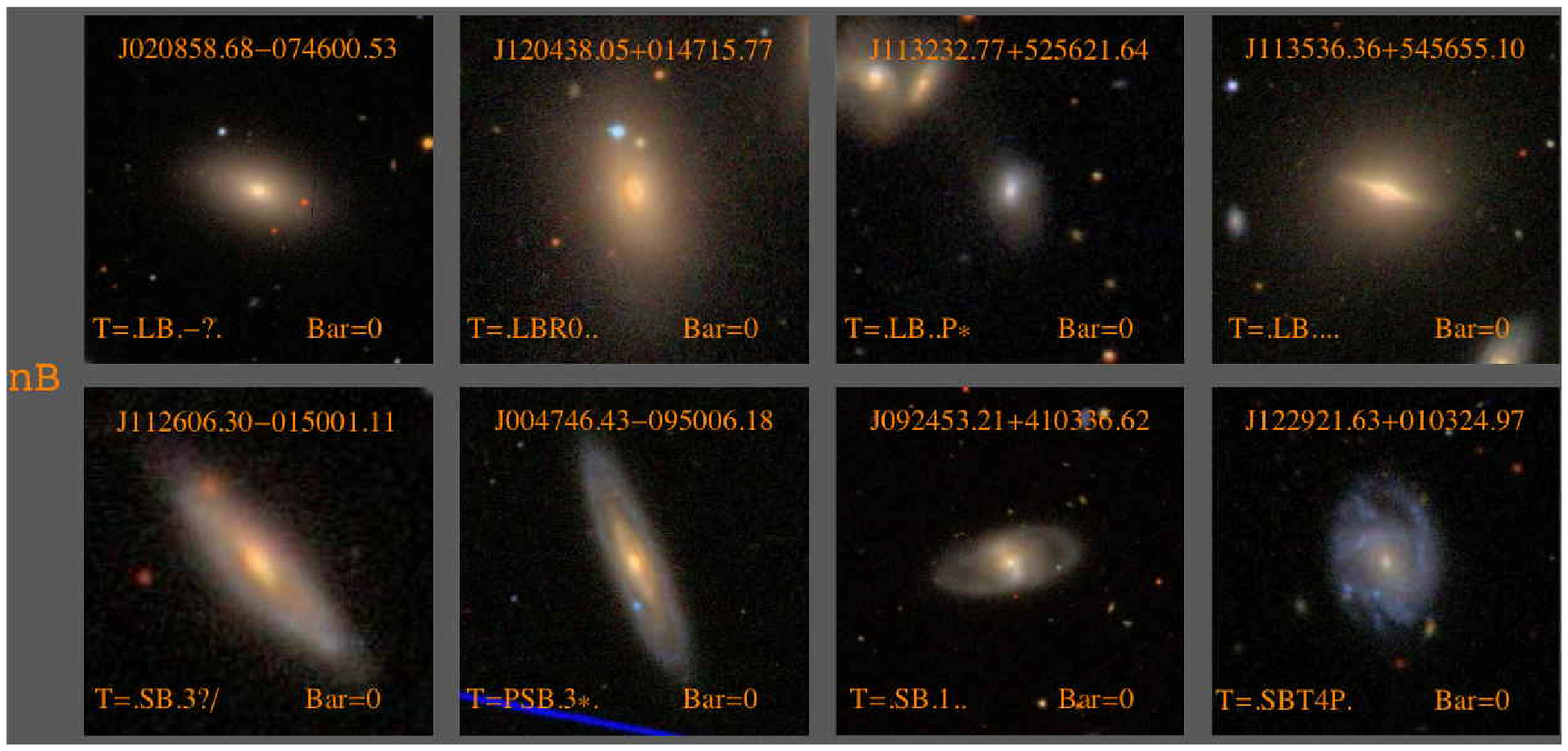}}}
\rotatebox{0}{\resizebox{16cm}{7.7cm}{\includegraphics{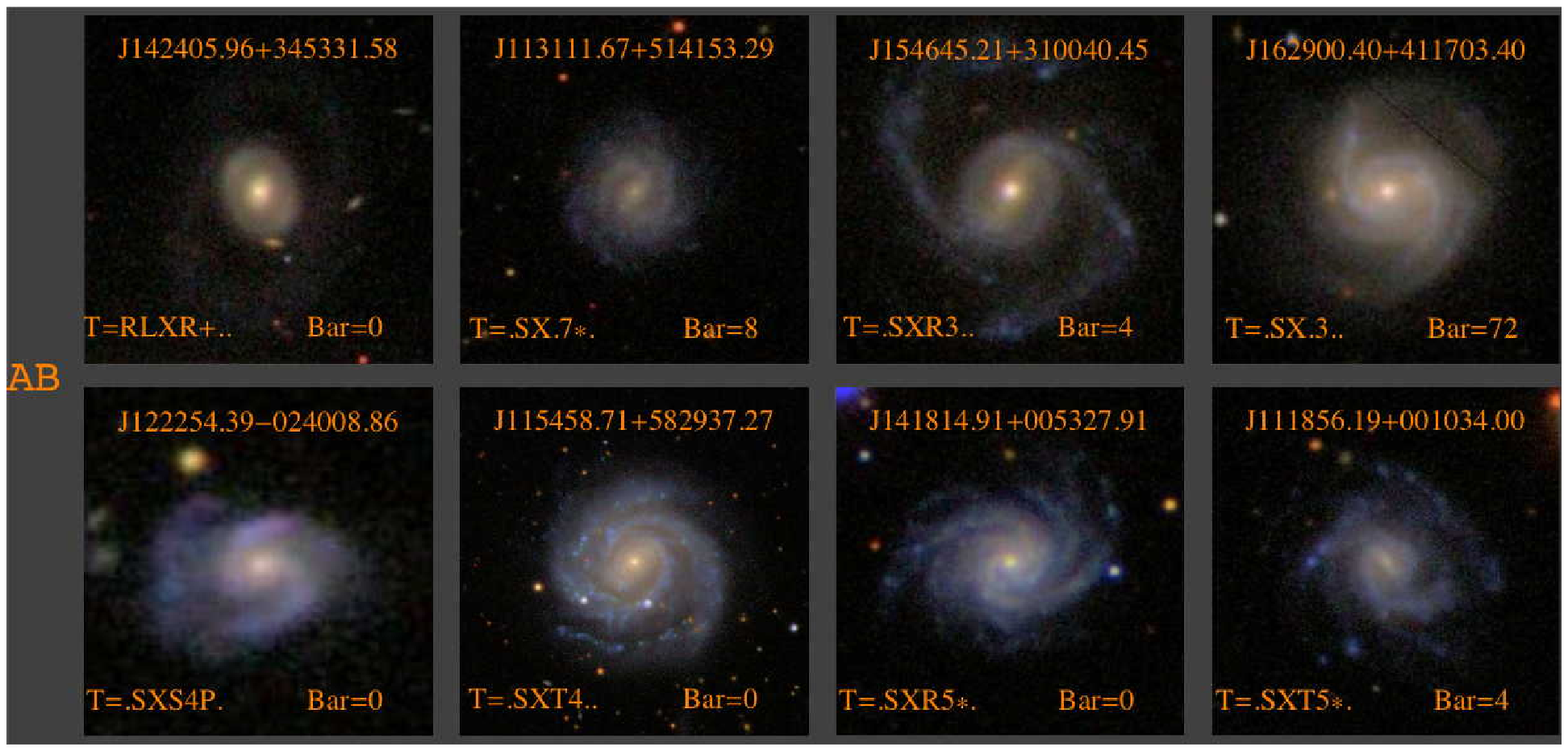}}}
\end{minipage}
\caption[Montage of Barred galaxies common with RC3 ]{\label{fig:RC3BarMontage} 
A montage of barred RC3 galaxies common with our sample split as follows: {\bf Top Panel}: Strong bars identified by both RC3 and this work (PN). {\bf Middle Panel}: Strong bars identified in RC3 but not by us and {\bf Bottom Panel}: Weak or AB bars identified by RC3 with our classification listed. Each stamp is 50 $h^{-1}$ kpc on a side. The J2000 object ID is listed at the top. The RC3 classification is listed at the bottom left hand corner and our (PN) bar flag classification in the right hand corner. Zero corresponds to no bar, 2 to a strong bar, 4 to an intermediate bar, 8 to a weak bar and any higher number to peanuts+ansae+unsure flags. As stated in the text, our bar classification cannot be equated directly to the RC3 bar classification. Our strong, intermediate and weak bar classes can be considered subdivisions of the RC3 strong bar class. See text for details. In summary, of the 71 objects defined as strongly barred in the RC3 sample, we find $\thicksim$ 66\% to be barred. Of the 25 objects considered to be weakly barred in the RC3, we find 44\% to be barred.
 }
% \end{center}
\end{figure*}

\begin{figure*}[htbp!]
%\begin{center}
\unitlength1cm
\hspace{1cm}
%\vspace{1.5cm}
\begin{minipage}[t]{4.0cm}
%\vspace{0.8cm}
\rotatebox{0}{\resizebox{16cm}{23cm}{\includegraphics{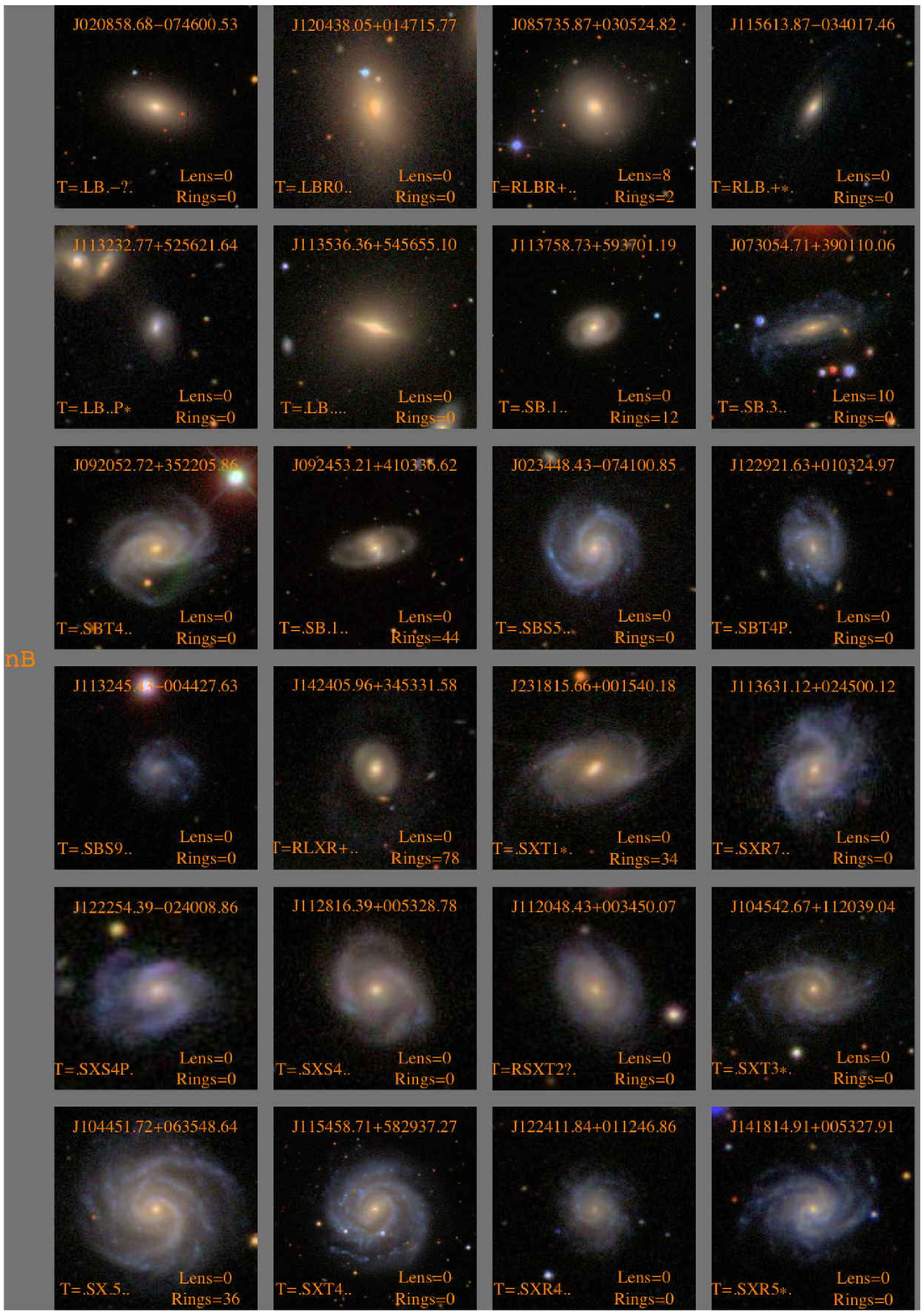}}}
\end{minipage}
\caption[RC3 missed face-on Montage of Barred galaxies ]{\label{fig:RC3MissedFaceOnBarMontage} 
A montage of face on ($b/a>0.5$) galaxies classified as strongly or weakly barred in RC3 but as unbarred in our classification. The J2000 object ID is listed at the top. The RC3 classification is listed at the bottom left hand corner and the PN ring and lens classifications in the right hand corner. 0 corresponds to no ring/lens, 2 to PN inner ring/lens, 4 to outer ring/lens and any higher number to rings or lens with partial, pseudo or unsure flags set. See text for details regarding classification designations. The galaxies are arranged in order of bar strength with the RC3 strong bars first followed by the RC3 weak bars. Each stamp is 50 $h^{-1}$ kpc on a side. 
 }
% \end{center}
\end{figure*}

As stated previously, our bar classifications are fundamentally different from RC3 bar classifications. We carefully examined all galaxies in common between our catalog and the RC3 in order to understand the differences, and conclude that our strong, intermediate and weak bar classes are probably best considered to be subdivisions of the RC3 strong bar class\footnote{Our initial comparison was with the SDSS DR2 release which matched 350 RC3 objects. The bar and ring comparison plots shown in this section are for the 350 matched RC3 galaxies, not the 1793 matched sample. Our results with the larger sample are consistent}. To illustrate this, Figure~\ref{fig:RC3BarMontage} shows a montage of SDSS galaxies which are defined as strongly (B) or weakly barred (X) in the RC3. In the top panel we show objects which are defined as barred by PN and strongly barred by RC3. Most of the bars are large in scale and there is no confusion in these cases. The middle panel shows objects classified as strong bars by RC3 but as being unbarred by us. With the exception of three galaxies, the others cannot be definitively stated to be barred from the SDSS images. Two of the galaxies which do appear barred (J004746.43-095006.18 and J092453.21+410336.62) are very weak and appear to be ansae. In the case of J004746.43-095006.18 the confusion arises due to inclination effects where the object could either be considered to have an inner ring or a weak bar. We opted for an inner ring. The third missed object, J113536+545655.10, was classified as a peculiar S0 galaxy as opposed to barred. The bottom panel in Figure~\ref{fig:RC3BarMontage} shows objects considered weakly barred in RC3 (denoted by X in the short designation on the bottom left of the images). In 4 of the 8 cases we classify the galaxy as barred or possibly barred (bar type$>$8, see \S 7). In three of the galaxies we cannot identify bars. J142405.96+345331.58 seems more likely to be an inner lens with an inner ring and possibly a nuclear bar. In J122254.39-024008.86, the bar is difficult to distinguish from the arms in the SDSS image. In J115458.71-582937.27, the galaxy could have been considered to host an intermediate bar by us and illustrates that although classifications have been carried out twice, there may be some intermediate/weak bars that have been missed. In summary, of the 71 objects defined as strongly barred in the RC3 sample, we find $\thicksim$ 66\% to be barred. Of the 25 objects considered to be weakly barred in the RC3, we find 44\% to be barred. Thus our detection rate of RC3 strong bars is statistically much higher than our detection of RC3 weak bars. 

It is apparent from Figure~\ref{fig:RC3BarMontage} that we may have missed some bars because of (a) inclination effects, or (b) because we may have classified some of the RC3 strong and weak bars as other fine features like lenses or rings, or (c) that in some cases our images may be too shallow to detect some bars, or (d) because our definition of a bar is very strict in comparison to that used in previous work, or (e) because the RC3 is in error. We consider the importance of each effect in turn. 

Applying an axial ratio cut of $b/a>0.5$ to our sample, we find our detection rate increases to 74\% for strongly barred RC3 galaxies and 50\% for weakly barred RC3 galaxies. Figure~\ref{fig:RC3MissedFaceOnBarMontage} shows the remaining 24 face-on ($b/a>0.5$) barred RC3 galaxies missed by our classification, arranged in order of RC3 bar strength. The first four galaxies classified by RC3 as strong bars do not appear to show a bar in the SDSS image. The same is true for deeper images of these galaxies taken from NED. There are 5 objects where confusion with an inner ring or lens may cause us to miss a weak bar or ansae. These are J113758.73+593701.19, J073054.71+390110.06, J092453.21+410336.62, J142405.96+345331.58, and J231815.66+001540.18 which have the `Rings' or `Lens' flags in our catalog set to greater than zero (see \S7). In three cases, J113536.36+545666.10, J113631.12+024500.12 and J115458.71-582937.27, we have missed possible bars. In the remaining galaxies the weak bars (if present) are better classified as twists in our images. After accounting for (arguably) incorrect RC3 classifications, our RC3 strong bar detection rate increases to $\sim80\%$ while the weak bar fraction detection rate is unchanged.

There are two possible reasons for the lower detection rate of RC3 weak bars: the SDSS exposures may not be of sufficient depth or our classifications may be too conservative. Because the SDSS integration times are less than a minute long, some of the bar classifications in RC3 and RSA were done with photographic images that have greater depth (and, in some cases, more information elements) than the corresponding SDSS data. This could lead to misclassifications, especially for the weak bars, although we emphasize that in general the SDSS observations are of comparable quality to the photographic data. 

Since bar classification is subjective, we attempted to better understand whether our bar classifications are particularly conservative by asking a well-known expert visual morphologist to classify a small set of objects independently. Prof. Debra Elmegreen kindly agreed to independently classify the galaxies seen in Figure~\ref{fig:RC3MissedFaceOnBarMontage} (which, we emphasize, were not specially chosen except for being in the RC3). In summary, only 3 of the 13 RC3 strongly barred galaxies were reclassified by Prof. Elmegreen as strongly barred, 5 objects were classified as unbarred and the remaining 5 objects were classified as weakly barred. Specifically, the first 4 objects were classified as `not-barred'. Prof. Elmegreen agreed with the RC3 classification of the second row but argued the third galaxy shows ansae but not necessarily a bar while the last galaxy is more uncertain, and given the `bar' lies along the major axis she would more likely call it an SAB galaxy. In the third row, the first galaxy was reclassified as unbarred while the remaining three are classified as SAB or weak bars. Prof. Elmegreen agreed with most of the RC3 classified weak bars (SAB systems). Specifically, her classification of the fourth row is unchanged from the RC3. In the remaining two rows, the objects were reclassified as weakly barred but with larger uncertainties. J115458.71-582937.27 was reclassified as a strong bar.We emphasize that this exercise does not necessarily mean the RC3 classifications are wrong, and it may simply mean that the data quality in the SDSS is inferior to that used to compile the RC3 in many cases, so that bona fide bars are invisible on the SDSS images. But in any case, it does mean that after incorporating these reclassifications, for the example shown our strong bar detection rate for face-on ($b/a>0.5$) galaxies common with RC3 rises to 93\% while the weak bar detection rate decreases to 41\%, which is the basis for our view that our bar classifications should best be considered to be similar to classifications in the RC3 as strong bars.

\subsection{Comparison of ring classifications}

As stated earlier the total inner plus outer ring fraction for disk galaxies (including lenticulars) in our sample is $25\% \pm 0.5\%$. From Table III in \cite{Buta:1996p2930}, the total face-on ring fraction based on 911 disk galaxies in RC3 is $\thicksim 54\%$, with $\thicksim 20\%$ for complete rings (r) and $\thicksim 35\% $ for the partial/pseudo (rs) ring variety. Based on 5247 galaxies in our sample with a similar inclination cut ($b/a>0.6$) we find a total face-on ring fraction of 30\%$\pm 0.5\%$ with 23 \%$\pm 1\%$ for complete rings(r) and 7\%$\pm 1\%$ for partial rings. Thus our total ring fraction estimate is significantly lower than \cite{Buta:1996p2930}. Partial rings seem to be significantly underestimated in our classification. It must be remembered that RC3 combines rings and lens classifications.
To investigate these differences, we compare our ring classifications with RC3 classifications for the 44 face-on ringed galaxies common to both samples. The recovery rate for RC3 full inner rings, partial inner rings, full outer rings and partial outer rings are 62\%$\pm10\%$, 38\%$\pm12\%$, 29\% $\pm17\%$ and 38\%$\pm17\%$ respectively. We miss a large fraction of partial inner-rings and outer rings in the RC3.

\begin{figure*}[htbp!]
\unitlength1cm
\hspace{1cm}
%\vspace{1.5cm}
\begin{minipage}[t]{4.0cm}
%\vspace{0.8cm}
\rotatebox{0}{\resizebox{16cm}{7cm}{\includegraphics{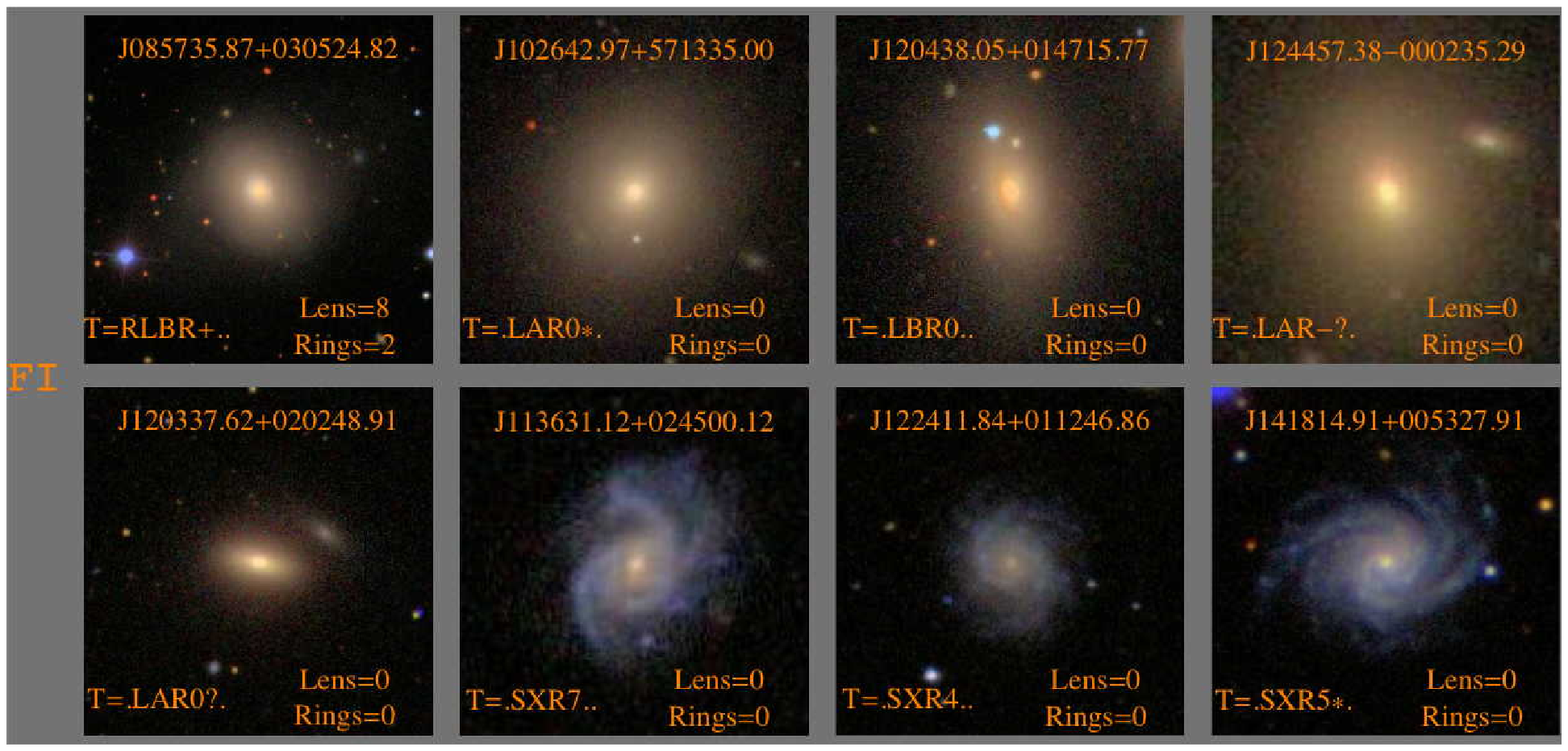}}}
\rotatebox{0}{\resizebox{16cm}{15cm}{\includegraphics{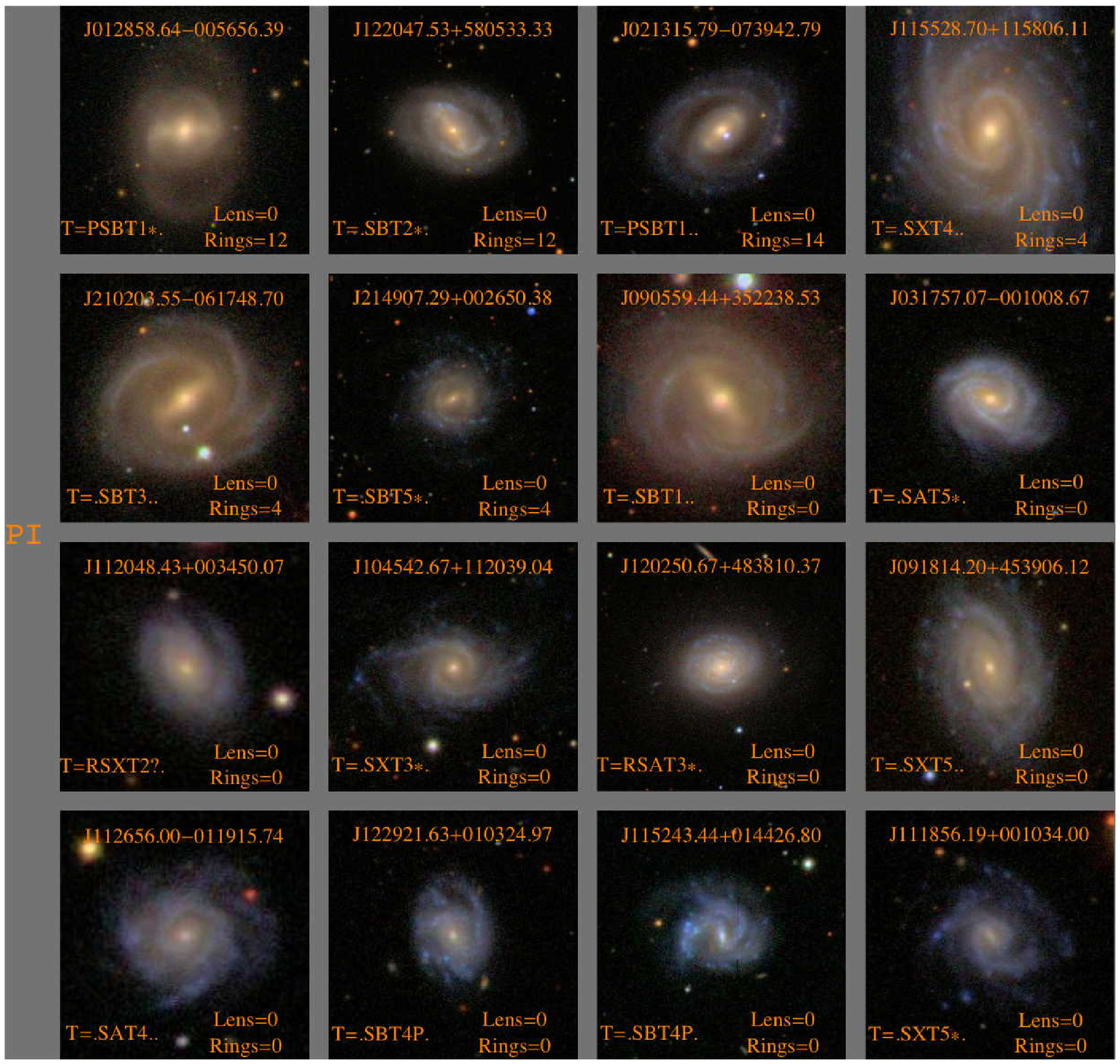}}}
\end{minipage}
\caption[RC3 face-on Montage of Outer Ringed galaxies ]{\label{fig:RC3FaceOnInnerRingMissingMontage} 
A montage of face on ($b/a>0.6$) galaxies with RC3 classifications of ({\bf Top Panel}) full inner rings and ({\bf Bottom Panel}) partial inner rings. The J2000 object ID is listed at the top. The RC3 classification is listed at the bottom left hand corner and the PN ring and lens classifications in the right hand corner. See \S7 for details regarding classification designations. The galaxies are arranged in order of ring type. Each stamp is 50 $h^{-1}$ kpc on a side. 
 }
\end{figure*}

Of the 21 galaxies in our sample classified by the RC3 as having full inner rings, eight are classified as not having an inner ring in our classification. These galaxies are shown in the top panel of Figure~\ref{fig:RC3FaceOnInnerRingMissingMontage}. In the first row we see no inner rings (ring type$>$2, see \S7), though the first galaxy may possess a nuclear ring. In the second row, we again feel no galaxy can be classified as having a full inner ring based on the SDSS images though J122411.84+011246.86 may have a partial inner ring. It is possible that the reason for disagreement may be because we cannot identify nuclear rings in the SDSS images. Thus we suspect that our low recovery rate of full inner rings is due to a combination of misclassifications in RC3 and/or insufficient depth of the SDSS image. Excluding the probable (arguable) RC3 misclassifications we find an inner ring recovery rate of 81\%(13/16).

In the bottom panel of Figure~\ref{fig:RC3FaceOnInnerRingMissingMontage} we show all the galaxies with RC3 partial inner ring classifications. In the first 6 of 16 galaxies we classify the objects as having inner rings, not partial inner rings. Thus in some cases our classification is more lenient than those listed in RC3. In the remaining 10 galaxies, we find that in retrospect at least 2 of the galaxies should have been classified by us as having a partial ring, specifically the last two galaxies in the second row, J090559.44+352238.53 and J031757.07-001008.67. In the third row the galaxies may have partial inner rings but they are very doubtful and hard to distinguish based on the SDSS g-band images. In the final row, the first two galaxies may possess a partial ring but the last two galaxies do not appear to host inner rings. Accounting for the two galaxies without partial rings our recovery rate increases to 43\% (6/14) and our total inner ring recovery rate increases to 63\% (19/30). 

\begin{figure*}[htbp!]
\unitlength1cm
\hspace{1cm}
\begin{minipage}[t]{4.0cm}
%\vspace{0.8cm}
\rotatebox{0}{\resizebox{16cm}{8cm}{\includegraphics{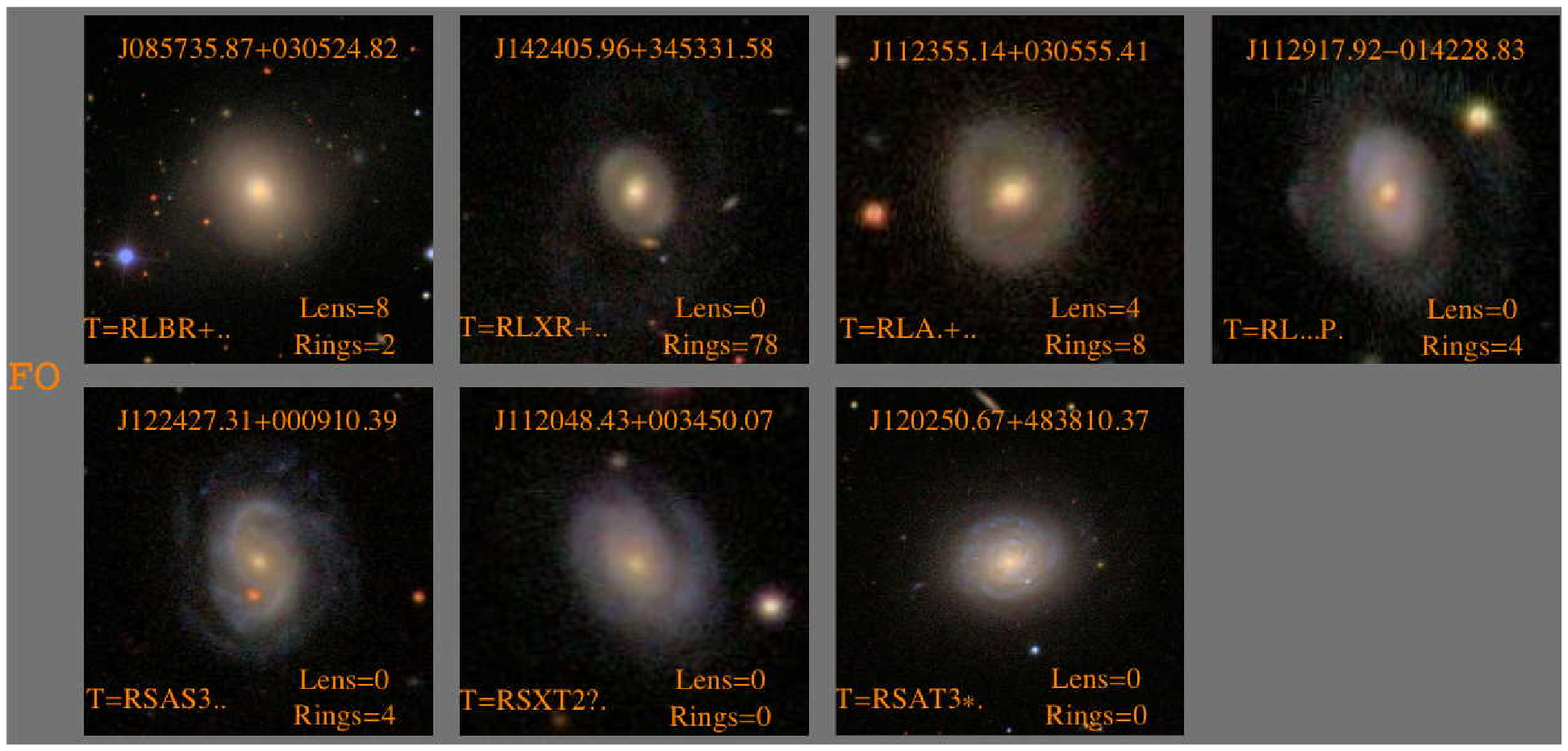}}}
\rotatebox{0}{\resizebox{16cm}{8cm}{\includegraphics{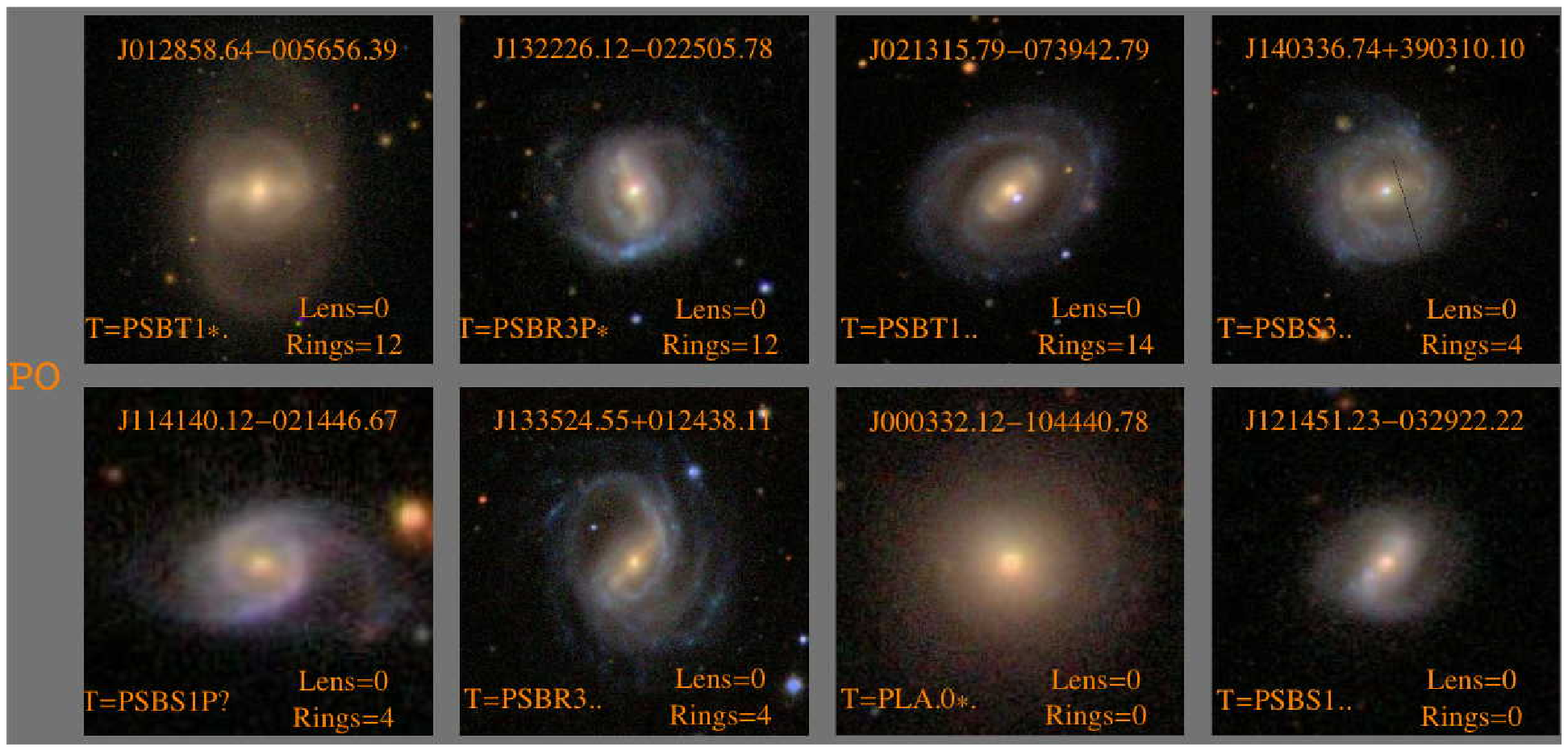}}}
\end{minipage}
\caption[RC3 missed face-on Montage of Inner Ringed galaxies ]{\label{fig:RC3FaceOnOuterRingMissingMontage} 
A montage of face on ($b/a>0.6$) galaxies classified as having (a) full outer rings and (b) partial outer rings in RC3. The J2000 object ID is listed at the top. The RC3 classification is listed at the bottom left hand corner and the PN ring and lens classifications in the right hand corner. See \S7 for details regarding classification designations. The galaxies are arranged in order of T-Type. Each stamp is 50 $h^{-1}$ kpc on a side. 
 }
\end{figure*}

Figure~\ref{fig:RC3FaceOnOuterRingMissingMontage} shows a montage of galaxies with outer rings.
 In the top panel we show the 7 galaxies classified as having a full outer ring by RC3. Our classifications agree for only 2 galaxies, J142405.96+345331.58 and J112355.14+030555.41. In the first galaxy we opted to identify the structure as a lens. In two galaxies, J112917.92-014228.83 and J122427.31+000910.39 we identified the inner rings but not outer rings. We classified the structure seen in J112917.92-014228.83 as an inner ring with a tail. The object could also be a collisional ring. J122427.31+000910.39 could be considered to have a partial outer ring. RC3 misses the inner ring in this case. In the last two galaxies, partial outer rings may exist though it is highly unsure. Accounting for lenses and misclassifications our full outer ring detection rate increases to 57\% (4/7). In the bottom panel, we show the 8 galaxies classified as having pseudo/partial outer rings. We identify the first three objects as outer rings and the next 2 as inner rings while RC3 identifies them as outer rings. The remaining three objects may have pseudo-rings though it seems highly doubtful in the last two cases. The outer structure in J000332.12-104440.78 in particular may be better classified as a shell. Our partial outer ring detection rate increases to 60\% (3/5) and our total outer ring detection rate increases to 58\% (7/12).

Thus our total ring detection rate is 62\% (26/42). The predominant cause for disagreement with previous work appears to be a simple difference in opinion as to what a particular fine feature should be called, e.g. an inner versus outer ring or outer ring versus outer lens. The best evidence for this is to note that there are only eight cases out of the 44 galaxies considered here where an object with an RC3 fine structure classification does not have a fine structure classification in our scheme. In four of those cases (J102642.97+571335.00, J120438.05+014715.77, J124457.38-000235.29 and J120337.62+020248.91 which are supposed to host RC3 inner rings) there is no bar or ring seen in deeper images available on NED in B band, so we have some confidence in attributing these to errors in the RC3. Thus we consider that our overall recovery rate of fine structures (bars+rings+lenses) is $\thicksim$ 91\%.

\begin{figure*}[htbp]
%\begin{center}
\unitlength1cm
\hspace{2cm}
\begin{minipage}[t]{4.0cm}
\includegraphics[width=5.5in]{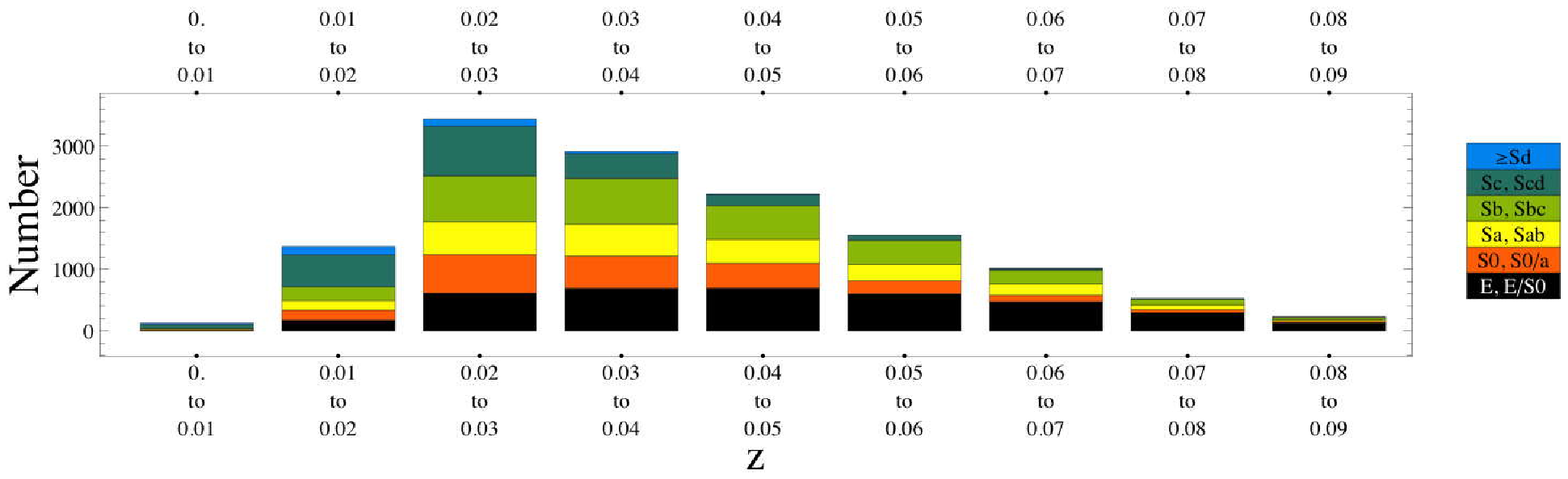}
\includegraphics[width=5.5in]{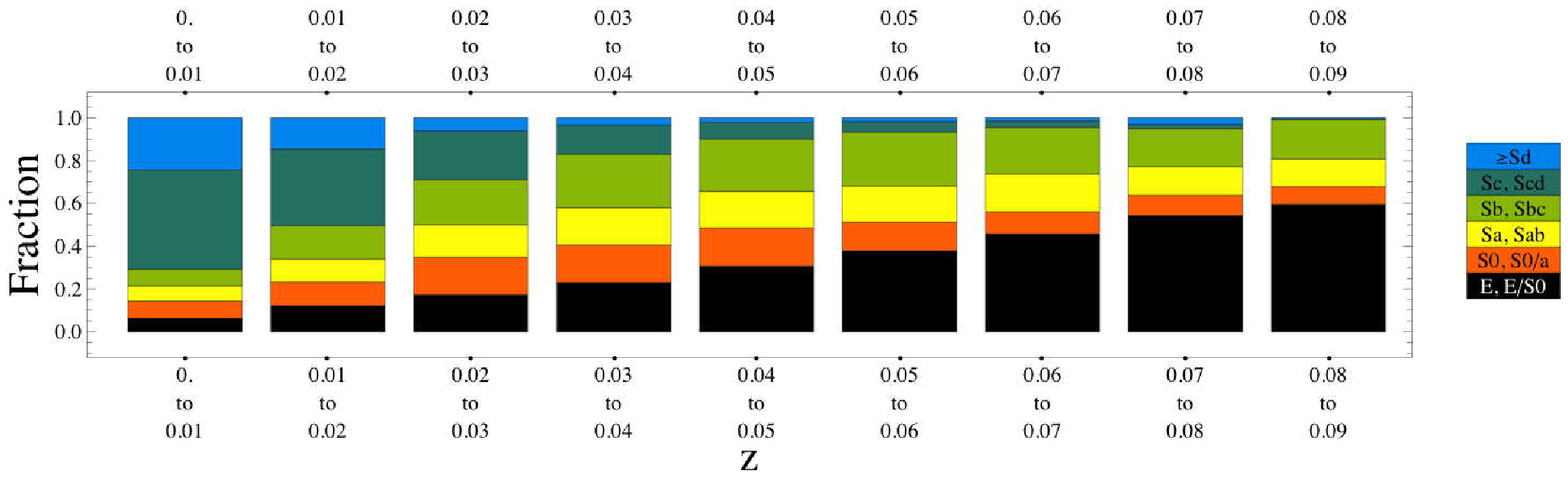}
\end{minipage}
\caption{\label{fig:TTypeSelectionEffects} {\bf [Top]} Histogram and {\bf [Bottom]} fractional distribution of T-Types as a function of redshift. Early type galaxies dominate the higher redshift bins while late types dominate the lower redshift bins. The median of the distribution is around $z\thicksim 0.03$. The sub-categories of galaxy types have been grouped into the following broad classes: E and E/S0 galaxies (black bars), S0 and S0/a galaxies (orange bars), Sa and Sab galaxies (yellow bars), Sb and Sbc galaxies (light green bars), Sc and Scd galaxies (dark green bars) and galaxies with T-Types later than Sd (blue bars.) }
%\end{center}
\end{figure*}

\begin{figure*}[htbp]
%\begin{center}
\unitlength1cm
\hspace{2cm}
\begin{minipage}[t]{4.0cm}
\includegraphics[width=5.5in]{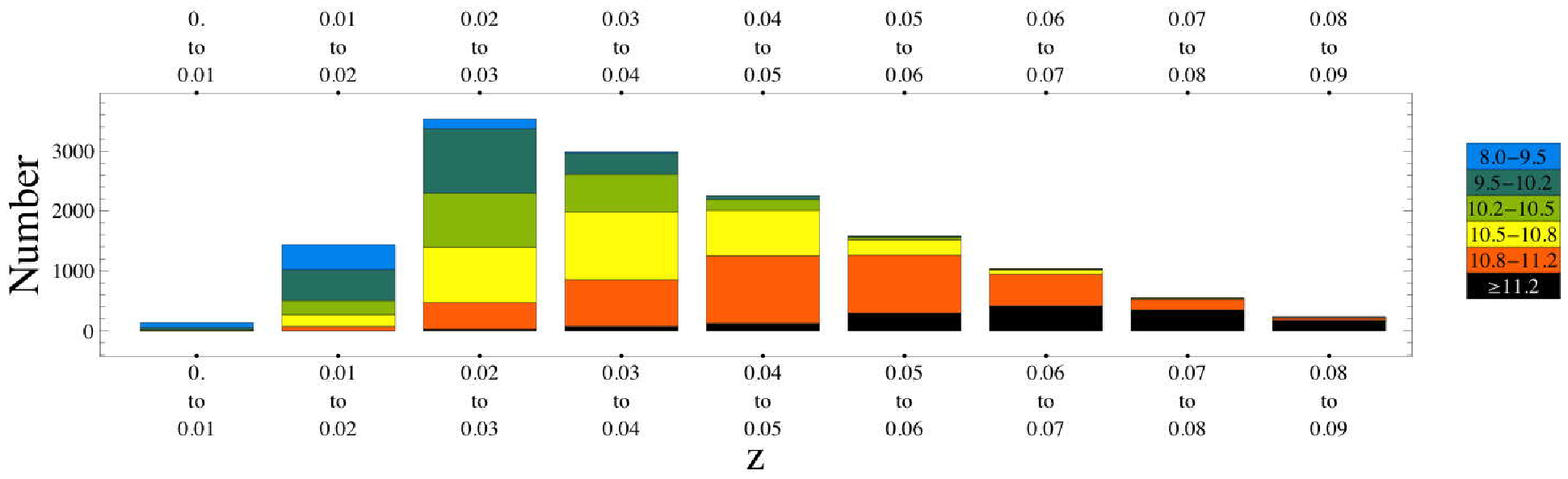}
\includegraphics[width=5.5in]{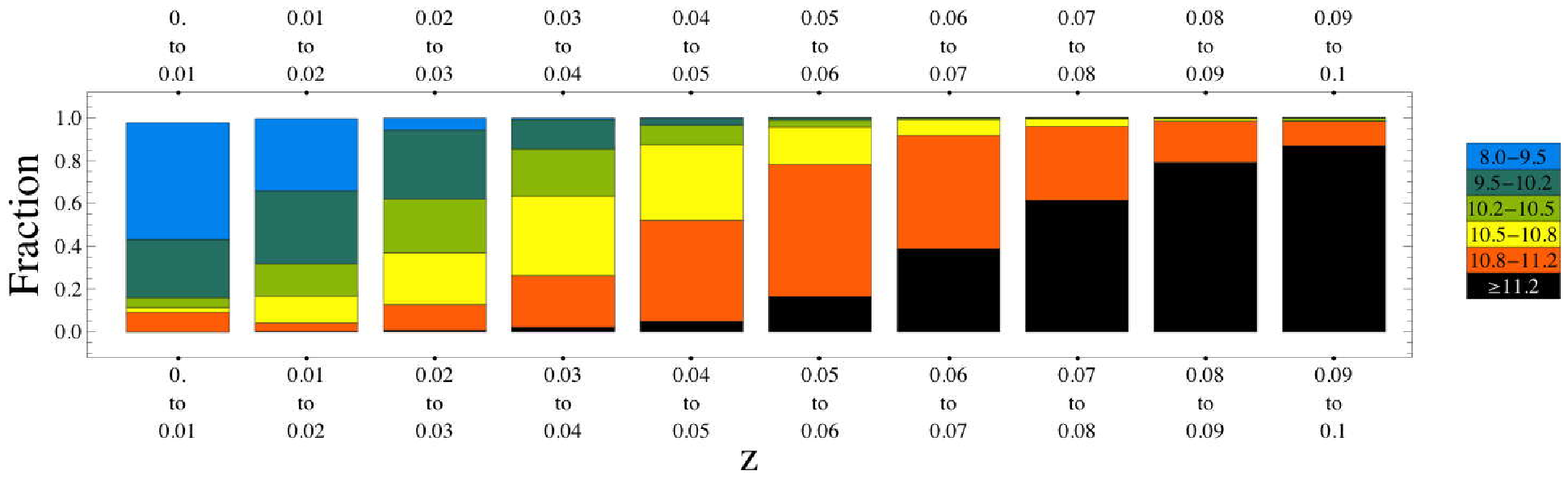}
\end{minipage}
\caption[Selection Effects]{\label{fig:MassSelectionEffects}
{\bf [Top]} Histogram and {\bf [Bottom]} fractional distribution of stellar mass as a function of redshift. The galaxies have been grouped into the following broad mass bins: Log M$>$ 11.2 (black bars), 10.8$<$Log M$<$ 11.2 (orange bars), 10.5$<$Log M$<$ 10.8 (yellow bars), 10.2$<$Log M$<$ 10.5 (light green bars), 9.5$<$Log M$<$ 10.2 (dark green bars) and 8.0$<$Log M$<$ 9.5 (blue bars). Massive galaxies dominate the higher redshift bins while low mass galaxies dominate the lower redshift bins. 
}
%\end{center}
\end{figure*}

\begin{figure*}[htbp]
%\begin{center}
\unitlength1cm
\centerline{\hspace{1cm} Bars \hspace{4cm} Rings \hspace{4cm} Lenses \hspace{1cm} }
\hspace{1cm}
\begin{minipage}[t]{4.0cm}
\rotatebox{0}{\resizebox{5.0cm}{5.0cm}{\includegraphics{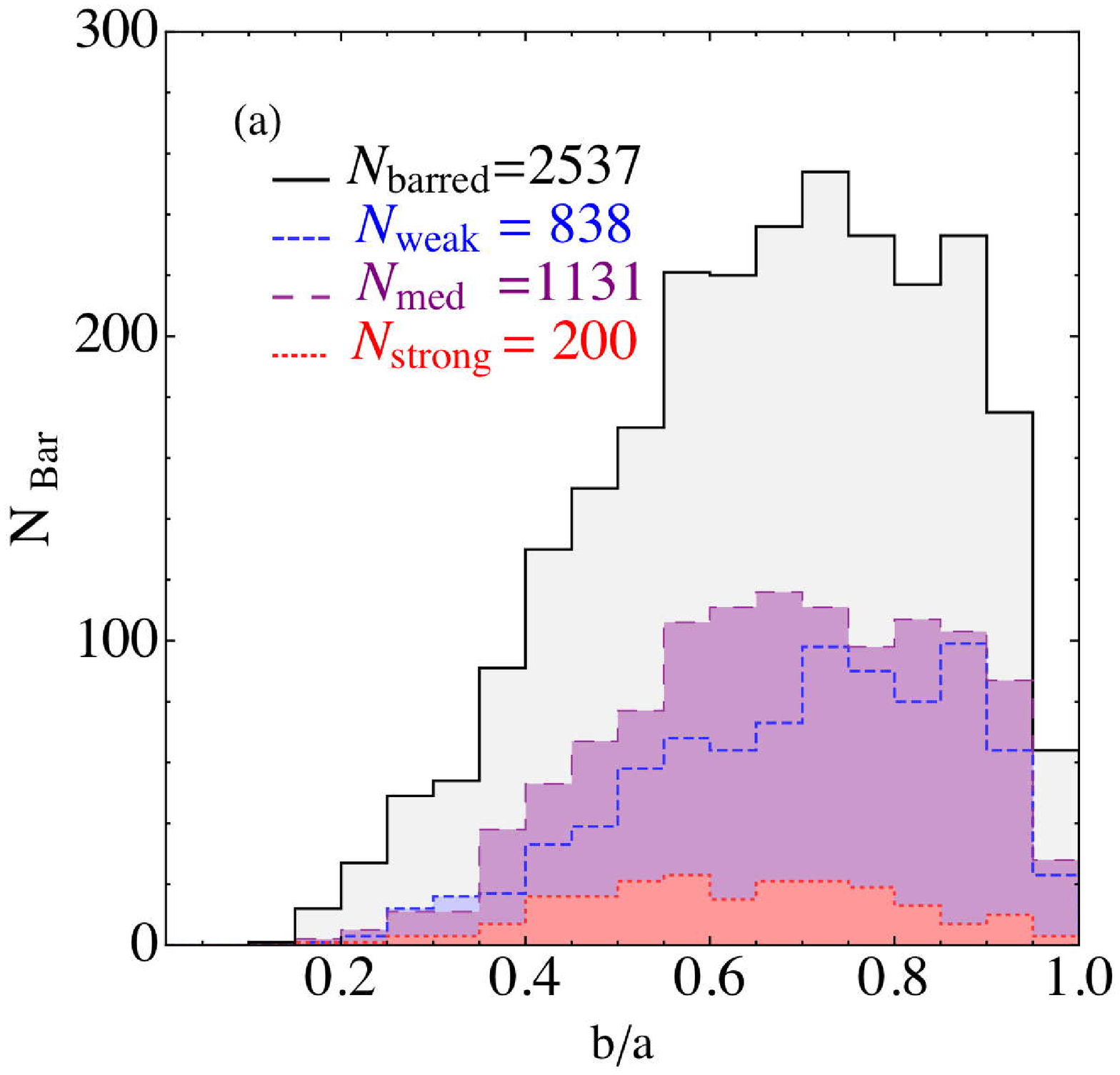}}}
\rotatebox{0}{\resizebox{5.0cm}{5.0cm}{\includegraphics{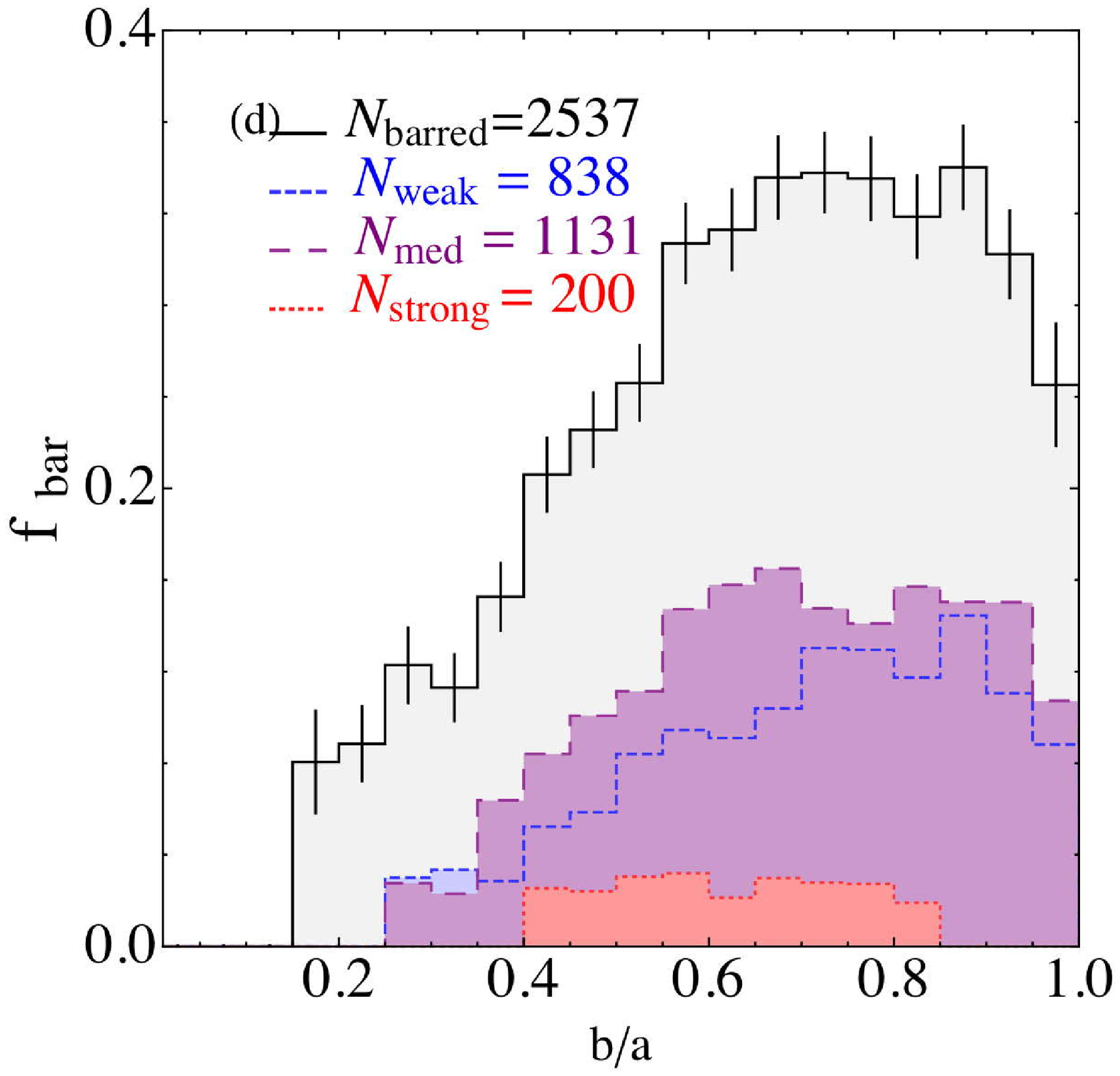}}}
\end{minipage}
\hspace{1cm}
\begin{minipage}[t]{4.0cm}
\rotatebox{0}{\resizebox{5.0cm}{5.0cm}{\includegraphics{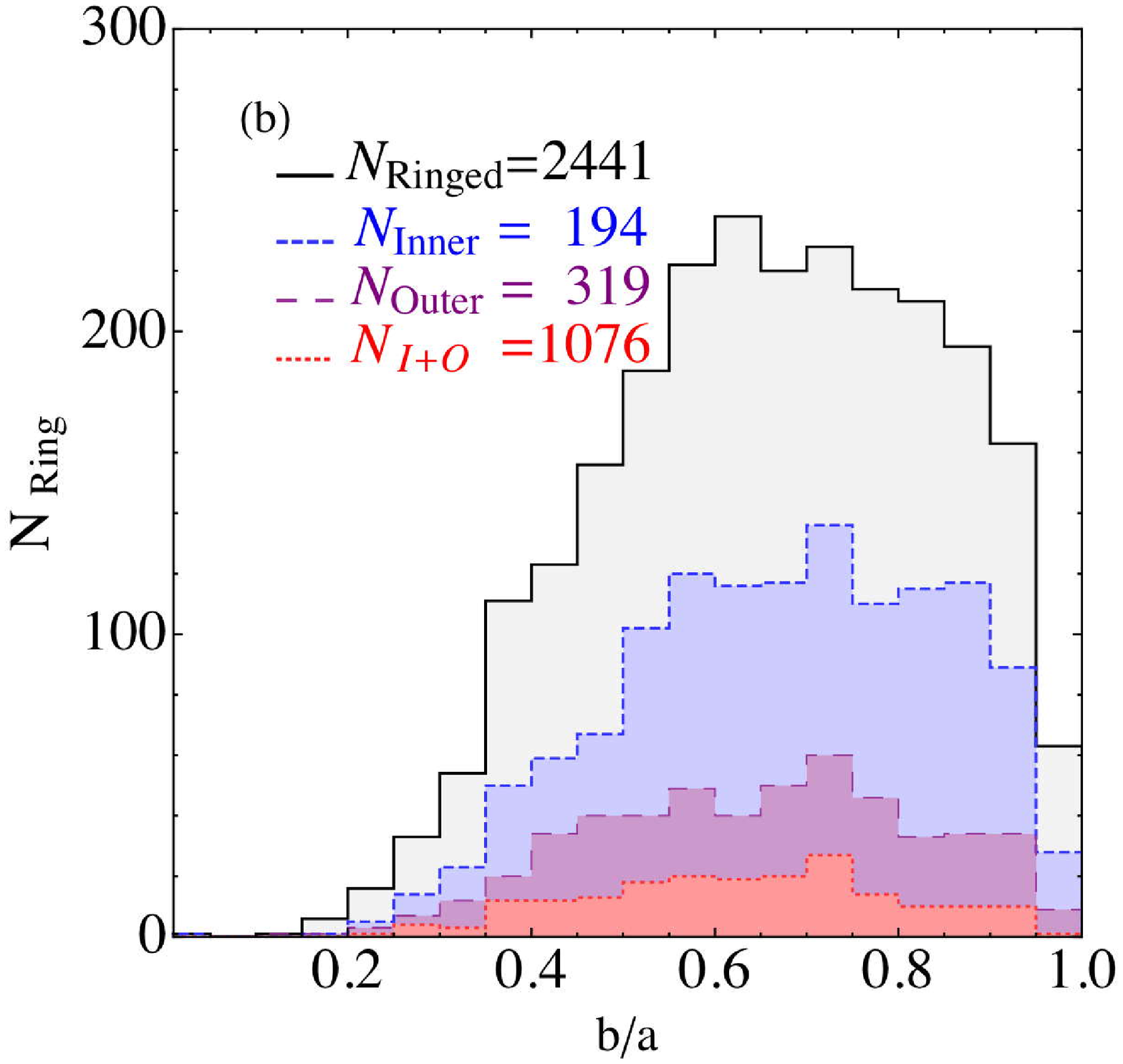}}}
\rotatebox{0}{\resizebox{5.0cm}{5.0cm}{\includegraphics{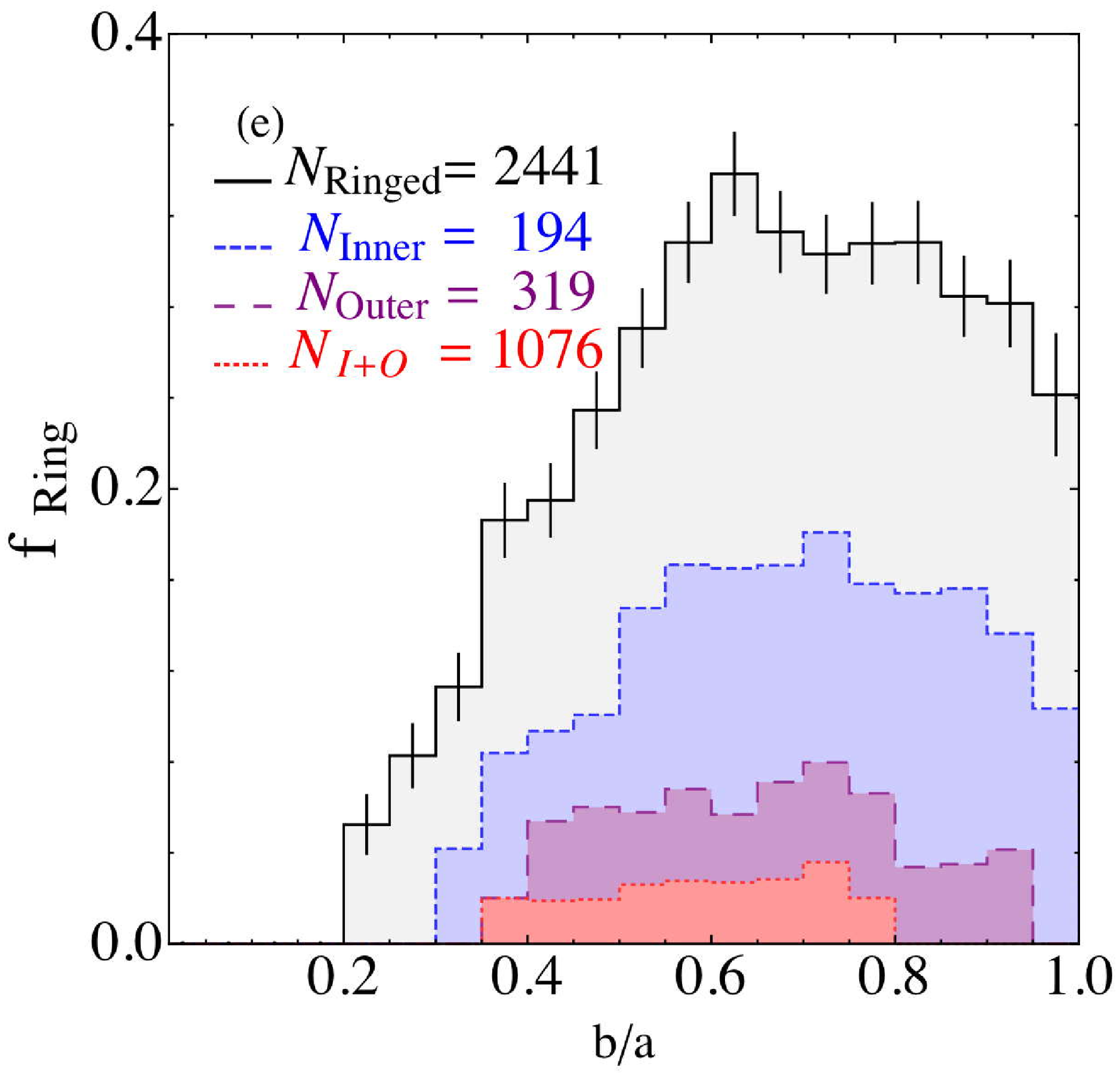}}}
\end{minipage}
\hspace{1cm}
\begin{minipage}[t]{4.0cm}
\rotatebox{0}{\resizebox{5.0cm}{5.0cm}{\includegraphics{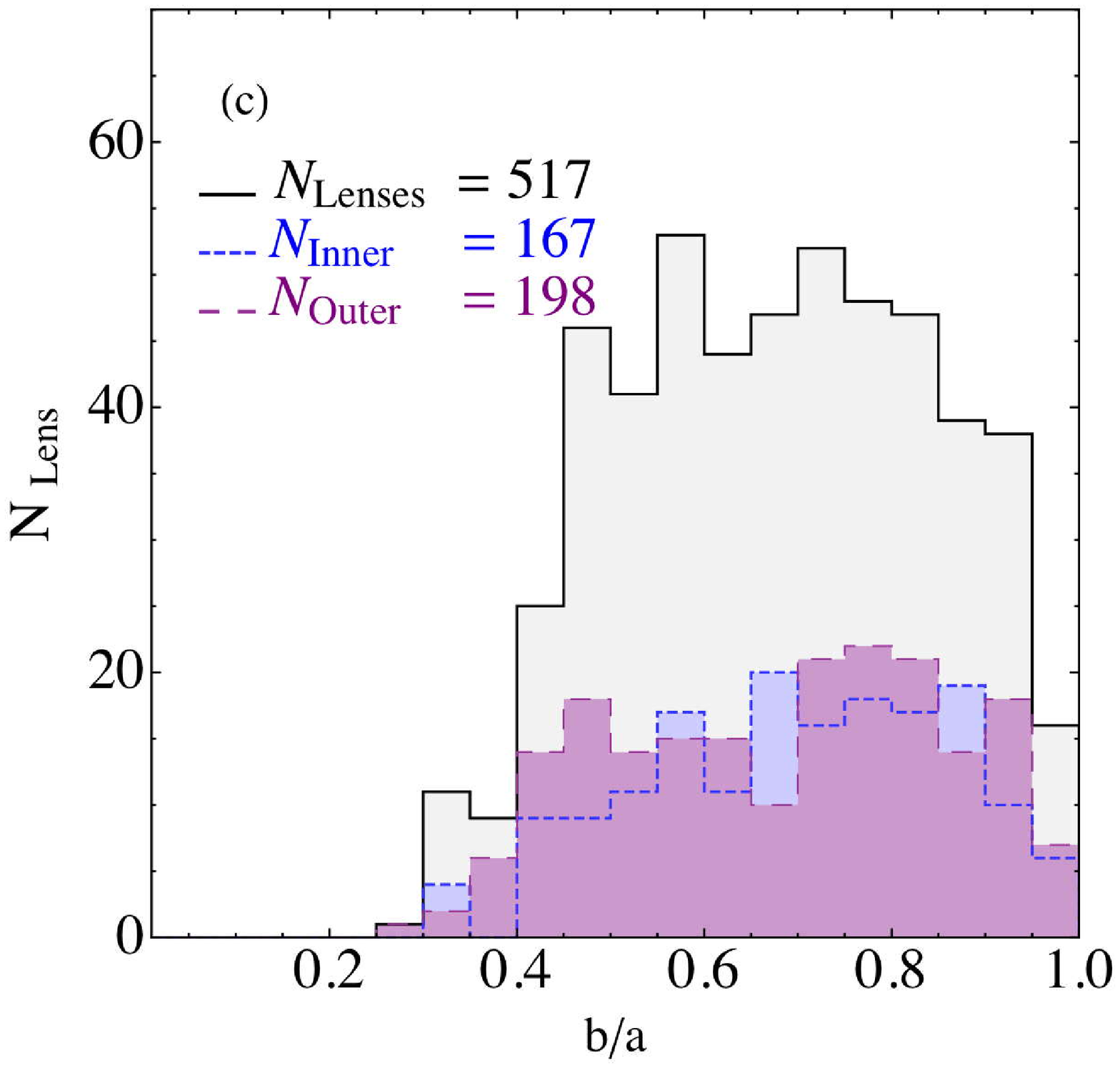}}}
\rotatebox{0}{\resizebox{5.0cm}{5.0cm}{\includegraphics{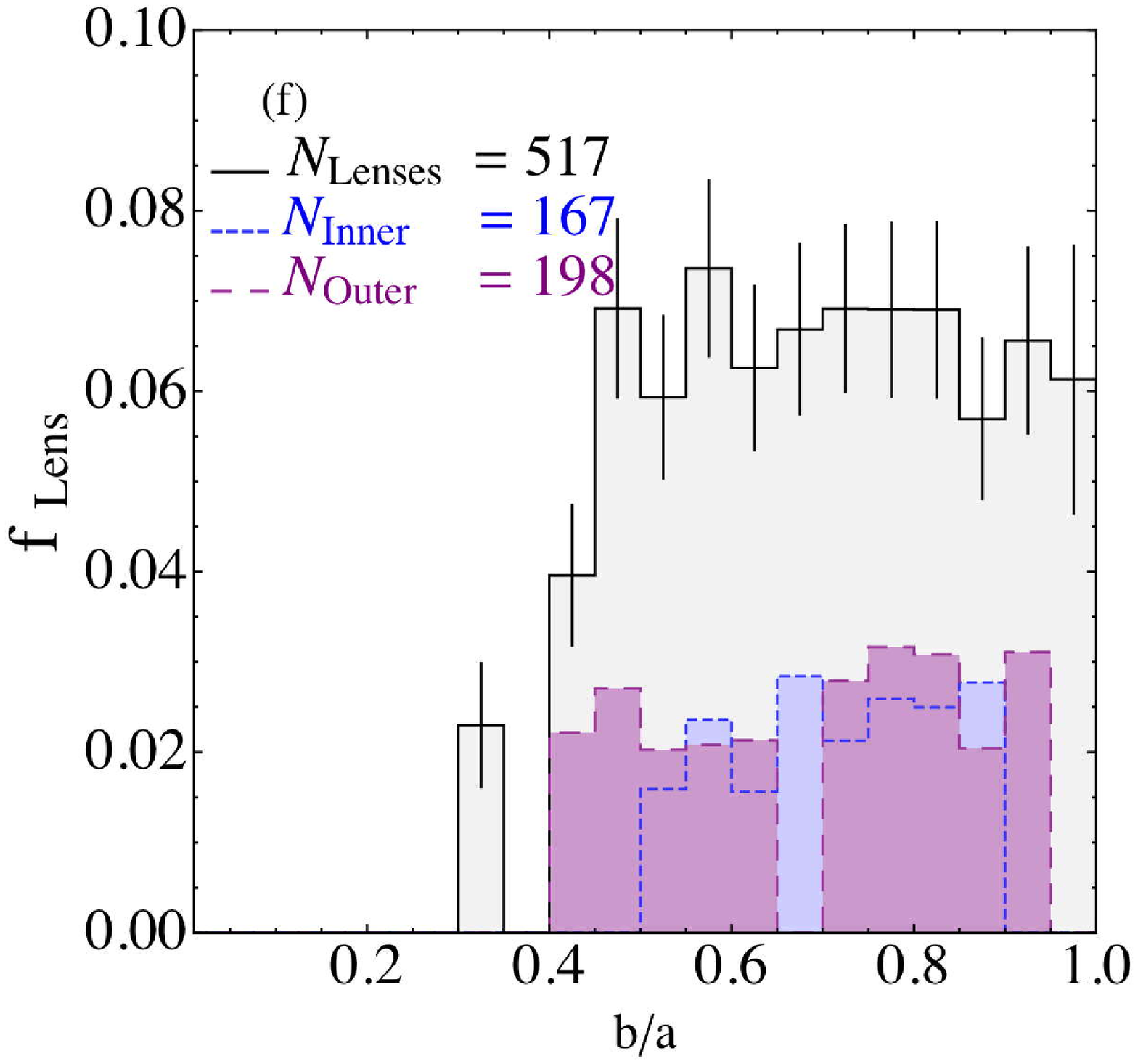}}}
\end{minipage}
\caption[Fine Fraction dependence on Axis Ratio]{\label{fig:BarRingLensFractionVsAxisRatio} Axial ratio dependence of fine fraction. {\bf Top}: Histogram distribution of axis ratios for (a) bars, (b) rings and (c) lenses. The histograms have not been corrected for volume effects. {\bf Bottom} : Fractional histogram for (d) Bars, (e) Rings and (f)Lenses as a function of $b/a$. For barred galaxies, the distribution of strong(red), intermediate(purple) and weak(blue) bars are shown. For ringed galaxies, inner (blue), outer (purple) and combination (red) ring distributions are shown. In galaxies with lenses, inner (blue) and outer (purple) lens distributions are shown. The grey region shows the total distribution. Error bars are shown for the total fractional distribution. The bar, ring and lens fractions are approximately constant for b/a$>$0.6.}
%\end{center}
\end{figure*}

\begin{figure*}[htbp!]
%\begin{center}
\unitlength1cm
\centerline{\hspace{1cm} Bars \hspace{4cm} Rings \hspace{4cm} Lenses \hspace{1cm} }
\hspace{1cm}
\begin{minipage}[t]{4.0cm}
\rotatebox{0}{\resizebox{5.0cm}{5.0cm}{\includegraphics{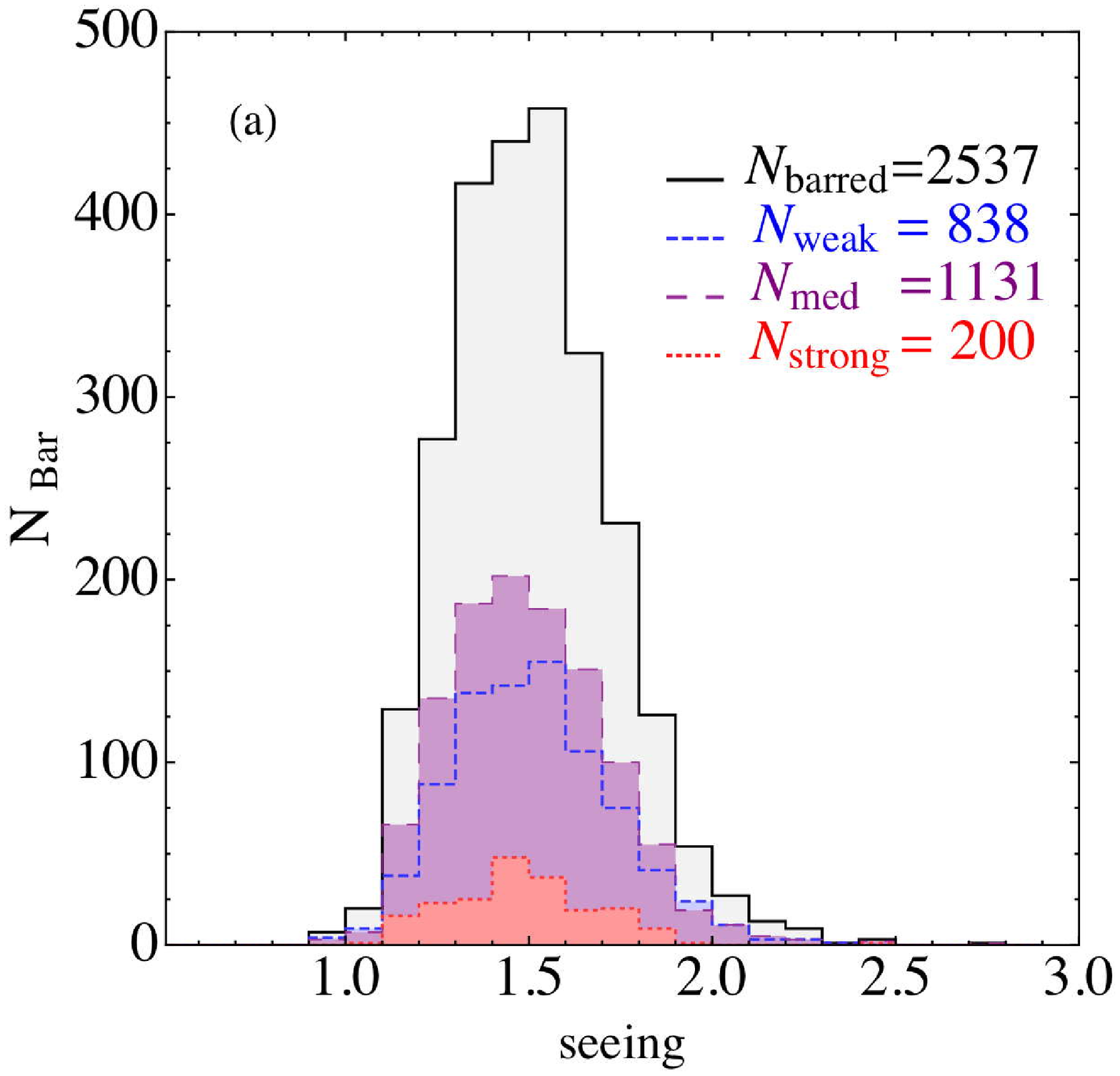}}}
\rotatebox{0}{\resizebox{5.0cm}{5.0cm}{\includegraphics{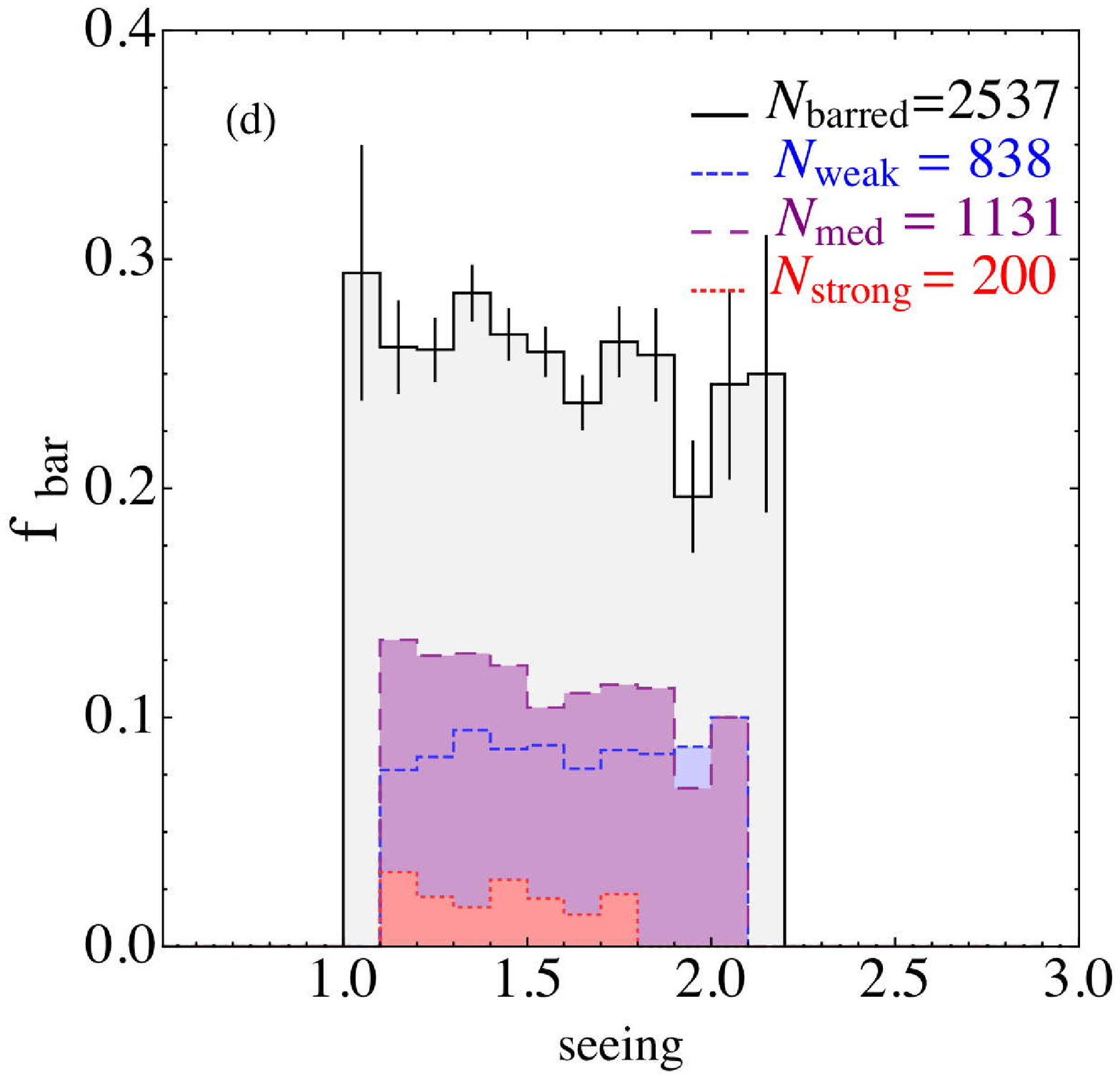}}}
\end{minipage}
\hspace{1cm}
\begin{minipage}[t]{4.0cm}
\rotatebox{0}{\resizebox{5.0cm}{5.0cm}{\includegraphics{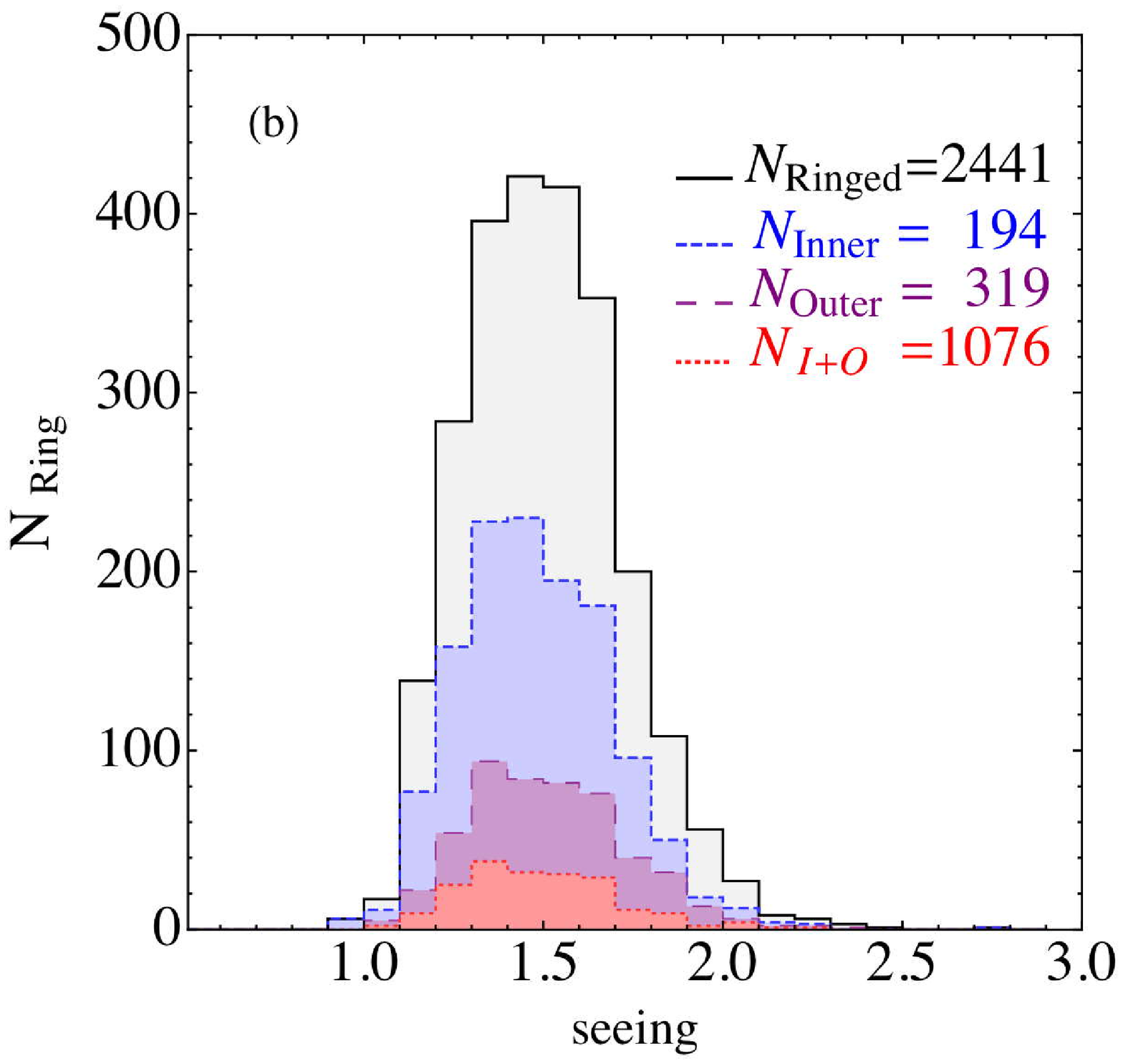}}}
\rotatebox{0}{\resizebox{5.0cm}{5.0cm}{\includegraphics{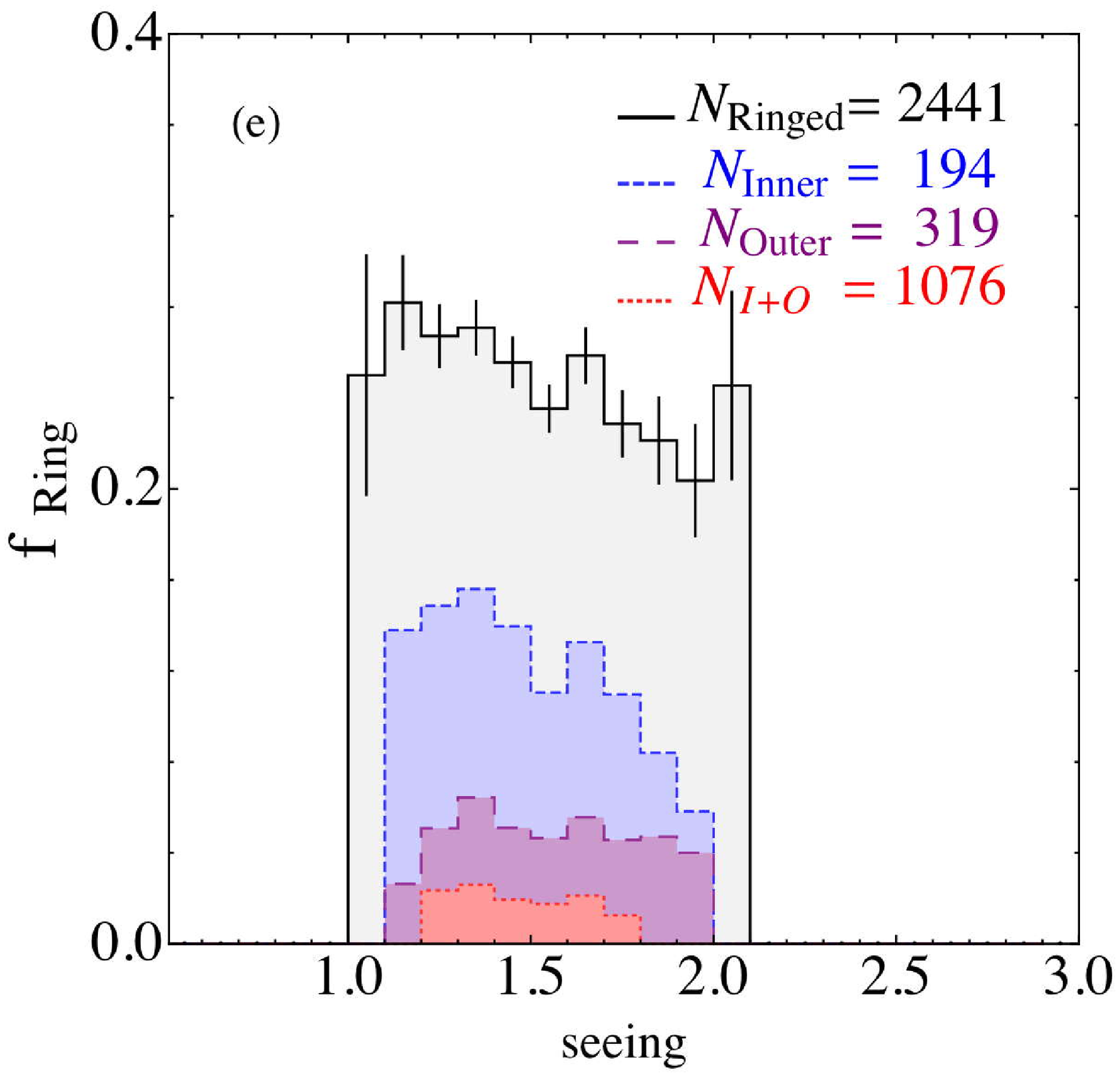}}}
\end{minipage}
\hspace{1cm}
\begin{minipage}[t]{4.0cm}
\rotatebox{0}{\resizebox{5.0cm}{5.0cm}{\includegraphics{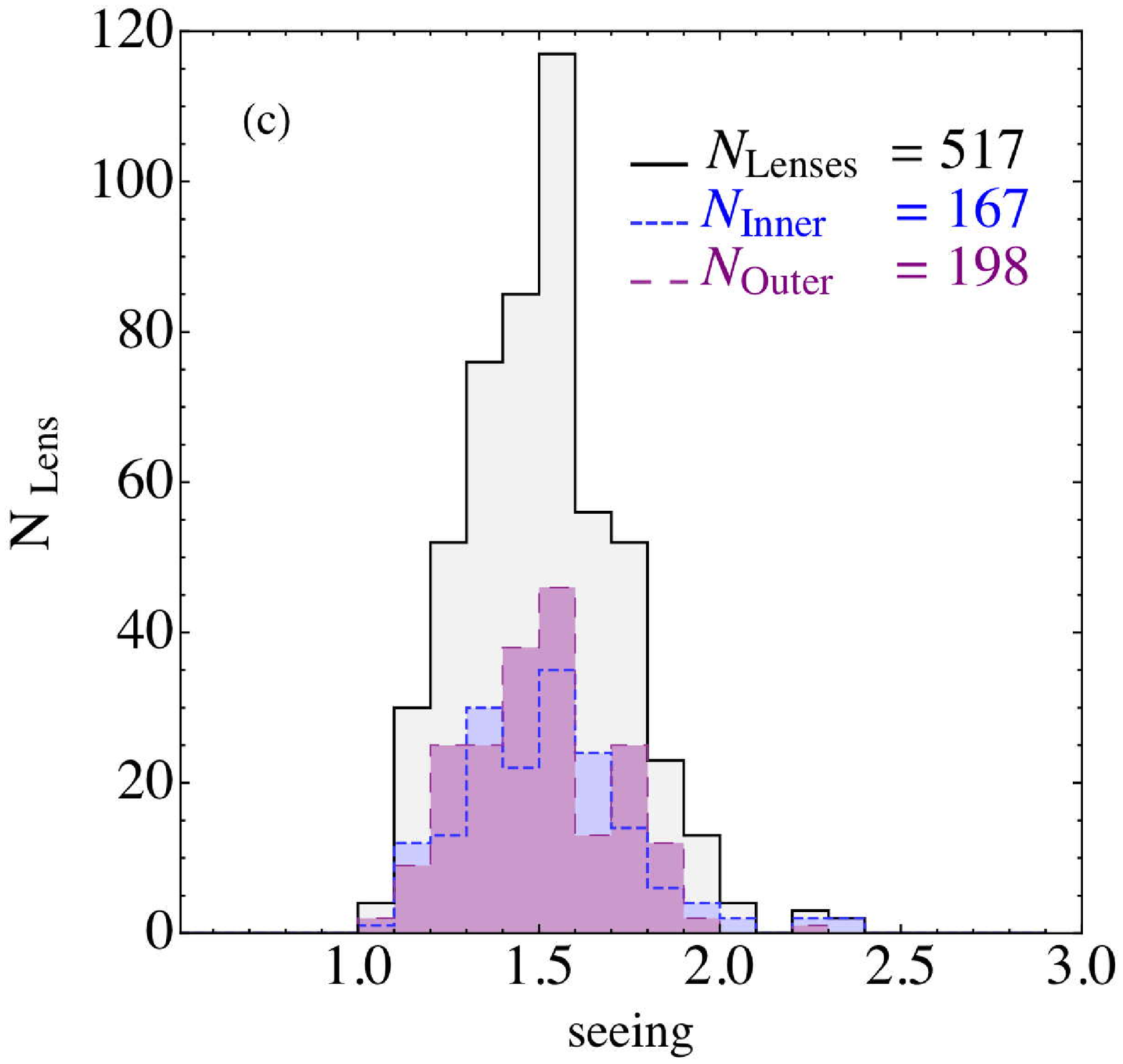}}}
\rotatebox{0}{\resizebox{5.0cm}{5.0cm}{\includegraphics{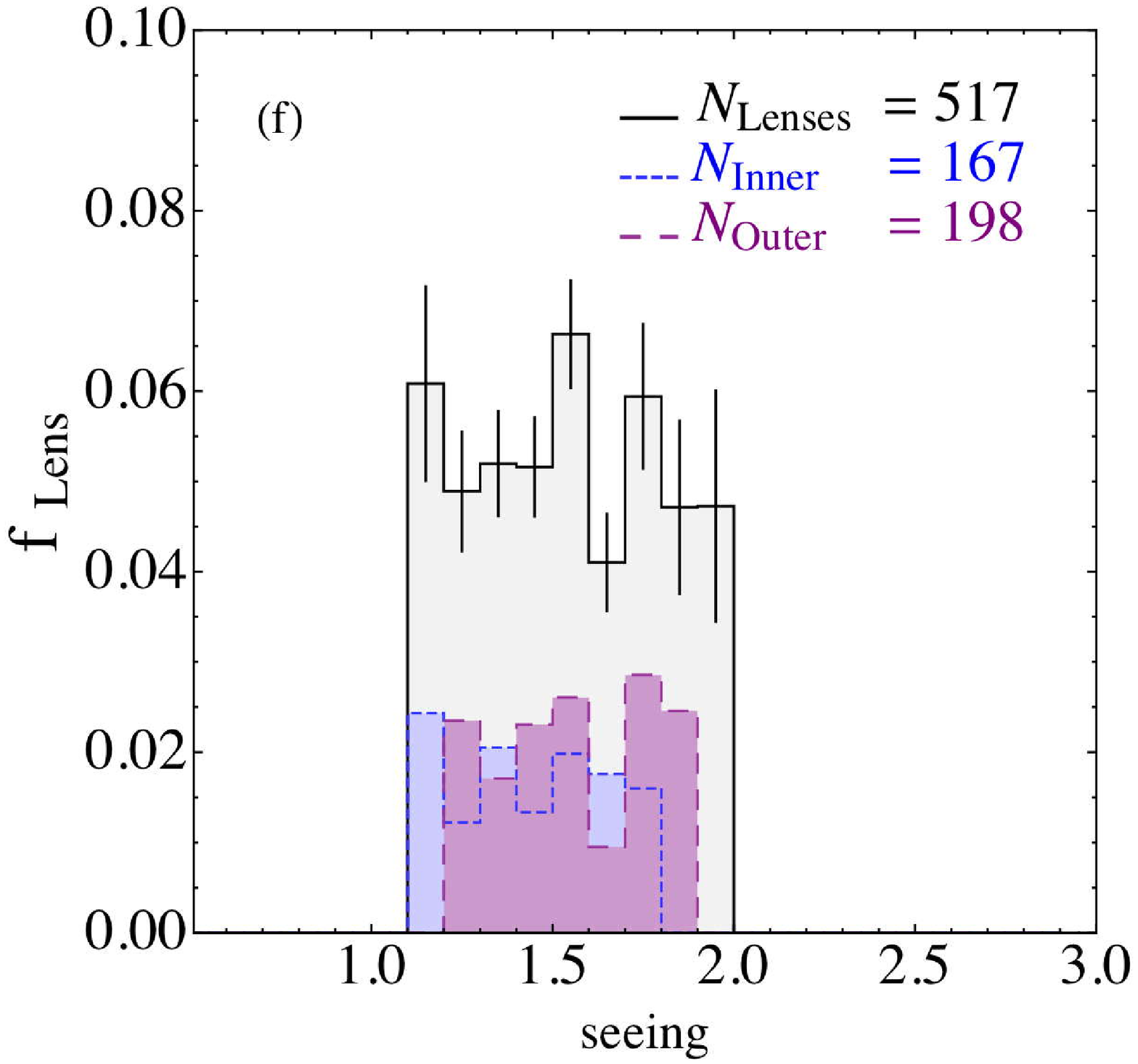}}}
\end{minipage}
\caption[Fine Fraction dependence on Seeing]{\label{fig:BarRingLensFractionVsSeeing} Dependence of fine fraction on
seeing. {\bf Top}: Histogram distribution of seeing (PSF FWHM in arcsec) for (a) bars, (b) rings and (c) lenses. The histograms have not been corrected for volume effects. {\bf Bottom} : Fractional histogram for (d) Bars, (e) Rings and (f)Lenses as a function of seeing. For barred galaxies, the distribution of strong(red), intermediate(purple) and weak(blue) bars are shown. For ringed galaxies, inner (blue), outer (purple) and combination (red) ring distributions are shown. In galaxies with lenses, inner (blue) and outer (purple) lens distributions are shown. The grey region shows the total distribution. 
}
%\end{center}
\end{figure*}

\begin{figure*}[t!]
%\begin{center}
\vspace{0.2cm}
\unitlength1cm
\centerline{\hspace{1cm} Bars \hspace{4cm} Rings \hspace{4cm} Lenses \hspace{1cm} }\hspace{1cm}
\begin{minipage}[t]{4.0cm}
\rotatebox{0}{\resizebox{5.0cm}{5.0cm}{\includegraphics{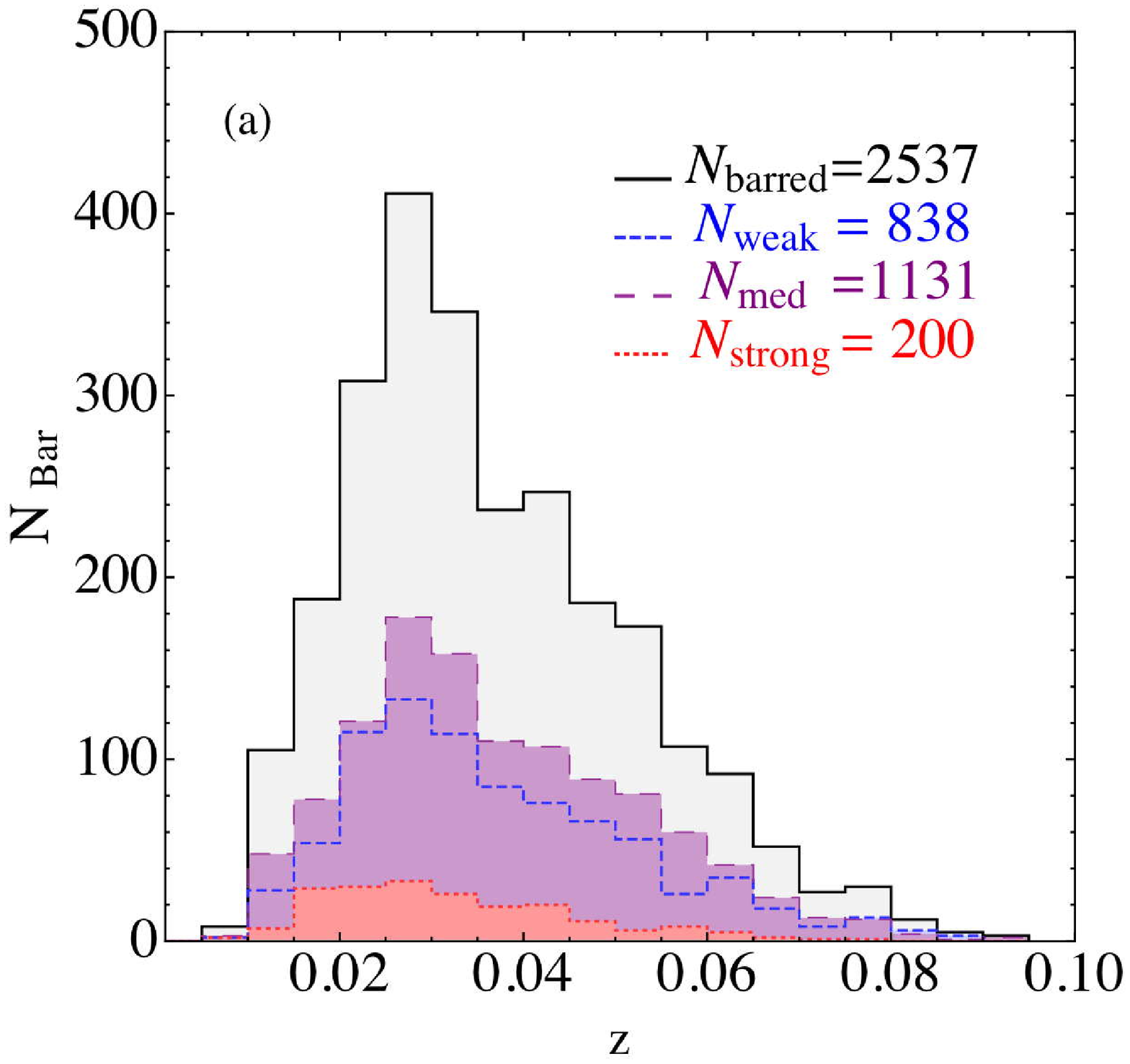}}}
\rotatebox{0}{\resizebox{5.0cm}{5.0cm}{\includegraphics{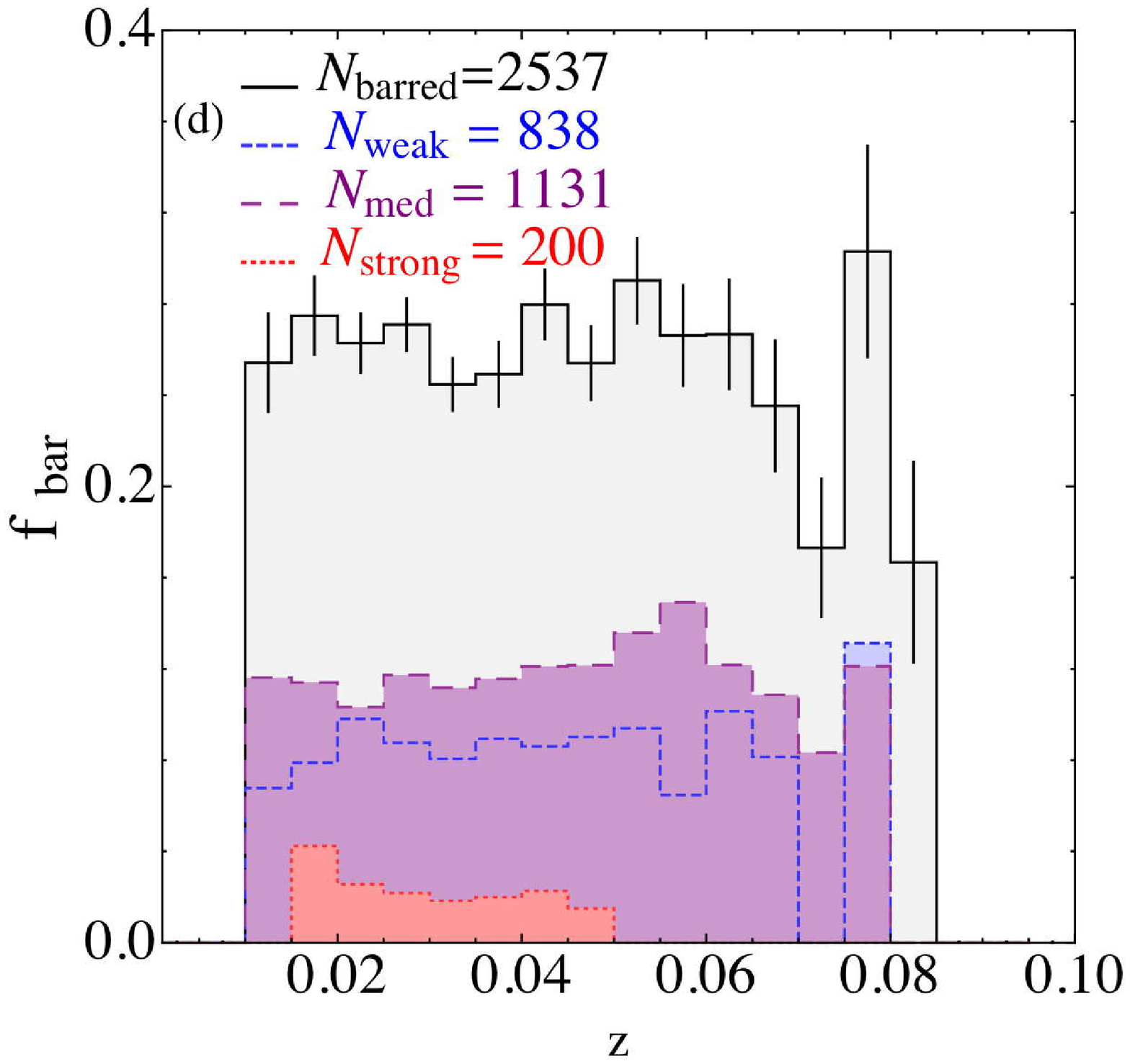}}}
\end{minipage}
\hspace{1cm}
\begin{minipage}[t]{4.0cm}
\rotatebox{0}{\resizebox{5.0cm}{5.0cm}{\includegraphics{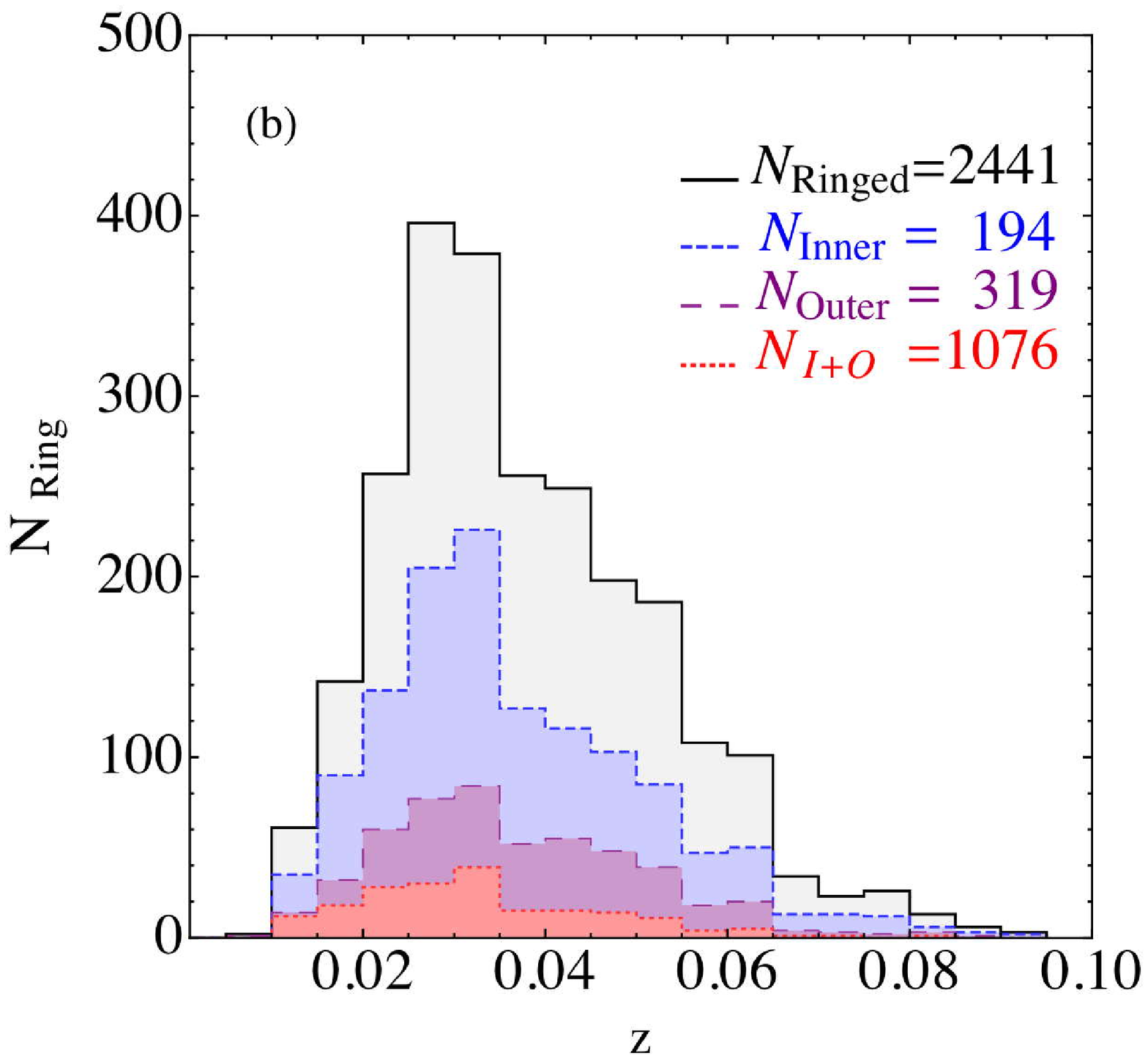}}}
\rotatebox{0}{\resizebox{5.0cm}{5.0cm}{\includegraphics{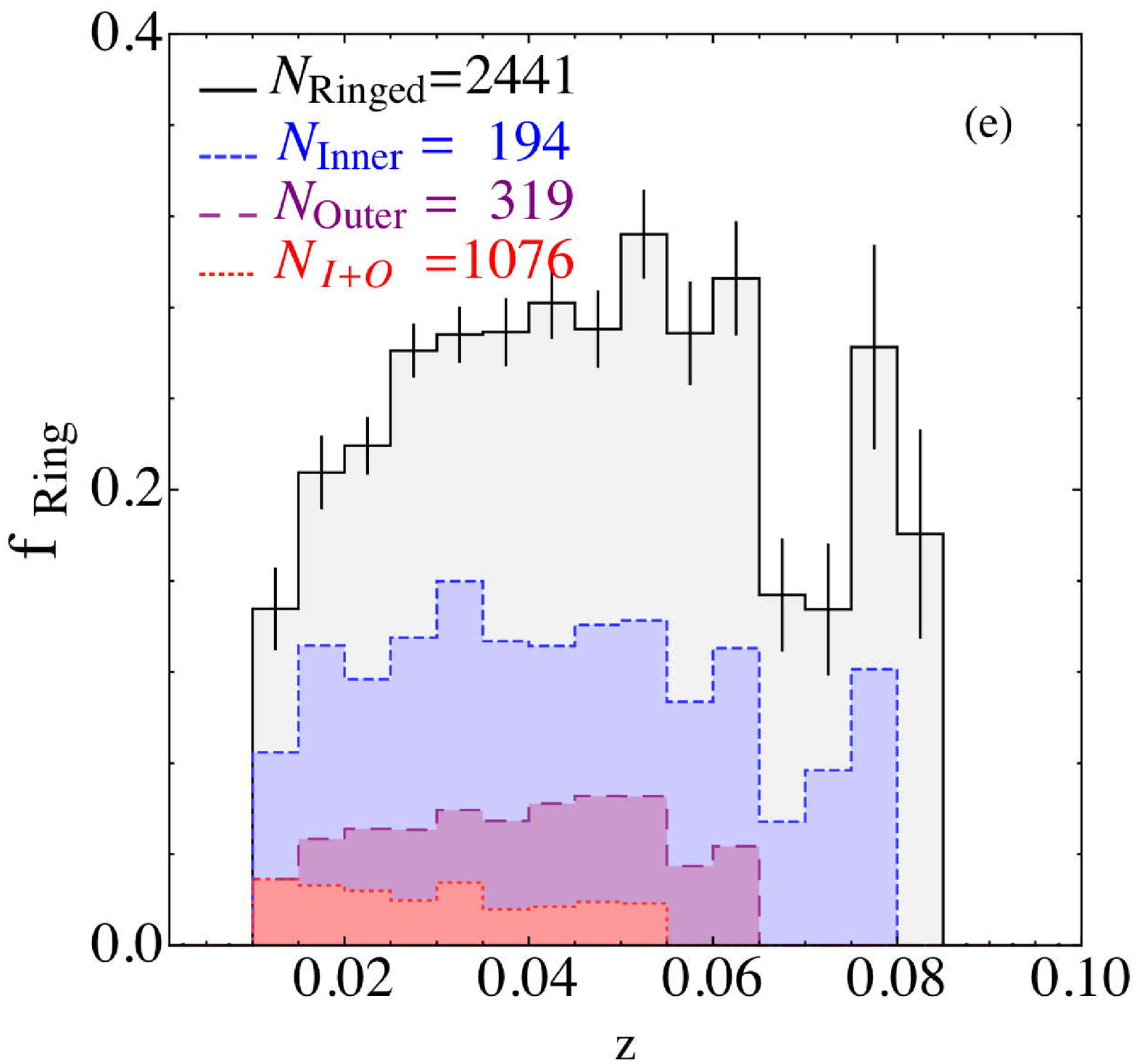}}}
\end{minipage}
\hspace{1cm}
\begin{minipage}[t]{4.0cm}
\rotatebox{0}{\resizebox{5.0cm}{5.0cm}{\includegraphics{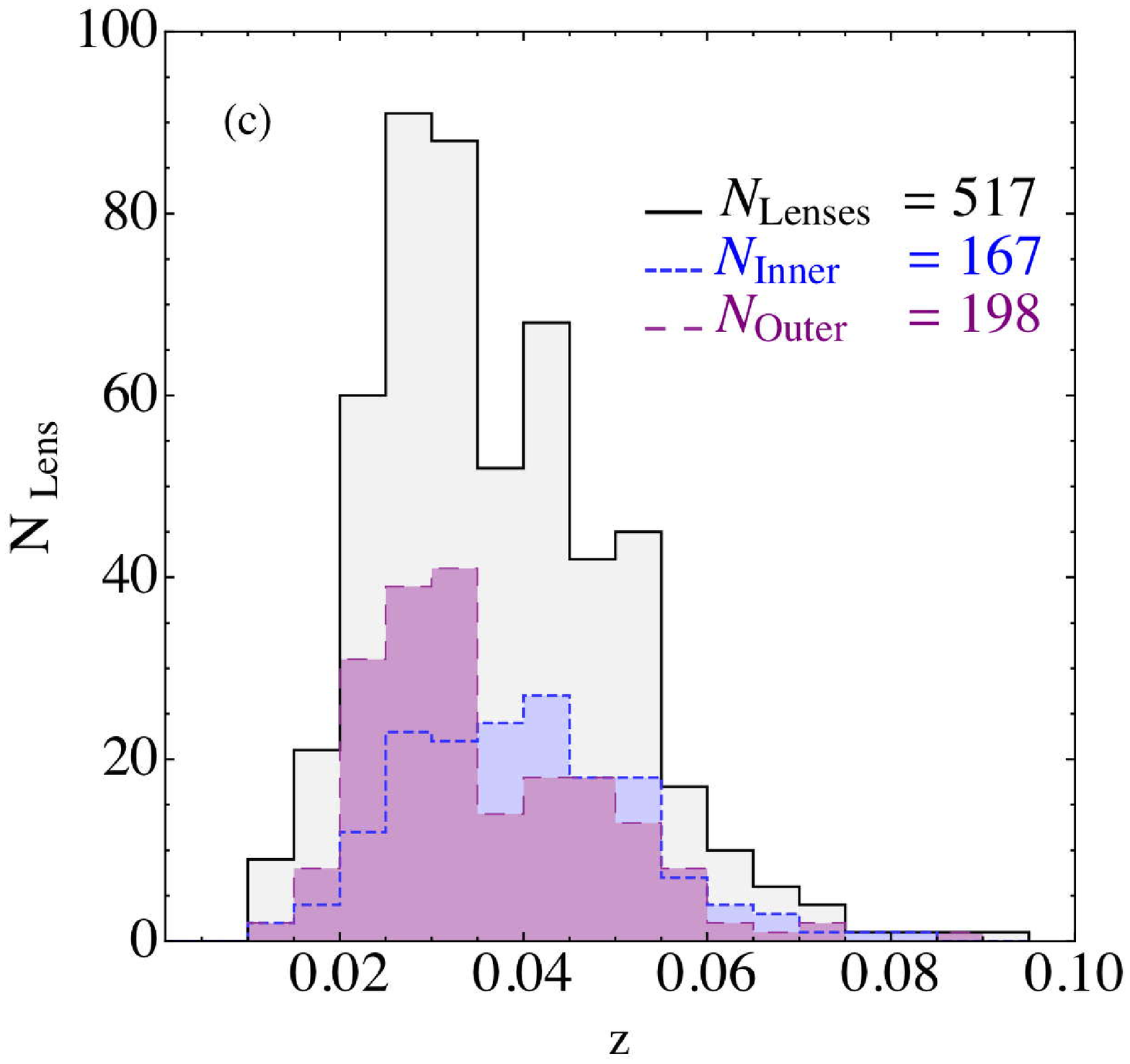}}}
\rotatebox{0}{\resizebox{5.0cm}{5.0cm}{\includegraphics{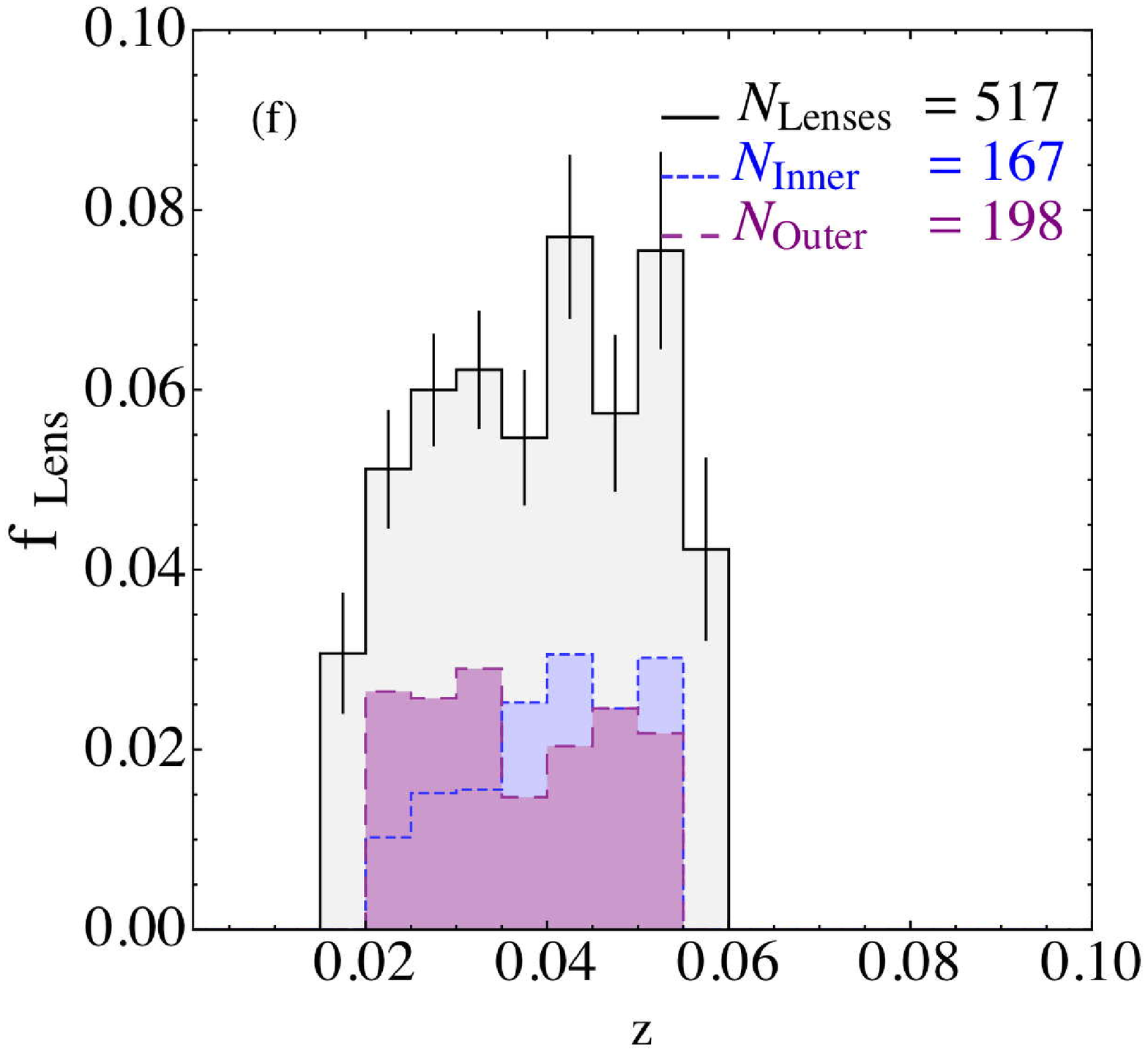}}}
\end{minipage}
\caption[Fine Fraction dependence on Redshift]{\label{fig:BarRingLensFractionVsRedshift} {\bf Redshift} dependence of Fine Fraction. {\bf Top}: Histogram distribution of redshifts for (a) bars, (b) rings and (c) lenses. The histograms have not been corrected for volume effects. {\bf Bottom} : Fractional histogram for (d) Bars, (e) Rings and (f)Lenses as a function of redshift. For barred galaxies, the distribution of strong(red), intermediate(purple) and weak(blue) bars are shown. For Ringed galaxies, inner (blue), outer (purple) and combination (red) ring distributions are shown. In galaxies with lenses, inner (blue) and outer (purple) lens distributions are shown. The grey region shows the total distribution. Error bars are shown for the total fractional distribution.We find bar fractions are nearly constant up to $z \thicksim 0.06$ beyond which the fraction drops. Ring and lens fractions are approximately constant between 0.03$<$z$<$0.06 beyond which only inner rings are detected. }
%\end{center}
\end{figure*}

\section{Selection Effects}

\subsection{Redshift-dependent T-Type selection}

Figure~\ref{fig:TTypeSelectionEffects} shows the histogram (top) and the fractional (bottom) distribution of Hubble types as a function of redshift. The sub-categories of galaxy types have been grouped into the following broad classes: E and E/S0 galaxies (black bars), S0 and S0/a galaxies (orange bars), Sa and Sab galaxies (yellow bars), Sb and Sbc galaxies (light green bars), Sc and Scd galaxies (dark green bars) and galaxies with T-Types later than Sd (blue bars.) The median redshift of our sample is around z$\thicksim$0.036. Galaxies between $0.02<z<0.04$ seem to span nearly the whole range in T-Types, though of course late type galaxies are preferentially seen in the lower redshift bins while early-type galaxies prefer the higher redshift bins. However as we described in our sample selection criteria, there is a mass dependence to our redshift selection as illustrated by Figure~\ref{fig:MassSelectionEffects}. Galaxies are shown in mass bins with the highest mass galaxies in black and the lowest mass galaxies in blue. We find the most massive galaxies in our sample prefer the higher redshift bins while the least massive galaxies are selected in the lowest redshift bin, as expected. Thus there is no redshift slice which can simultaneously sample the whole range in mass and the whole range in T-Types in our sample, re-emphasizing the importance of using a $V_{max}$ formalism to convert the numbers in our catalog into space densities. 

\subsection{Selection effects important for fine fraction recovery}
\subsubsection{Inclination}
Figure~\ref{fig:BarRingLensFractionVsAxisRatio} illustrates the effect of axis ratio cuts on bar, ring and lens distributions and fractions. The distributions have not been corrected for volume effects. For barred galaxies, the distribution of strong (red), intermediate (purple) and weak (blue) bars are shown. For ringed galaxies, inner (blue), outer (purple) and combination (red) ring distributions are shown. In galaxies with lenses, inner (blue) and outer (purple) lens distributions are shown. Figure~\ref{fig:BarRingLensFractionVsAxisRatio}(d) shows the bar-fraction as a function of axis ratio is nearly constant for $b/a>0.6$ (except in the highest bin) but decreases steeply below this threshold. Strong bars are not affected by inclination effects above $b/a>0.4$. In the case of ring fractions, shown in Figure~\ref{fig:BarRingLensFractionVsAxisRatio}(e), we find a similar result of fairly constant fraction above $b/a>0.6$ with a sharp decrease below this threshold. Lens fractions are more robust to inclination effects above $b/a>0.4$ and are hard to detect in more inclined systems, though it should be noted that the error bars are larger due to smaller sample sizes. Thus orientation effects can lead to a decrease in the observed optical bar/ring fraction in the local universe for more inclined systems, as expected. For example, for objects with $b/a>0.6$ the average bar(ring) fraction is 32\%(30\%) whereas for objects with $b/a<0.6$ the average bar(ring) fraction is 19\%(19\%). 

\subsubsection{Seeing}
Figure~\ref{fig:BarRingLensFractionVsSeeing} illustrates the effect of seeing on bar, ring and lens distributions and fractions. The color coding is the same as in the previous figure. We find in the case of bars and rings there is a slight dependence of the overall fractions on seeing with the fraction recovered decreasing as seeing increases. Inner rings are more strongly affected by poor seeing conditions. Lens fractions are roughly constant within the error bars. Overall, the effects are small, but the reader using our catalog may find it worthwhile to apply appropriate seeing cuts to the data in the master table, depending on the intended purpose.

\subsubsection{Redshift}

Figure~\ref{fig:BarRingLensFractionVsRedshift} illustrates the effect of redshift selection on bar, ring and lens distributions and fractions. The color coding is the same as in Figure~\ref{fig:BarRingLensFractionVsAxisRatio}. We find bar fractions are nearly constant up to $z \thicksim 0.06$ beyond which the fraction drops, though the error bars are larger. Total ring fractions, as well as individual ring types, are also nearly constant with redshift from z $\thicksim$ 0.03 to 0.06 and decrease thereafter. The decrease in the total ring fraction with redshift may be due to a lack of outer rings beyond z$\thicksim$0.06. A similar trend is seen with lenses. These declines are unlikely to be physical effects, and are likely due to biases introduced by the changing population mix as a function of redshift (see Figures 20 and 21). We will revisit this issue in subsequent papers in which we analyze the space densities of the populations.

\section{The Catalog}
\label{sec:catalog}

%\begin{landscape}
\begin{deluxetable*}{lrrrrrrrrrrrr}
\tabletypesize{\scriptsize}
\tablecaption{T-Type Classification Schemes  \label{tab:ClassificationCatalog}}
\tablecolumns{19}
\tablewidth{0pt}
\tabletypesize{\scriptsize}
\tablehead{
\colhead{JID } &
\colhead{RA} &
\colhead{DEC } &
\colhead{z} &
\colhead{zconf } &
\colhead{g} &
\colhead{r} &
\colhead{$M_{g}$ } &
\colhead{$L_{g}$} &
\colhead{Rp} &
\colhead{Rp50} &
\colhead{Rp90} \\
\colhead{spID } &
\colhead{$Mass$} &
\colhead{$Age$} &
\colhead{$Color$} &
\colhead{$SFRT$} &
\colhead{$SFRM$} &
\colhead{$\mu_{g}$} &
\colhead{$\mu_{M}$} &
\colhead{$M/L$} &
\colhead{$Area$} &
\colhead{$b/a$} &
\colhead{$seeing_{g}$} \\
\colhead{$ObjectTag_{NYU}$} &
\colhead{$n_{g}$} &
\colhead{$n_{r}$} &
\colhead{$chi^{2}_{g}$} &
\colhead{$chi^{2}_{r}$} &
\colhead{$R50_{n}$} &
\colhead{$R90_{n}$} &
\colhead{$V_{disp}$} &
\colhead{$V_{disp} Err$} &
\colhead{$Vmax$} &
\colhead{$AGN_{1}$} &
\colhead{$AGN_{2}$} \\
\colhead{$Group_{id}$} &
\colhead{$N_{group}$} &
\colhead{$gal_{bright}$} &
\colhead{$gal_{mass}$} &
\colhead{$gp_{lum}$} &
\colhead{$gp_{mass}$} &
\colhead{$h_{mass1}$} &
\colhead{$h_{mass2}$} &
\colhead{$rho_{nyu}$} &
\colhead{$rho$} &
\colhead{$rho_{4}$} &
\colhead{$rho_{5}$} \\
\colhead{T-Type} &
\colhead{Bar  } &
\colhead{Ring} &
\colhead{$Ring_{flag}$} &
\colhead{Lens } &
\colhead{flag} &
\colhead{Pairs} &
\colhead{$Pair_{flag}$} &
\colhead{dist} &
\colhead{tails} &
\colhead{RC3} &
\colhead{Tt} &
}
\startdata

J155341.74-003422.84 & 238.42 & -0.57  & 0.08 & 1. & 15.82 & 15.06 & -22.09  & 10.76 & 18.68 & 8.81 & 17.72 \\ 
343-51692-265 & 11.08 & 4.46  & 0.64 & 1.01 & -9.96 & 22.63 & 8.25 & 0.19 & 674.81  & 0.79 & 1.61  \\  
427500. & 2.93 & 3.56 & 1452.15 & 1457.84 & 5.84 & 18.65  & 143.68 & 7.89  & 0.81 &  999999 &  999999  \\  
148014 & 1. & 1. & 10.59 & 10.80 & 12.95 & 12.50 & 6.08 & -0.06 & -0.08 & -0.05 & 1.  \\ 
 3. & 0. & 36. & 0. & 0. & 0. & 0. & 0. & 0. & 0. & 999999 &  999999\\
 \\
J155146.83-000618.62 & 237.94 & -0.10 & 0.05 & 1. & 15.512 & 14.606 & -21.566 & 10.625 & 12.185 & 4.578 & 14.42  \\  343-51692-304 & 11.25  & 7.11 & 0.83  & 0.89 & -10.94 & 22.44 & 8.57 & 0.33 & 478.97 & 0.95  & 1.56  \\  
51013. & 3.32  & 3.45 & 476.32 & 280.61 & 4.24 & 13.93 & 204.81 & 5.36 & 0.529 & 3. & 3.  \\ 
10223 & 1. & 1. & 10.56  & 11.09 & 12.89 & 13.02 & 1.96 & -0.24 & -0.25 & -0.24  & 2.  \\
 -5. & 0. & 0. & 0. & 0. & 0. & 0. & 0. & 0. & 0. & 999999 &  999999 \\
 \\
J154453.22+002415.48 & 236.22 & 0.41 & 0.03 & 1. & 15.63 & 14.84  & -20.28 & 10.07  & 6.12 & 2.56 & 7.45  \\  
342-51691-381 & 10.41 & 4.29 & 0.746 & -0.01 & -10.84 & 22.47 & 8.15 & 0.09 & 179.24 & 0.85 & 1.37  \\  
87811. & 2.81 & 3.05  & 82.04 & 116.52 & 3.61 & 11.33 & 129.97 & 5.46 & 0.61 & 0. & 1. \\
45472 & 1. & 1. & 9.99 & 10.45 & 11.88 & 11.98 & 0. & -0.62 & -0.76 & -0.24  & 1. \\  
-2.  & 0. & 0. & 0. & 0. & 0. & 0. & 0. & 0. & 0. & 999999 &  999999 \\
 \\
J154711.32+002424.81 & 236.79 & 0.41  & 0.03 & 0.96 & 15.72  & 15.16 & -20.11 & 10.47 & 11.09 & 6.61 & 8.96  \\  342-51691-493 & 10.16 & 1.90 & 0.61 & 1.11 & -9.93 & 22.63 & 7.85 & 0.05 & 203.26 & 0.85 & 1.38   \\  
51451. & 2.55  & 3.13 & 2755.89 & 3844.55 & 5.69 & 17.56 & 45.25 & 12.27 & 0.71 &  999999 &  999999  \\ 
35234 & 1. & 1. & 9.94 & 10.14 & 11.80 & 11.69  &  2.16 & -0.14 & -0.26 & 0.26 & 1. \\ 
  4. & 0. & 0. & 0. & 0. & 0. & 0. & 0. & 0. & 0. & 999999 &  999999 \\
 \\
J154514.39+004619.89 & 236.31 & 0.77 & 0.01  & 1. & 15.34  & 14.96 & -18.38 & 9.11 & 6.92 & 2.55 & 4.87  \\  
342-51691-430 & 9.17 & 1.89 & 0.57  & -1.74 & -10.29 & 22.53 & 7.33 & -0.08 & 69.22 & 0.33 & 1.43  \\ 
 92158. & 2.32 & 2.27 & 63.38 & 21.12 & 12.79  & 38.48  & 89.32  & 14.59 & 0.38 &  999999 &  999999  \\ 
46837 & 1. & 1. & -99. & -99. & 0. & 0. &   0. & -0.89 & -1.83 & 0.93 & 1. &  \\  
  5. & 0. & 0. & 0. & 0. & 2. & 4. & 0. & 0. & 0. & .SAS7?. &  999999 \\
 \\
J155255.43+004304.87 & 238.23 & 0.72 & 0.03 & 0.99 & 15.86  & 15.08 & -19.99 & 9.96 & 3.46 & 1.55 & 4.33  \\  
342-51691-637 & 10.48 & 6.84  & 0.74 & 0.44 & -11.138 & 22.13 & 8.43 & 0.27 & 112.71 & 0.76 & 1.58  \\ 
 51997. & 2.63 & 2.52 & 17.24  & 50.02 & 2.1 & 6.54 & 188.82 & 5.21 & 0.84 & 0. & 1.  \\ 
35379 & 1. & 1. & -99. & -99. & 0. & 0.  &   0. & -0.26 & -0.57  & 2.43 & 1.\\ 
   -5. & 0. & 0. & 0. & 0. & 0. & 0. & 0. & 0. & 0. & 999999 &  999999 \\
 \\
J155357.40+004117.11 & 238.49 & 0.69  & 0.04 & 0.99 & 15.78 & 15.15 & -20.43  & 10.21 & 6.88  & 3.29 & 7.58   \\
 343-51692-377 & 10.63 & 2.00 & 0.58 & 0.46 & -9.98 & 22.17 & 8.30 & 0.27 & 210.29 & 0.49 & 1.51  \\ 
 52004. & 1.22 & 1.42 & 72.62 & 76.61 & 4.4 & 10.84 & 102.61 & 7.84 & 0.75 &  999999 &  999999  \\ 
35380 & 1. & 1. & -99. & -99. & 0. & 0.  &  0. & -0.74 & -1.09 & 1.75 & 1.\\ 
 1. & 4. & 4. & 0. & 0. & 0. & 0. & 0. & 0. & 0. & 999999 &  999999 \\
 \\
J110122.00-010824.89 & 165.34 & -1.14  & 0.07 & 0.99 & 15.59  & 14.63 & -22.24 & 10.87 & 17.78 & 6.81 & 21.95  \\  277-51908-126 & 11.44 & 7.49  & 0.8 & 0.176 & -11.33 & 22.463 & 8.63 & 0.28 & 650.04 & 0.76 & 1.95  \\ 
 53000. & 5.11 & 4.95 & 181.61 & 278.13 & 4. & 13.89 & 285.78  & 6.63 & 0.58 &  999999 &  999999  \\ 
5436 & 1. & 1. & 10.901 & 11.52 & 13.49 & 13.64 &  7.91 &  999999 &  999999 &  999999 & 3.  \\ 
 -5. & 0. & 0. & 0. & 0. & 0. & 0. & 0. & 0. & 0. & 999999 & 0.  \\
 \\
J112000.06-010711.96 & 170.00 & -1.12 & 0.03 & 0.99 & 15.97 & 15.89 & -19.16 & 9.62 & 3.13  & 1.55 & 3.98  \\  
280-51612-241 & 9.12 & 0.50 & 0.45 & 0.17 & -9.03 & 21.06 & 7.26 & -0.39 & 72.29 & 0.56 & 2.3   \\ 
 53158. & 2. & 2.49 & 33.87 & 45.99 & 2.6 & 7.52  & 40.34 & 17.84 & 1. &  999999 &  999999  \\ 
35434 & 1. & 1. & -99. & -99. & 0. & 0. &   4.06  &  999999 &  999999 &  999999 & 1.\\ 
99. & 0. & 0. & 0. & 0. & 0. & 0. & 0. & 0. & 0. & 999999 &  999999  \\
 \\
J112408.63-010927.83 & 171.04 & -1.16 & 0.03 & 0.99 & 14.85 & 14.27 & -20.74 & 10.28 & 6.91 & 2.49 & 7.31 \\  
280-51612-50 & 10.65 &  999999 & 0.55 & 0.63 & -9.67  & 22.00 & 8.31 &  999999 & 216.57 & 0.62 & 2.03   \\  53181. & 2.79 & 2.53  & 309.12 & 200.16 & 3.4 & 10.71 & 155.22 & 5.03 & 0.21 &  999999 &  999999  \\ 
35443 & 1. & 1. & 10.16 & 10.39  & 12.13 & 11.90  &  0. &  999999 &  999999 &  999999 & 1. \\ 
  0. & 0. & 72. & 0. & 0. & 0. & 0. & 0. & 0. & 0. & .LXR0.. & 2.  \\
 \\ 
J113057.91-010851.06 & 172.74 & -1.15 & 0.05 & 1. & 15.95 & 15.20 & -20.79 & 10.35 & 7.72 & 3.52  & 8.11  \\  
282-51658-286 & 10.51 & 2.86 & 0.67 & 0.53 & -10.32 & 22.36 & 8.03 & -0.05 & 299.45 & 0.95  & 2.31  \\  
53226. & 1.86 & 2.11 & 780.5 & 847.45 & 3.09 & 8.77 & 99.97 & 4.55 & 0.94  & 3. & 3.  \\  
 5702 & 1. & 1. & 10.24 & 10.58 & 12.27 & 12.16  & 4.16 &  999999 &  999999 &  999999 & 3. \\ 
 -2. & 0. & 0. & 0. & 0. & 0. & 0. & 0. & 0. & 0. & 999999 & 0.5  \\
 \\
J113833.27-011104.16 & 174.64 & -1.18 & 0.02 & 1. & 14.24 & 13.58 & -20.57 & 10.27 & 8.99 & 4.14 & 8.83  \\  
282-51658-49 & 10.74 &  999999 & 0.56 & 0.63 & -9.48 & 22.35 & 8.25 &  999999 & 309.23 & 0.67 & 2.43  \\  
53246. & 1.95 & 2.09 & 1192.19 & 1221.29 & 11.8 & 33.85 & 82.05 & 6.2 & 0.097 &  999999 &  999999  \\ 
35473 & 1. & 1. & 10.09  & 10.28  & 12.01 & 11.79  & 0. &  999999 &  999999 &  999999 & 1.\\ 
 2. & 0. & 36. & 0. & 0. & 0. & 0. & 0. & 0. & 0. & .SXR2?. & 3.5 \\
 \\
\enddata
\end{deluxetable*}
%\end{landscape}

The full catalog of visual classifications presented in this paper will be available in the electronic edition on ApJS as well as at the following site: \url{http://www.bo.astro.it/~nair/Morphology/}. To orient the reader with respect to the information included in the catalog,
Table~\ref{tab:ClassificationCatalog} presents a small sample of the full catalog. 
The catalog contains 60 columns for 14034 galaxies and has the following information:\\

\begin{figure*}[t!]
\begin{center}
\unitlength1cm
\hspace{-2cm}
\begin{minipage}[t]{4.0cm}
\rotatebox{0}{\resizebox{6cm}{6cm}{\includegraphics{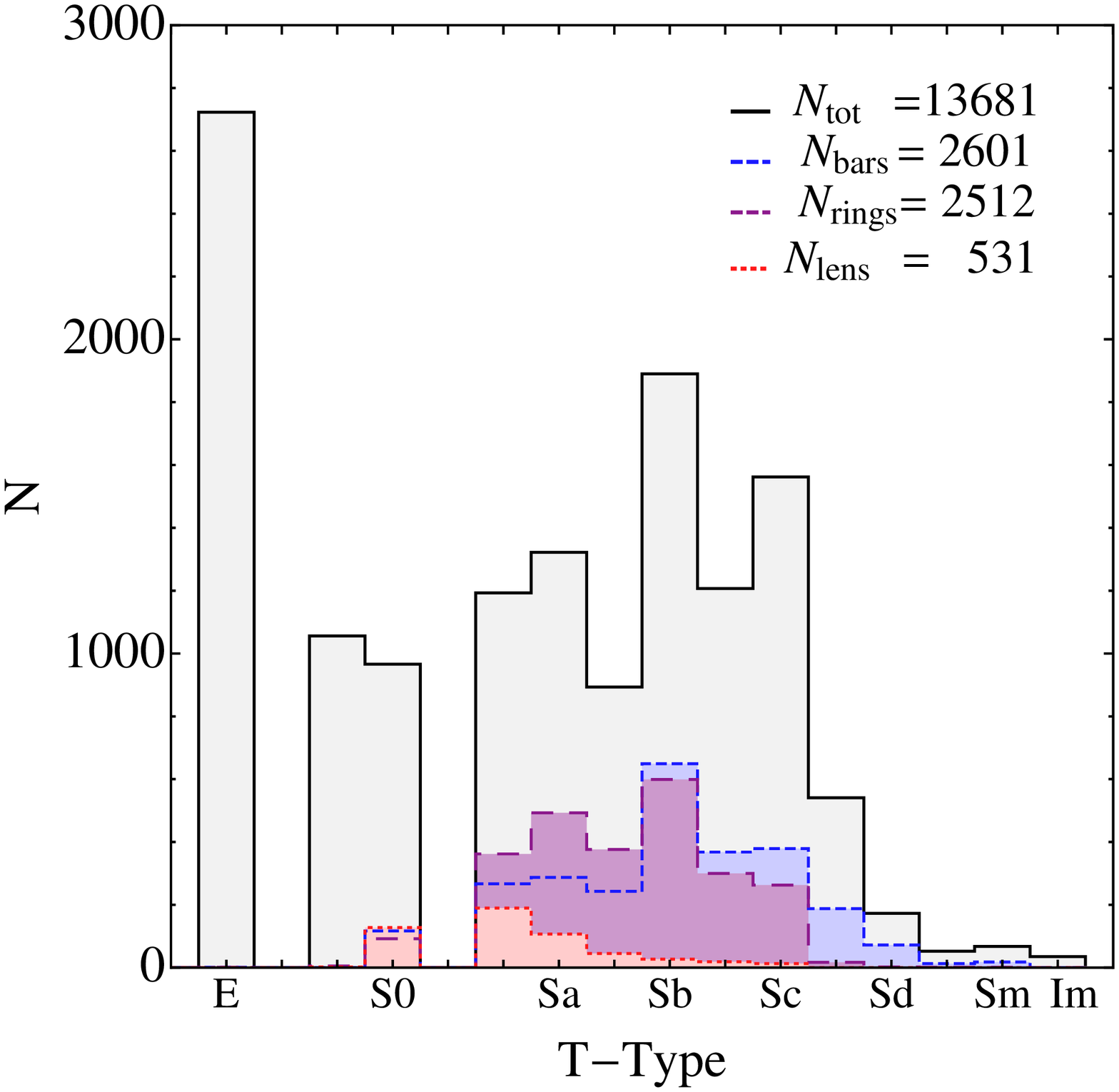}}}
\end{minipage}
\hspace{3.0cm}
\begin{minipage}[t]{4.0cm}
\rotatebox{0}{\resizebox{6cm}{6cm}{\includegraphics{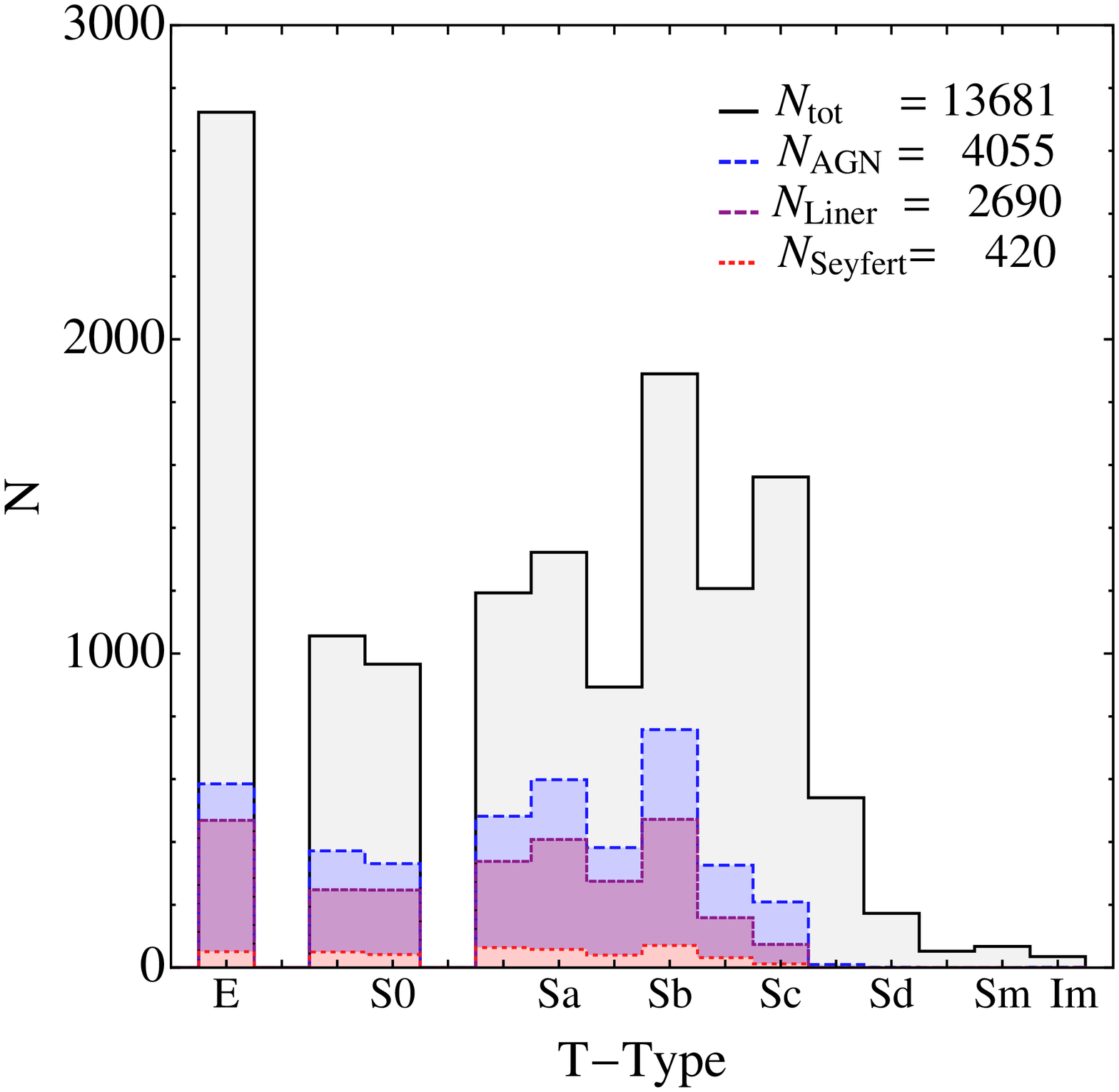}}}
\end{minipage}
\caption[Histograms of T-Types]{\label{fig:LocalDistributionsT} Histograms of T-Types for (a) Bars (blue, short dash), Rings (purple, long dash), Lenses (red, dotted), and (b) AGN distribution in blue (short dash), LINER subtype in purple (long dash) and Seyferts in red (dotted). The grey distributions are for the entire sample. Objects with T-Type $=$ 99 are not shown. See text for details.}
\end{center}
\end{figure*}

\begin{deluxetable*}{lrrrrrrrr}
\tabletypesize{\scriptsize}
\tablecaption{Summary of Distributions \label{TableStats}}
\tablecolumns{18}
\tablewidth{0pt}
\tabletypesize{\small}
\tablehead{
\colhead{Class } & 
\colhead{No. } &
\colhead{Bars} &
\colhead{Rings } &
\colhead{Lenses } &
\colhead{AGNs } &
\colhead{Liners} &
\colhead{Seyferts } &
\colhead{Dist.}
}
\startdata
E & 2723 & 0 & 0 & 0 & 585 (21\%)& 469 (17\%)& 51 (2\%)& 90\\
ES0 & 1056 & 0 & 6 & 2 & 372 (35\%) & 248 (23\%)& 50 (5\%)& 164\\
S0 & 966 & 117 (12\%)& 92 (10\%)& 127 (13\%)& 331 (34\%)& 247(26\%)& 42 (5\%)& 71\\
S0a & 1193 & 267 (22\%)& 362 (30\%)& 190 (16\%)& 482 (40\%)& 338 (28\%)& 64(5\%)& 81\\
Sa & 1322 & 287 (22\%)& 493 (37\%)& 107 (8\%)& 598 (45\%)& 408 (31\%)& 58 (4\%)& 91\\
Sab & 893 & 243 (27\%) & 376 (42\%)& 45 (5\%)& 382 (43\%)& 275 (31\%)& 40 (5\%)& 78\\
Sb & 1890 & 649 (34\%)& 599 (32\%)& 27 (14\%)& 758 (40\%)& 472 (25\%)& 71 (4\%)& 93\\
Sbc & 1207 & 368 (30\%)& 300 (25\%)& 19 (2\%)& 326 (27\%)& 159 (13\%)& 32 (3\%)& 54\\
Sc & 1562 & 379 (24\%)& 263 (17\%)& 13 (1\%)& 209 (13\%)& 74 (5\%)& 12 & 49\\
Scd & 541 & 188 (35\%)& 17 (3\%)& 0 & 10 (2\%)& 0 & 0 & 25\\
Sd & 173 & 72 (42\%)& 2 & 0 & 1 & 0 & 0 & 16\\
Sdm & 52 & 13 (25\%)& 1 & 0 & 0 & 0 & 0 & 7\\
Sm & 68 & 18 (26\%)& 1 & 0 & 0 & 0 & 0 & 10\\
Im & 35 & 0 & 0 & 0 & 1 & 0 & 0 & 6\\
Un & 353 & 11 (3\%)& 8 (2\%)& 1 & 57 (16\%)& 29 (8\%)& 1 & 134
\\\hline
Total & 14034 & 2612 (26\%)& 2520 (25\%) & 532 (5\%) & 4112 (29\%)& 2719 (19\%)& 421 (3\%)& 969(7\%) 
\enddata
\end{deluxetable*}

\begin{deluxetable*}{lrrrrrr}
\tabletypesize{\scriptsize}
\tablecaption{Summary of Disturbed Objects \label{Disturbed}}
\tablecolumns{18}
\tablewidth{0pt}
\tabletypesize{\small}
\tablehead{
\colhead{Class } & 
\colhead{Shells } &
\colhead{Short Tails} &
\colhead{Inter Tails} &
\colhead{Long Tails } &
\colhead{Pairs} &
\colhead{General} }
\startdata
E &  16 &  3 &  0 &  2 &  24  &  44 \\
ES0 &  30 &  22 &  9 &  7 &  38  &  77 \\
S0 &  7 &  9 &  10 &  7 &  25 &  28  \\
S0a &  10 &  10 &  8 &  7 &  13 &  39   \\
Sa &  8 &  12 &  6 &  6 &  10  &  50   \\
Sab &  1 &  6 &  7 &  8 &  9  &  50   \\
Sb &  0 &  4 &  12 &  10 &  13  &  59   \\
Sbc &  1 &  3 &  6 &  7 &  19 &  23  \\
Sc &  1 &  4 &  4 &  7 &  17 &  22   \\
Scd &  0 &  0 &  0 &  4 &  5 &  15   \\
Sd &  1 &  0 &  3 &  2 &  1  &  10   \\
Sdm &  0 &  1 &  1 &  2 &  0  &  3   \\
Sm &  0 &  2 &  1 &  3 &  2  &  3   \\
Im &  1 &  0 &  1 &  0 &  1  &  3   \\
Un &  3 &  20 &  32 &  33 &  79  &  28
\\\hline
Total & 79 &  96 &   100 &  105  &  256 &  454
%AGN & 17\% &  17\% &   28\% &  27\% &  23\% &  23\% \\
\enddata
\end{deluxetable*}

%The catalog contains 34 columns for 14034 galaxies and has the following information:

\noindent
Column  1   :  J2000 ID. The format preferred by the SDSS collaboration for object identification.\\
Column   2   :  Right Ascension (J2000) in degrees.\\
Column   3   :  Declination (J2000) in degrees\\
Column   4   :  Spectroscopic Redshift \\
Column   5   :  Confidence level in redshift measurement\\
Column   6   :  {\em g'} apparent magnitude (extinction corrected)\\
Column   7   :  {\em r'} apparent magnitude (extinction corrected)\\\
Column  8  :  {\em g'}-band absolute magnitude corrected to z=0 using the {\tt kcorrect} code of Blanton et al 2003.\\
Column  9  :  Luminosity in {\em g'}-band in solar units\\
Column  10  :  Petrosian Radius \citep{Petrosian:1976p6531}\\
Column  11  :  Rp50 in kiloparsec\\
Column  12  :  Rp90 in kiloparsec\\
Column   13   :  Spectra ID made up of the MJD, Plate and Fibre number \\
Column  14  :  Mass in log units \citep{Kauffmann:2003p97}\\
Column  15  :  Age in Gyr \citep{Kauffmann:2003p97}\\
Column 16   :  {\em g'-r'} color\\
Column  17  :  Total star formation rate \citep{Brinchmann:2004p3060}\\
Column  18  :  Total star formation rate per unit mass \citep{Brinchmann:2004p3060}\\
Column  19  :  Surface Brightness in g, corrected for galactic extinction and internal extinction as prescribed by RC3. \\
Column  20  :  Surface Mass Density\\
Column  21  :  M/L in g\\
Column  22  :  Area of the galaxy in arcsec square  \\
Column  23  :  Axis Ratio (b/a)\\
Column 24   :  Seeing in g-band\\
Column 25   :  OBJECT\_TAG from the NYU value added galaxy catalog (VAGC) for the DR4 release \citep{Blanton:2005p79}. This is provided to aid catalog matching with the NYU database.\\
Column 26   :  Single component Sersic index in g-band calculated by NYU-VAGC group \citep{Blanton:2003p3068}. \\
Column 27   : Single component Sersic index in r-band calculated by NYU-VAGC group \citep{Blanton:2003p3068}.\\
Column 28   :  Chi-sq for single component sersic index in g-band calculated by NYU-VAGC group \citep{Blanton:2003p3068}.\\
Column 29   : Chi-sq for single component sersic index in r-band calculated by NYU-VAGC group \citep{Blanton:2003p3068}.\\
Column 30   : Single component sersic petrosian half light radii R50 in g-band calculated by NYU-VAGC group \citep{Blanton:2003p3068}.\\
Column 31   : Single component sersic petrosian R90 radii in g-band calculated by NYU-VAGC group \citep{Blanton:2003p3068}.\\
Column  32  : Velocity dispersion from SDSS in km/s.\\
Column  33  : Error in velocity dispersion from SDSS in km/s.\\
Column 34   : V/Vmax.\\
Column  35  :  AGN types using the demarcation line of \cite{Kauffmann:2003p23}. The flag can take the following possible values :
0: Star forming galaxy, 1: Transition/mixed AGN, 2: Seyfert only, 3: LINER only. The total AGN sample consists of objects classified as 1, 2 and 3.\\
Column  36  :  AGN types using the demarcation line of \cite{Kewley:2001p21444}. The flag can take the following possible values :
0: not AGN, 1: Transition/mixed AGN, 2: Seyfert only, 3: LINER only. The total AGN sample consists of objects classified as 1, 2 and 3.\\
Column 37 : Group ID  from \cite{Yang:2007p19054} (YA07) group catalog. Briefly, \cite{Yang:2007p19054} use an iterative halo-based group finder on the NYU-VAGC SDSS catalog identifying tentative group members using a modified friends-of-friends algorithm. The group members were used to determine the group center, size, mass and velocity dispersion. New group memberships were determined iteratively based on the halo properties. The final catalog yields additional information identifying the brightest galaxy in the group (BCG), the most massive galaxy in the group (both used as proxies for central galaxies), estimated group mass, group luminosity and halo mass.\\
Column 38 : Number of galaxies in the group (from YA07).\\
Column 39 : Flag indicating if galaxy is the most massive galaxy in the group. 1: most massive galaxy. 2: satellite galaxies.
It is important to note that groups with only 1 member have the massive galaxy flag set (see YA07).\\
Column 40 : Flag indicating if galaxy is the most luminous galaxy in the group. 1: most massive galaxy. 2: satellite galaxies.
It is important to note that groups with only 1 member have the luminous galaxy flag set (see YA07).\\
Column 41 : Group luminosity (see YA07).\\
Column 42 : Group mass (see YA07).\\
Column 43 : Group halo mass estimate (see YA07).\\
Column 44 : Group halo mass estimate (see YA07).\\
Column 45 : Environmental over-density (rho) estimated by NYU-VAGC group \citep{Hogg:2004p294,Blanton:2005p2271}. For each galaxy, `neighbors within a cylinder of 1 $h^{-1}$ Mpc and a comoving half-length of 8 $h^{-1}$ are counted and the results divided by the mean predicted from the luminosity function' \citep{Blanton:2003p2277}. Terms are missing for 434 galaxies and rho is set to 999999 in those instances.\\
Column 46 : The average environmental density from \cite{Baldry:2006p103} . It is defined as  $\Sigma = N/ (\pi d_{N}^{2})$,
where $d_{N}$ is the projected comoving distance (in Mpc) to the Nth nearest neighbour (within $\pm$1000 km/s if a spectroscopic redshift was available or else with photometric redshift errors within the 95 percent confidence limit). A best estimate density (to account for spectroscopic incompleteness) was obtained by calculating the average density for N=4 and N=5 with spectroscopically confirmed members only and with the entire sample. \\
Column 47 : Nth nearest neighbor environment estimate from \cite{Baldry:2006p103}  with N=4.\\
Column 48 : Nth nearest neighbor environment estimate from \cite{Baldry:2006p103}  with N=5.\\
Column  49  :  T-Type classification using the modified RC3 classifiers as specified in the previous section.\\
Column  50  :  Bar Type encoded as  $\sum_{i}2^{i}$  where $i$ flags the following possible values :\\
1: strong bar, 2: intermediate, 3: weak bar, 4: ansae, 5 : peanut, 6: nuclear bar 7: bar unsure. Thus if a large scale strong bar and a nuclear bar is present the bar type will be $2^{1}+2^{6} = 66$ . \\
Column  51  : Ring Types encoded as $\sum_{i}2^{i}$ where $i$ flags the following possible values-
1: nuclear ring, 2: inner ring, 3: outer ring.\\
Column  52  : Ring Flags encoded as  $\sum_{i}2^{i}$  where $i$ flags the following possible values :\\
1: partial ring,  2: pseudo outer ring R1, 3: pseudo outer ring R2.\\
Column  53  :  Lens Type, encoded as  $\sum_{i}2^{i}$ where $i$ flags the following possible values :\\
 1: inner lens is present, 2: outer lens is present\\
Column  54  :  Flags:  T-Type flags are 0:No flag set, 1: Doubtful, 2: Highly Doubtful, 3: Unknown, 4: peculiar.  \\
Column 55  : Pair Flag encoded as $\sum_{i}2^{i}$ where $i$ takes the following possible values : \\
  1: Close Pair, 2: Projected Pair, 3: Adjacent Pair, 4: Overlapping Pair\\
Column 56  : Pair Type Flag encoded as $\sum_{i}2^{i}$ where $i$ takes the following possible values : \\
  1: Star, 2: Compact object (small spheroid like), 3: small fuzzy blob (morphology unclear), 4: Elliptical/S0 galaxy, 5: Disk galaxy, 6: Irregular/Peculiar galaxy\\
Column 57  :  Interaction Types encoded as $\sum_{i}2^{i}$ where $i$ takes the following possible values :\\ 1: No Interaction Signature, 2: Disturbed, 3: Warp, 4: Shells, 5: Short Tail, 6: Medium Tail, 7: Long Tail, 8: Bridge\\
Column  58  : Number of tails; 1 Tail,  2 Tails, 3+Tails, Bunny.\\
Column  59 : RC3 full visual classification when available. Reader is referred to the RC3 for details on notations.The field is set to 999999 when no classification is available.\\
Column 60 : \cite{Fukugita:2007p3031} `Tt' classification when available. Reader is referred to the original paper for details. The field is set to 999999 when no classification is available.\\

Columns 50 to 58 are using binary mask encoding to record the existence of various flags using a single integer. This format is convenient because it allows multiple flags to be set simultaneously. For example, consider a galaxy that has both an inner and an outer ring. As noted in the explanatory text for column 27, an inner ring is denoted by a flag with a value of 2, and an outer ring by a flag with a value of 3. The existence of both an inner and an outer ring is denoted by giving column 27 a value of 12, which is $2^2 + 2^3$. To decode this information, this decimal number should be converted to a binary number (i.e. 12 in the example is `1100'). Determining which flags are set is trivially implemented in code by using two's complement masking. One can determine which flags are set visually by scanning the binary number from right-to-left (preceding zero positions must be counted), and noting which positions contain a `1'. 

\section{Summary statistics}

As has already been noted, the main analysis of the trends in this catalog will be presented in a series of follow-up papers. For present purposes, we confine ourselves to summarizing the overall statistical properties of the sample, noting along the way a few obvious trends that provide consistency checks, and which allow for general comparisons to be made between our catalog and previously published work.

\begin{figure*}[htbp!]
%\begin{center}
\unitlength1cm
\hspace{2cm}
\begin{minipage}[t]{4.0cm}
\includegraphics[width=5.5in]{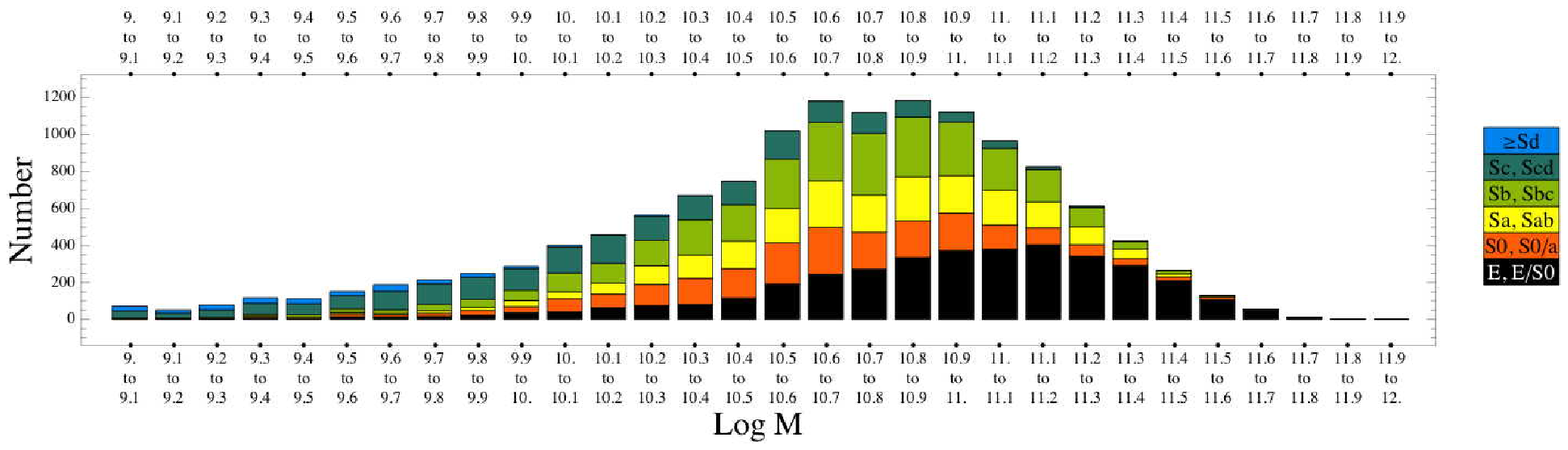}
\includegraphics[width=5.5in]{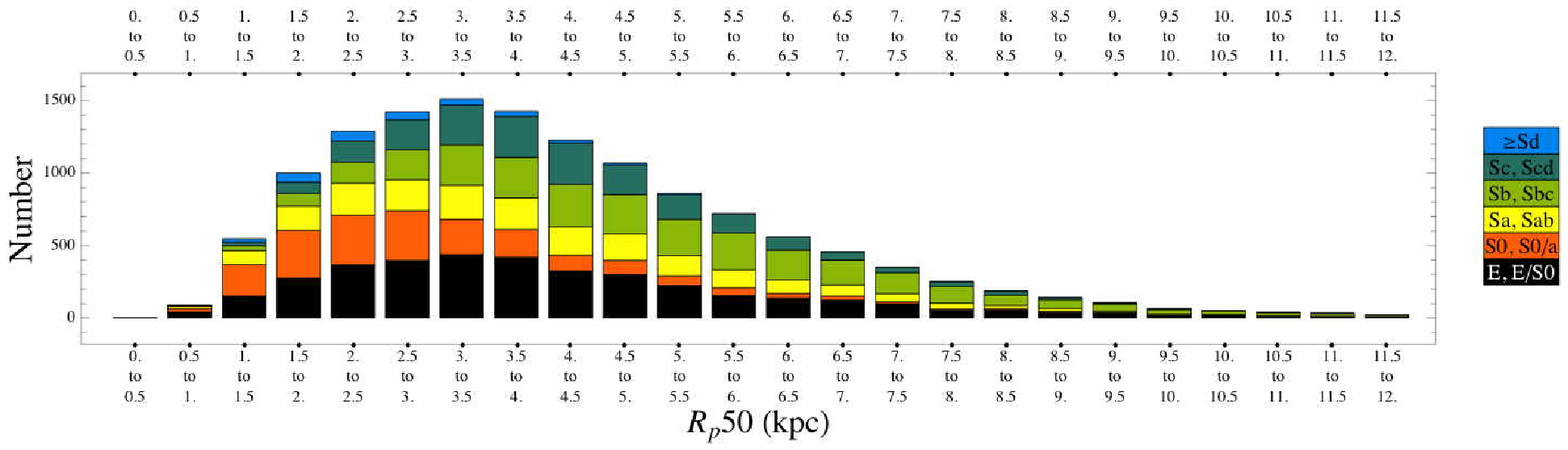}
\end{minipage}
\caption[Histograms of mass and size keyed to T-Type]{\label{fig:MassT} Histogram of various Hubble Types as a function of (a) Mass in log units and (b) Petrosian half light radius $R_{p}50$ in kpc. The sub-categories of galaxy types have been grouped into the following broad classes: E and E/S0 galaxies (black bars), S0 and S0/a galaxies (orange bars), Sa and Sab galaxies (yellow bars), Sb and Sbc galaxies (light green bars), Sc and Scd galaxies (dark green bars) and galaxies with T-Types later than Sd (blue bars.) See text for details.
}
%\end{center}
\end{figure*}

\begin{figure*}[htbp!]
%\begin{center}
\unitlength1cm
\hspace{2cm}
\begin{minipage}[t]{4.0cm}
\includegraphics[width=5.5in]{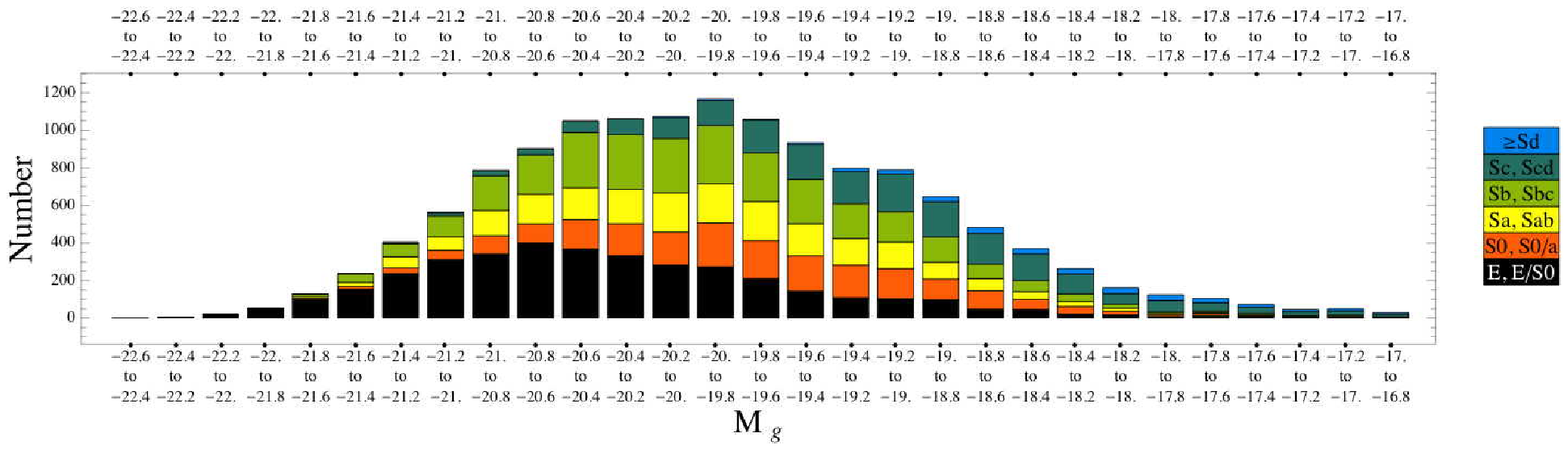}
\includegraphics[width=5.5in]{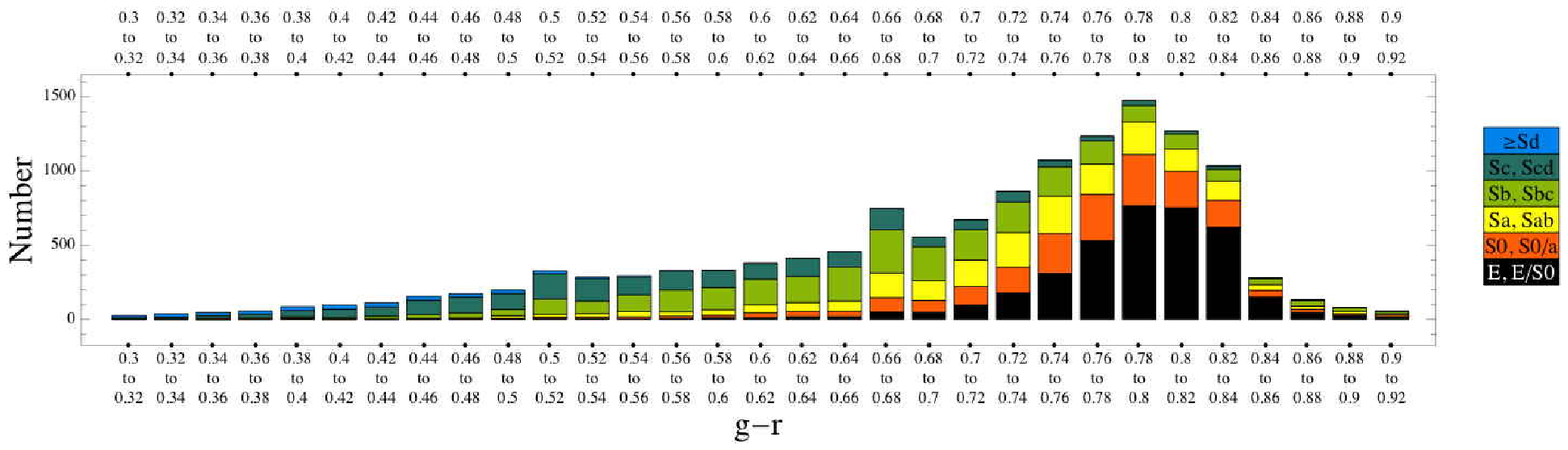}
\end{minipage}
\caption[Histogram of g-band absolute magnitude keyed to T-Type]{\label{fig:MagT} Histogram of various Hubble Types as a function of (a) g-band absolute magnitude and (b) g-r color. The sub-categories of galaxy types have been grouped into the following broad classes: E and E/S0 galaxies (black bars), S0 and S0/a galaxies (orange bars), Sa and Sab galaxies (yellow bars), Sb and Sbc galaxies (light green bars), Sc and Scd galaxies (dark green bars) and galaxies with T-Types later than Sd (blue bars.) See text for details.
}
%\end{center}
\end{figure*}

\subsection{Local Statistics}

A summary of the statistics of our sample is given in Table~\ref{TableStats}. The T-Type distribution clearly shows that elliptical and classical spirals are well represented but objects later than Scd are not in comparison. This is of course entirely as expected, since any apparent magnitude-limited sample has an absolute magnitude distribution peaking around $M_\star$. Elliptical and S0 galaxies account for $34\%$ of our sample, classical spirals $61\%$ and very late types, including peculiars/mergers approximately $5\%$. We postpone the discussion on abundances of fine classes (bars, rings, lenses) and interacting objects to a forthcoming paper in the series along with the correlations seen with AGN activity but provide a brief overview here. Figure~\ref{fig:LocalDistributionsT} shows the distribution of (a) galaxies with definite bars, rings and lenses and, (b) galaxies defined as AGN, pure seyferts and pure LINERs, as in \cite{Kauffmann:2003p23}\footnote{Seyfert galaxies are defined to have $[OIII]/H\beta>3$ and $[NII]/H\alpha>0.6$ while LINERs have $[OIII]/H\beta<3$ and $[NII]/H\alpha>0.6$}. For objects later than ES0 but with no inclination cut, we find bars, rings and lenses are 26\% $\pm<0.5\%$, 25\% $\pm<0.5\%$ and 5\% $\pm<0.5\%$ of our sample population respectively. We find rings and lenses are located nearly entirely in classical spirals (classes earlier than Scd), though there is a strong T-Type dependence as can be seen in Table~\ref{TableStats} with a ring fraction peak of 42\% for Sa galaxies. Bars are distributed through all disk T-Types as expected. Figure~\ref{fig:LocalDistributionsT}(b) shows the AGN in our sample are dominated by pure LINERs (19\%) with far fewer pure Seyferts(3\%). The total AGN fraction is approximately 29\% in our sample, though again, from Table~\ref{TableStats}, there may be a T-Type dependence. We find AGN fraction is highest for classical spirals, with 45\% of Sa galaxies being active.

\subsection{Interacting galaxies}
 
Table~\ref{Disturbed} provides a summary of the different types of interacting objects in our sample, specifically objects with tidal tails or shells as well as objects identified in close pairs with another galaxy. Galaxies under the `general' column are disturbed but have not been placed into any of the previous categories. Objects listed as pairs are objects with a nearby interacting companion and include both the early and late stages of interaction. Interaction classifications are not mutually exclusive, and there is overlap between some of the columns (galaxies with shells or tails can also be in pairs). In total, there are 969 (7\%) interacting objects in our sample. $30\% \pm 2\%$ of the interacting objects host an AGN, similar to $29\% \pm 0.5\%$ of non-disturbed galaxies. %Objects in pairs (which includes both early and late stages of interaction) show an AGN fraction of 23\%. 
However, the AGN fraction differs among the different classes of interacting objects. Considering only those close pairs which are at an early stage of interaction (171) we find a slightly reduced AGN fraction of $25\%\pm 4\%$ (though we have not tried to account for projected pairs). Objects with shells represent the end stage of the merger process and have an AGN fraction of ($22\% \pm 5\%$). We find objects undergoing an interaction with medium or long tails have a higher AGN fraction ($37\% \pm 4\%$) than other disturbed objects. Galaxies with short tails have a lower AGN fraction ($26\% \pm 4\%$). This may imply that major interactions which yield larger tails are more likely to trigger an AGN than a minor interaction which leads to a short tail.

\subsection{Distribution of T-Types with physical properties}
%XXX:
Figure~\ref{fig:MassT} shows the distribution of Hubble types with (a) mass and (b) petrosian half light radius. We find E+E/S0 have the highest masses as expected, though again they span a wide range. At the highest mass end there are no objects later than S0 galaxies. Classical spirals peak in the middle of the stellar mass distribution, $\log(M) \thicksim 10.7 M_\sun$ with later types dominating the low mass end. The histogram distribution for size shows an interesting trend. E+E/S0 have small petrosian half light radii and Hubble types Sa onwards have larger sizes on average. However S0 galaxies peak earlier than ellipticals. This may be because we do not distinguish between massive cD galaxies or dwarf ellipticals. Hence the range in sizes spanned by E/ES0 is larger than S0 galaxies.

Figure~\ref{fig:MagT} shows the histogram distribution of Hubble types with (a) $g$-band absolute magnitude and (b) $g-r$ color. 
As expected we see elliptical galaxies dominate the bright end of the magnitude distribution, classical spirals dominate the mid-range and late-type spirals dominate the low luminosity end of the distribution. With respect to color, E+E/S0 are the reddest galaxies as expected but have a large tail in their distribution. There is significant overlap with early classical spirals. Late type spirals dominate the blue end of the spectrum. Color cuts are also frequently used to identify ellipticals. Using a $g-r$ color-cut of 0.7, we find the contamination by disk galaxies (Sa and later) is 36\%. Using a stringent axis ratio cut of $b/a>0.6$ reduces the contamination to 19\%.

\begin{figure*}[htbp!]
\unitlength1cm
%\vspace{-2cm}
\hspace{2cm}
\begin{minipage}[t]{4.0cm}
\includegraphics[width=5.5in]{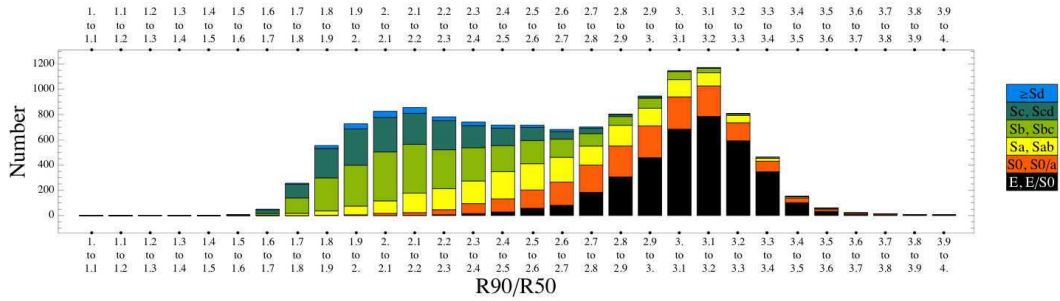} 
\end{minipage}
\caption[Concentration Histogram]{\label{fig:ConcentrationHistogram} 
Histogram of various Hubble Types as a function of central concentration R90/R50 as used by SDSS collaboration. The sub-categories of galaxy types have been grouped into the following broad classes: E and E/S0 galaxies (black bars), S0 and S0/a galaxies (orange bars), Sa and Sab galaxies (yellow bars), Sb and Sbc galaxies (light green bars), Sc and Scd galaxies (dark green bars) and galaxies with T-Types later than Sd (blue bars.) See text for details.
}
\end{figure*}

\begin{figure*}[htbp!]
\unitlength1cm
%\vspace{-2cm}
\hspace{2cm}
\begin{minipage}[t]{4.0cm}
\includegraphics[width=5.5in]{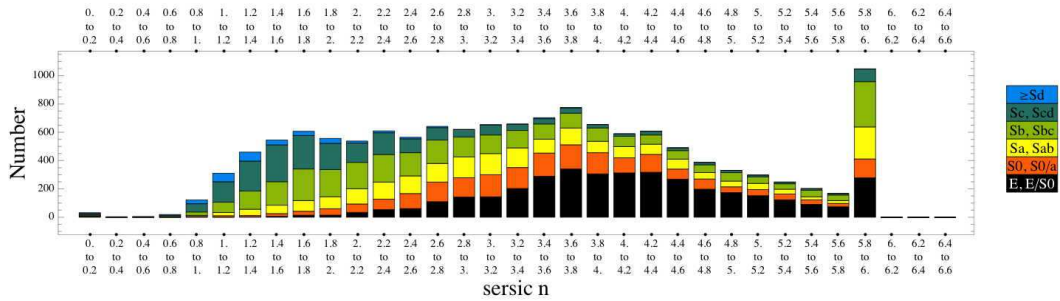} 
\end{minipage}
\caption[Sersic Histogram]{\label{fig:SersicHistogram} 
Histogram of various Hubble Types as a function of g-band sersic index calculated by the NYU value-added catalog group \citep{Blanton:2005p79}. The sub-categories of galaxy types have been grouped into the following broad classes: E and E/S0 galaxies (black bars), S0 and S0/a galaxies (orange bars), Sa and Sab galaxies (yellow bars), Sb and Sbc galaxies (light green bars), Sc and Scd galaxies (dark green bars) and galaxies with T-Types later than Sd (blue bars.) See text for details.
}
\end{figure*}

\subsection{Comparison with other measures of morphology}
The previous section illustrates the usefulness and pitfalls of color as a tool for morphological segregation.  In addition to color, concentration, sersic index and location of galaxies on color-magnitude, color-color or color-gradient vs color spaces are used to segregate populations.

In Figure~\ref{fig:ConcentrationHistogram} we present a histogram showing the number of galaxies grouped by Hubble type and binned in central concentration. We use the SDSS definition of the ratio of the radii enclosing 90$\%$ and 50$\%$ of the flux of a galaxy. The overall properties of the galaxies as a function of Hubble type will be described in a forthcoming paper but for present purposes, it suffices to simply note a few points which can be inferred from Figure~\ref{fig:ConcentrationHistogram} :

\noindent {\em (i) Elliptical and E/S0 galaxies are the most highly centrally concentrated systems} (as expected) and the concentration of later Hubble types decline smoothly with T-Type. 

\noindent {\em (ii) No single concentration cut isolates a complete sample of E+E/S0 galaxies which is free from contamination by spiral galaxies.} As expected, the degree (and nature) of contamination depends on the specific concentration cut chosen.

\noindent {\em (iii) A surprisingly broad range of spiral galaxies exhibit central concentrations which overlap with those of elliptical galaxies.} While most are highly concentrated early spirals (S0,Sa), a significant number of intermediate-late type spirals (Sb,Sbc) have central concentrations comparable with those of low-concentration ellipticals. 

The data in these figures is useful for determining how well parameters like concentration can be used
as a proxy for Hubble stage. For example, \cite{Shen:2003p70} use a $R90/R50$ concentration cut of 2.86 as defined by \cite{Nakamura:2003p18391} to distinguish between early type and late type galaxies. This selects a large range of E, E/S0 and S0 galaxies ($85\%$ of all such galaxies) but is also significantly contaminated (22$\%$) by Sa,Sab,Sb and Sbc galaxies. Restricting the sample to-face on objects with $b/a>0.6$ reduces the contamination to 15\%.

Figure~\ref{fig:SersicHistogram} shows the number of galaxies grouped by Hubble type and binned in sersic index, as calculated by the NYU value added catalog group \citep{Blanton:2005p79,Blanton:2003p3068}. Sersic index, which is also used as a proxy for B/T, also shows a similar trend to concentration such that there is no sersic index cut which will select only elliptical or E/S0 galaxies without significant contamination by early type spiral galaxies or without the loss of the tail of the elliptical distribution. A more detailed study of the relationship between visual morphology and quantitative morphology and the role of selection effects is postponed to a later paper in this series.

\section{Conclusion}

We have presented a catalog of detailed visual classifications for 14034 galaxies in the Sloan Digital Sky Survey (SDSS) Data Release 4 (DR4) in the redshift range $0.01<z<0.1$ and with $g<16$ mag. In addition to T-Types we record the existence of bars, rings, lenses, tails, warps, dust lanes, arm flocculence and multiplicity. 1793 galaxies in our sample are also contained in the RC3, and a comparison of our classifications to those in the RC3 shows good agreement in T-Types, as does a comparison between our classifications and those of Fukugita et al. (2007), for the 450 galaxies in common. The fraction of systems classified as barred in our sample (26\%) is substantially lower than the fraction in RC3, which we attribute to our more stringent definition for classifying a galaxy as weakly barred. Our ring fraction is 26\%, though it peaks at 42\% for Sa galaxies.
Our lens fraction is 5\%, though it is higher for early type galaxies. 
These numbers suggest a possible
%In further papers in this series we will investigate the possible 
dependence of bar, ring, and lens frequency on T-Type.
%, stellar mass, color and AGN activity. 
We will investigate this further in later papers in this series along with their dependence on stellar mass, color, AGN activity and environment. 
%The second paper in this series will investigate the size-luminosity relation as a function of T-Type and environment.

With regard to quantitative measurements related to morphological classifications, we find that although
concentration and sersic indices are correlated with visual classifications, they are not equivalent. This is by design, as Hubble based his classification scheme on multiple parameters (bulge-to-disk ratio, spiral arm pitch angle, and arm resolution) which are correlated, but not perfectly. Hence a single quantitative parameter cannot replace T-Type. In Paper V in this series we will present the quantitative morphology catalog corresponding to this data set and investigate the number of parameters required to account for the wide diversity in morphology seen in the local universe.
% and will release a catalog of quantitative parameters corresponding to this visual catalog.

\acknowledgments
\noindent{\em Acknowledgments}

We are grateful to Debra Elmegreen for agreeing to visually classify a number of candidate barred galaxies for us, which greatly helped us to understand the systematics of our sample. Jarle Brinchmann, Guinevere Kauffmann, Michael Blanton, David Hogg, Ivan Baldry and Xiaohu Yang deserve our thanks for allowing us to incorporate information from their papers as columns in our catalog, thus saving users of this catalog from much tedious object matching. We thank Luc Simard for commenting on early versions of many of the figures which appeared in PN's thesis, and we thank Sidney van den Bergh for many useful and enjoyable discussions regarding galaxy morphology.

Funding for the SDSS and SDSS-II has been provided by the Alfred P. Sloan Foundation, the Participating Institutions, the National Science Foundation, the U.S. Department of Energy, the National Aeronautics and Space Administration, the Japanese Monbukagakusho, the Max Planck Society, and the Higher Education Funding Council for England. The SDSS Web Site is http://www.sdss.org/.

The SDSS is managed by the Astrophysical Research Consortium for the Participating Institutions. The Participating Institutions are the American Museum of Natural History, Astrophysical Institute Potsdam, University of Basel, University of Cambridge, Case Western Reserve University, University of Chicago, Drexel University, Fermilab, the Institute for Advanced Study, the Japan Participation Group, Johns Hopkins University, the Joint Institute for Nuclear Astrophysics, the Kavli Institute for Particle Astrophysics and Cosmology, the Korean Scientist Group, the Chinese Academy of Sciences (LAMOST), Los Alamos National Laboratory, the Max-Planck-Institute for Astronomy (MPIA), the Max-Planck-Institute for Astrophysics (MPA), New Mexico State University, Ohio State University, University of Pittsburgh, University of Portsmouth, Princeton University, the United States Naval Observatory, and the University of Washington.

%\bigskip
%\appendix
%\bigskip
%%\centerline{\Large\bf Appendices}


\begin{thebibliography}{36}
\expandafter\ifx\csname natexlab\endcsname\relax\def\natexlab#1{#1}\fi

\bibitem[{Abraham {et~al.}(1994)Abraham, Valdes, Yee, \& van~den
  Bergh}]{Abraham:1994p438}
Abraham, R.~G., Valdes, F., Yee, H. K.~C., \& van~den Bergh, S. 1994,
  Astrophysical Journal, 432, 75

\bibitem[{Adelman-McCarthy {et~al.}(2006)Adelman-McCarthy, Ag{\"u}eros, Allam,
  Anderson, Anderson, Annis, Bahcall, Baldry, Barentine, Berlind, Bernardi,
  Blanton, Boroski, Brewington, Brinchmann, Brinkmann, Brunner, Budav{\'a}ri,
  Carey, Carr, Castander, Connolly, Csabai, Czarapata, Dalcanton, Doi, Dong,
  Eisenstein, Evans, Fan, Finkbeiner, Friedman, Frieman, Fukugita, Gillespie,
  Glazebrook, Gray, Grebel, Gunn, Gurbani, de~Haas, Hall, Harris, Harvanek,
  Hawley, Hayes, Hendry, Hennessy, Hindsley, Hirata, Hogan, Hogg, Holmgren,
  Holtzman, Ichikawa, Ivezi{\'c}, Jester, Johnston, Jorgensen, Juri{\'c}, Kent,
  Kleinman, Knapp, Kniazev, Kron, Krzesinski, Kuropatkin, Lamb, Lampeitl, Lee,
  Leger, Lin, Long, Loveday, Lupton, Margon, Mart{\'\i}nez-Delgado, Mandelbaum,
  Matsubara, McGehee, McKay, Meiksin, Munn, Nakajima, Nash, Neilsen, Newberg,
  Newman, Nichol, Nicinski, Nieto-Santisteban, Nitta, O'Mullane, Okamura, Owen,
  Padmanabhan, Pauls, Peoples, Pier, Pope, Pourbaix, Quinn, Richards, Richmond,
  Rockosi, Schlegel, Schneider, Schroeder, Scranton, Seljak, Sheldon,
  Shimasaku, Smith, Smol{\v c}i{\'c}, Snedden, Stoughton, Strauss, SubbaRao,
  Szalay, Szapudi, Szkody, Tegmark, Thakar, Tucker, Uomoto, Berk, Vandenberg,
  Vogeley, Voges, Vogt, Walkowicz, Weinberg, West, White, Xu, Yanny, Yocum,
  York, Zehavi, Zibetti, \& Zucker}]{AdelmanMcCarthy:2006p4486}
Adelman-McCarthy, J.~K., Ag{\"u}eros, M.~A., Allam, S.~S., Anderson, K. S.~J.,
  Anderson, S.~F., Annis, J., Bahcall, N.~A., Baldry, I.~K., Barentine, J.~C.,
  Berlind, A., et al. 2006,
  The Astrophysical Journal Supplement Series, 162, 38

\bibitem[{Baldry {et~al.}(2006)Baldry, Balogh, Bower, Glazebrook, Nichol,
  Bamford, \& Budavari}]{Baldry:2006p103}
Baldry, I.~K., Balogh, M.~L., Bower, R.~G., Glazebrook, K., Nichol, R.~C.,
  Bamford, S.~P., \& Budavari, T. 2006, \mnras, 373, 469
  
  \bibitem[{Baldry {et~al.}(2004)Baldry, Glazebrook, Brinkmann, Ivezi{\'c},
  Lupton, Nichol, \& Szalay}]{Baldry:2004p43}
Baldry, I.~K., Glazebrook, K., Brinkmann, J., Ivezi{\'c}, {\v Z}., Lupton,
  R.~H., Nichol, R.~C., \& Szalay, A.~S. 2004, \apj, 600,
  681

\bibitem[{Bertin \& Arnouts(1996)}]{Bertin:1996p6150}
Bertin, E. \& Arnouts, S. 1996, Astronomy and Astrophysics Supplement, 117, 393

\bibitem[{Blanton {et~al.}(2003)Blanton, Hogg, Bahcall, Baldry, Brinkmann,
  Csabai, Eisenstein, Fukugita, Gunn, Ivezi{\'c}, Lamb, Lupton, Loveday, Munn,
  Nichol, Okamura, Schlegel, Shimasaku, Strauss, Vogeley, \&
  Weinberg}]{Blanton:2003p3068}
Blanton, M.~R., Hogg, D.~W., Bahcall, N.~A., Baldry, I.~K., Brinkmann, J.,
  Csabai, I., Eisenstein, D., Fukugita, M., Gunn, J.~E., Ivezi{\'c}, {\v Z}.,
  Lamb, D.~Q., Lupton, R.~H., Loveday, J., Munn, J.~A., Nichol, R.~C., Okamura,
  S., Schlegel, D.~J., Shimasaku, K., Strauss, M.~A., Vogeley, M.~S., \&
  Weinberg, D.~H. 2003, The Astrophysical Journal, 594, 186

\bibitem[{Blanton {et~al.}(2003)Blanton, Hogg, Bahcall, Brinkmann, Britton,
  Connolly, Csabai, Fukugita, Loveday, Meiksin, Munn, Nichol, Okamura, Quinn,
  Schneider, Shimasaku, Strauss, Tegmark, Vogeley, \&
  Weinberg}]{Blanton:2003p2277}
Blanton, M.~R., Hogg, D.~W., Bahcall, N.~A., Brinkmann, J., Britton, M.,
  Connolly, A.~J., Csabai, I., Fukugita, M., Loveday, J., Meiksin, A., Munn,
  J.~A., Nichol, R.~C., Okamura, S., Quinn, T., Schneider, D.~P., Shimasaku,
  K., Strauss, M.~A., Tegmark, M., Vogeley, M.~S., \& Weinberg, D.~H. 2003, The
  Astrophysical Journal, 592, 819

\bibitem[{Blanton {et~al.}(2005)Blanton, Schlegel, Strauss, Brinkmann,
  Finkbeiner, Fukugita, Gunn, Hogg, Ivezi{\'c}, Knapp, Lupton, Munn, Schneider,
  Tegmark, \& Zehavi}]{Blanton:2005p79}
Blanton, M.~R., Schlegel, D.~J., Strauss, M.~A., Brinkmann, J., Finkbeiner, D.,
  Fukugita, M., Gunn, J.~E., Hogg, D.~W., Ivezi{\'c}, {\v Z}., Knapp, G.~R.,
  Lupton, R.~H., Munn, J.~A., Schneider, D.~P., Tegmark, M., \& Zehavi, I.
  2005, The Astronomical Journal, 129, 2562
  
  \bibitem[{Blanton {et~al.}(2005{\natexlab{a}})Blanton, Eisenstein, Hogg,
  Schlegel, \& Brinkmann}]{Blanton:2005p2271}
Blanton, M.~R., Eisenstein, D., Hogg, D.~W., Schlegel, D.~J., \& Brinkmann, J.
  2005{\natexlab{a}}, The Astrophysical Journal, 629, 143

\bibitem[{Brinchmann {et~al.}(2004)Brinchmann, Charlot, White, Tremonti,
  Kauffmann, Heckman, \& Brinkmann}]{Brinchmann:2004p3060}
Brinchmann, J., Charlot, S., White, S. D.~M., Tremonti, C., Kauffmann, G.,
  Heckman, T., \& Brinkmann, J. 2004, Monthly Notices of the Royal Astronomical
  Society, 351, 1151

\bibitem[{Bruzual \& Charlot(2003)}]{Bruzual:2003p2874}
Bruzual, G. \& Charlot, S. 2003, Monthly Notices of the Royal Astronomical
  Society, 344, 1000

\bibitem[{Buta(1990)}]{Buta:1990p8843}
Buta, R. 1990, (Galactic models; Proceedings of the 4th Florida Workshop on
  Nonlinear Dynamics, 596, 58

\bibitem[{Buta \& Combes(1996)}]{Buta:1996p2930}
Buta, R. \& Combes, F. 1996, Fundamentals of Cosmic Physics, 17, 95

\bibitem[{Charlot \& Fall(2000)}]{Charlot:2000p6396}
Charlot, S. \& Fall, S.~M. 2000, The Astrophysical Journal, 539, 718

\bibitem[{de~Vaucouleurs(1963)}]{deVaucouleurs:1963p9392}
de~Vaucouleurs, G. 1963, Astrophysical Journal Supplement, 8, 31

\bibitem[{de~Vaucouleurs {et~al.}(1991)de~Vaucouleurs, de~Vaucouleurs, Corwin,
  Buta, Paturel, \& Fouque}]{deVaucouleurs:1991p4597}
de~Vaucouleurs, G., de~Vaucouleurs, A., Corwin, H.~G., Buta, R.~J., Paturel,
  G., \& Fouque, P. 1991, Volume 1-3

\bibitem[{Driver {et~al.}(2006)Driver, Allen, Graham, Cameron, Liske, Ellis,
  Cross, Propris, Phillipps, \& Couch}]{Driver:2006p17183}
Driver, S.~P., Allen, P.~D., Graham, A.~W., Cameron, E., Liske, J., Ellis,
  S.~C., Cross, N. J.~G., Propris, R.~D., Phillipps, S., \& Couch, W.~J. 2006,
  \mnras, 368, 414

\bibitem[{Elmegreen {et~al.}(2005)Elmegreen, Elmegreen, Rubin, \&
  Schaffer}]{Elmegreen:2005p117}
Elmegreen, D.~M., Elmegreen, B.~G., Rubin, D.~S., \& Schaffer, M.~A. 2005, The
  Astrophysical Journal, 631, 85

\bibitem[{Frei {et~al.}(1996)Frei, Guhathakurta, Gunn, \&
  Tyson}]{Frei:1996p4556}
Frei, Z., Guhathakurta, P., Gunn, J.~E., \& Tyson, J.~A. 1996, Astronomical
  Journal v.111, 111, 174

\bibitem[{Fukugita {et~al.}(2007)Fukugita, Nakamura, Okamura, Yasuda,
  Barentine, Brinkmann, Gunn, Harvanek, Ichikawa, Lupton, Schneider, Strauss,
  \& York}]{Fukugita:2007p3031}
Fukugita, M., Nakamura, O., Okamura, S., Yasuda, N., Barentine, J.~C.,
  Brinkmann, J., Gunn, J.~E., Harvanek, M., Ichikawa, T., Lupton, R.~H.,
  Schneider, D.~P., Strauss, M.~A., \& York, D.~G. 2007, The Astronomical
  Journal, 134, 579
  
 \bibitem[{Hogg {et~al.}(2004)Hogg, Blanton, Brinchmann, Eisenstein, Schlegel,
  Gunn, McKay, Rix, Bahcall, Brinkmann, \& Meiksin}]{Hogg:2004p294}
Hogg, D.~W., Blanton, M.~R., Brinchmann, J., Eisenstein, D.~J., Schlegel,
  D.~J., Gunn, J.~E., McKay, T.~A., Rix, H.-W., Bahcall, N.~A., Brinkmann, J.,
  \& Meiksin, A. 2004, The Astrophysical Journal, 601, L29


\bibitem[{Kauffmann {et~al.}(2003{\natexlab{a}})Kauffmann, Heckman, Tremonti,
  Brinchmann, Charlot, White, Ridgway, Brinkmann, Fukugita, Hall, Ivezi{\'c},
  Richards, \& Schneider}]{Kauffmann:2003p23}
Kauffmann, G., Heckman, T.~M., Tremonti, C., Brinchmann, J., Charlot, S.,
  White, S. D.~M., Ridgway, S.~E., Brinkmann, J., Fukugita, M., Hall, P.~B.,
  Ivezi{\'c}, {\v Z}., Richards, G.~T., \& Schneider, D.~P. 2003{\natexlab{a}},
  Monthly Notices of the Royal Astronomical Society, 346, 1055

\bibitem[{Kauffmann {et~al.}(2003{\natexlab{b}})Kauffmann, Heckman, White,
  Charlot, Tremonti, Brinchmann, Bruzual, Peng, Seibert, Bernardi, Blanton,
  Brinkmann, Castander, Cs{\'a}bai, Fukugita, Ivezic, Munn, Nichol,
  Padmanabhan, Thakar, Weinberg, \& York}]{Kauffmann:2003p97}
Kauffmann, G., Heckman, T.~M., White, S. D.~M., Charlot, S., Tremonti, C.,
  Brinchmann, J., Bruzual, G., Peng, E.~W., Seibert, M., Bernardi, M., Blanton,
  M., Brinkmann, J., Castander, F., Cs{\'a}bai, I., Fukugita, M., Ivezic, Z.,
  Munn, J.~A., Nichol, R.~C., Padmanabhan, N., Thakar, A.~R., Weinberg, D.~H.,
  \& York, D. 2003{\natexlab{b}}, Monthly Notice of the Royal Astronomical
  Society, 341, 33

\bibitem[{Kauffmann {et~al.}(2003{\natexlab{c}})Kauffmann, Heckman, White,
  Charlot, Tremonti, Peng, Seibert, Brinkmann, Nichol, SubbaRao, \&
  York}]{Kauffmann:2003p7199}
Kauffmann, G., Heckman, T.~M., White, S. D.~M., Charlot, S., Tremonti, C.,
  Peng, E.~W., Seibert, M., Brinkmann, J., Nichol, R.~C., SubbaRao, M., \&
  York, D. 2003{\natexlab{c}}, Monthly Notice of the Royal Astronomical
  Society, 341, 54
  
\bibitem[{Kewley {et~al.}(2001)Kewley, Dopita, Sutherland, Heisler, \&
  Trevena}]{Kewley:2001p21444}
Kewley, L.~J., Dopita, M.~A., Sutherland, R.~S., Heisler, C.~A., \& Trevena, J.
  2001, The Astrophysical Journal, 556, 121  

\bibitem[Nakamura et al.(2003)]{Nakamura:2003p18391} Nakamura, O., 
Fukugita, M., Yasuda, N., Loveday, J., Brinkmann, J., Schneider, D.~P., 
Shimasaku, K., \& SubbaRao, M.\ 2003, \aj, 125, 1682 


\bibitem[{Petrosian(1976)}]{Petrosian:1976p6531}
Petrosian, V. 1976, Astrophysical Journal, 209, L1

\bibitem[{Sandage(1961)}]{Sandage:1961p5247}
Sandage, A. 1961, Washington: Carnegie Institution

\bibitem[{Sandage \& Bedke(1994)}]{Sandage:1994p4888}
Sandage, A. \& Bedke, J. 1994, Washington

\bibitem[{Sandage \& Tammann(1981)}]{Sandage:1981p5045}
Sandage, A. \& Tammann, G.~A. 1981, Washington: Carnegie Institution

\bibitem[Shen et al.(2003)]{Shen:2003p70} Shen, S., Mo, H.~J., 
White, S.~D.~M., Blanton, M.~R., Kauffmann, G., Voges, W., Brinkmann, J., 
\& Csabai, I.\ 2003, \mnras, 343, 978 

\bibitem[{Stoughton {et~al.}(2002)Stoughton, Lupton, Bernardi, Blanton, Burles,
  Castander, Connolly, Eisenstein, Frieman, Hennessy, Hindsley, Ivezi{\'c},
  Kent, Kunszt, Lee, Meiksin, Munn, Newberg, Nichol, Nicinski, Pier, Richards,
  Richmond, Schlegel, Smith, Strauss, SubbaRao, Szalay, Thakar, Tucker, Berk,
  Yanny, Adelman, Anderson, Anderson, Annis, Bahcall, Bakken, Bartelmann,
  Bastian, Bauer, Berman, B{\"o}hringer, Boroski, Bracker, Briegel, Briggs,
  Brinkmann, Brunner, Carey, Carr, Chen, Christian, Colestock, Crocker, Csabai,
  Czarapata, Dalcanton, Davidsen, Davis, Dehnen, Dodelson, Doi, Dombeck,
  Donahue, Ellman, Elms, Evans, Eyer, Fan, Federwitz, Friedman, Fukugita, Gal,
  Gillespie, Glazebrook, Gray, Grebel, Greenawalt, Greene, Gunn, de~Haas,
  Haiman, Haldeman, Hall, Hamabe, Hansen, Harris, Harris, Harvanek, Hawley,
  Hayes, Heckman, Helmi, Henden, Hogan, Hogg, Holmgren, Holtzman, Huang, Hull,
  Ichikawa, Ichikawa, Johnston, Kauffmann, Kim, Kimball, Kinney, Klaene,
  Kleinman, Klypin, Knapp, Korienek, Krolik, Kron, Krzesi{\'n}ski, Lamb, Leger,
  Limmongkol, Lindenmeyer, Long, Loomis, Loveday, MacKinnon, Mannery, Mantsch,
  Margon, McGehee, McKay, McLean, Menou, Merelli, Mo, Monet, Nakamura,
  Narayanan, Nash, Neilsen, Newman, Nitta, Odenkirchen, Okada, Okamura,
  Ostriker, Owen, Pauls, Peoples, Peterson, Petravick, Pope, Pordes, Postman,
  Prosapio, Quinn, Rechenmacher, Rivetta, Rix, Rockosi, Rosner, Ruthmansdorfer,
  Sandford, Schneider, Scranton, Sekiguchi, Sergey, Sheth, Shimasaku, Smee,
  Snedden, Stebbins, Stubbs, Szapudi, Szkody, Szokoly, Tabachnik, Tsvetanov,
  Uomoto, Vogeley, Voges, Waddell, Walterbos, i~Wang, Watanabe, Weinberg,
  White, White, Wilhite, Wolfe, Yasuda, York, Zehavi, \&
  Zheng}]{Stoughton:2002p1611}
Stoughton, C., Lupton, R.~H., Bernardi, M., Blanton, M.~R., Burles, S.,
  Castander, F.~J., Connolly, A.~J., Eisenstein, D.~J., Frieman, J.~A.,
  Hennessy, G.~S., et al. 2002, The Astronomical Journal, 123, 485

\bibitem[{Strauss {et~al.}(2002)Strauss, Weinberg, Lupton, Narayanan, Annis,
  Bernardi, Blanton, Burles, Connolly, Dalcanton, Doi, Eisenstein, Frieman,
  Fukugita, Gunn, Ivezi{\'c}, Kent, Kim, Knapp, Kron, Munn, Newberg, Nichol,
  Okamura, Quinn, Richmond, Schlegel, Shimasaku, SubbaRao, Szalay, Berk,
  Vogeley, Yanny, Yasuda, York, \& Zehavi}]{Strauss:2002p6172}
Strauss, M.~A., Weinberg, D.~H., Lupton, R.~H., Narayanan, V.~K., Annis, J.,
  Bernardi, M., Blanton, M., Burles, S., Connolly, A.~J., Dalcanton, J. et al. 2002, The
  Astronomical Journal, 124, 1810

\bibitem[{van~den Bergh {et~al.}(2002)van~den Bergh, Abraham, Whyte,
  Merrifield, Eskridge, Frogel, \& Pogge}]{vandenBergh:2002p9582}
van~den Bergh, S., Abraham, R.~G., Whyte, L.~F., Merrifield, M.~R., Eskridge,
  P.~B., Frogel, J.~A., \& Pogge, R. 2002, \aj, 123, 2913

\bibitem[{Yang {et~al.}(2007)Yang, Mo, van~den Bosch, Pasquali, Li, \&
  Barden}]{Yang:2007p19054}
Yang, X., Mo, H.~J., van~den Bosch, F.~C., Pasquali, A., Li, C., \& Barden, M.
  2007, \apj, 671, 153
  
\bibitem[{York {et~al.}(2000)York, Adelman, Anderson, Anderson, Annis, Bahcall,
  Bakken, Barkhouser, Bastian, Berman, Boroski, Bracker, Briegel, Briggs,
  Brinkmann, Brunner, Burles, Carey, Carr, Castander, Chen, Colestock,
  Connolly, Crocker, Csabai, Czarapata, Davis, Doi, Dombeck, Eisenstein,
  Ellman, Elms, Evans, Fan, Federwitz, Fiscelli, Friedman, Frieman, Fukugita,
  Gillespie, Gunn, Gurbani, de~Haas, Haldeman, Harris, Hayes, Heckman,
  Hennessy, Hindsley, Holm, Holmgren, Hao Huang, Hull, Husby, Ichikawa,
  Ichikawa, Ivezi{\'c}, Kent, Kim, Kinney, Klaene, Kleinman, Kleinman, Knapp,
  Korienek, Kron, Kunszt, Lamb, Lee, Leger, Limmongkol, Lindenmeyer, Long,
  Loomis, Loveday, Lucinio, Lupton, MacKinnon, Mannery, Mantsch, Margon,
  McGehee, McKay, Meiksin, Merelli, Monet, Munn, Narayanan, Nash, Neilsen,
  Neswold, Newberg, Nichol, Nicinski, Nonino, Okada, Okamura, Ostriker, Owen,
  Pauls, Peoples, Peterson, Petravick, Pier, Pope, Pordes, Prosapio,
  Rechenmacher, Quinn, Richards, Richmond, Rivetta, Rockosi, Ruthmansdorfer,
  Sandford, Schlegel, Schneider, Sekiguchi, Sergey, Shimasaku, Siegmund, Smee,
  Smith, Snedden, Stone, Stoughton, Strauss, Stubbs, SubbaRao, Szalay, Szapudi,
  Szokoly, Thakar, Tremonti, Tucker, Uomoto, Berk, Vogeley, Waddell, i~Wang,
  Watanabe, Weinberg, Yanny, \& Yasuda}]{York:2000p3192}
York, D.~G., Adelman, J., Anderson, J.~E., Anderson, S.~F., Annis, J., Bahcall,
  N.~A., Bakken, J.~A., Barkhouser, R., Bastian, S., Berman, E., et al.  2000, The Astronomical Journal, 120, 1579

\end{thebibliography}
\end{document}